\documentclass[12pt, letterpaper]{article}

\usepackage[includeheadfoot, marginratio={1:1,2:3}, width=412pt, height=688pt,]{
    geometry
}
\usepackage{tikz}
\usepackage{amsmath}
\usepackage{amsfonts}
\usepackage{amsthm}
\usepackage{amssymb}
\usepackage{graphicx}
\usepackage{booktabs}
\usepackage{empheq}
\usepackage{paralist}
\usepackage{cite}
\usepackage[hidelinks]{hyperref}
\usepackage{xcolor}
\usepackage{tensor}
\usepackage{adjustbox}
\usepackage{subcaption}
\usepackage{braket}
\usepackage{tikz-cd}
\usepackage{yfonts}
\usepackage{enumitem}
\usepackage{physics}
\usepackage{mdframed}

\usetikzlibrary{decorations.pathreplacing, calc}
\newcommand{\nc}{\newcommand}
\nc{\lb}{\llbracket}
\nc{\rb}{\rrbracket}
\nc{\gl}{\llbracket}
\nc{\gr}{\rrbracket}
\nc{\bbR}{\mathbb{R}}
\nc{\bbC}{\mathbb{C}}
\nc{\bbZ}{\mathbb{Z}}
\nc{\cA}{\mathcal{A}}
\nc{\cO}{\mathcal{O}}
\nc{\cS}{\mathcal{S}}
\nc{\cM}{\mathcal{M}}
\nc{\cE}{\mathcal{E}}
\nc{\cT}{\mathcal{T}}
\nc{\cX}{\mathcal{X}}
\nc{\cQ}{\mathcal{Q}}
\nc{\cD}{\mathcal{D}}
\nc{\cC}{\mathcal{C}}
\nc{\cG}{\mathcal{G}}
\nc{\cU}{\mathcal{U}}
\nc{\cF}{\mathcal{F}}
\nc{\cI}{\mathcal{I}}
\nc{\cL}{\mathcal{L}}
\nc{\pd}{\partial}
\nc{\la}{\lambda}
\newcommand{\beq}{\begin{equation}}
\newcommand{\eeq}{\end{equation}}
\nc{\del}{\partial}
\nc{\tri}{\hspace{-31pt}\vartriangle\hspace{-31pt}}
\nc{\blacktri}{\blacktriangle}
\nc{\eq}[1]{\begin{equation}\begin{split}#1\end{split}\end{equation}}
\nc{\ul}{\underline}
\nc{\ov}{\overline}
\nc{\fa}{\hat}
\nc{\fb}{\MakeUppercase}
\nc{\fc}{\tilde }
\nc{\Lie}{{\cal L}}
\nc{\lambdabar}{{\mkern0.75mu\mathchar '26\mkern -9.75mu\lambda}}
\newcommand{\R}{\mathbb{R}}

\newcommand{\C}{\mathbb{C}}

\renewcommand{\L}{\mathcal{L}}

\newcommand{\A}{\mathcal{A}}
\newcommand{\G}{\mathcal{G}}
\newcommand{\N}{\mathbb{N}}
\newcommand{\Z}{\mathbb{Z}}

\newcommand{\F}{\mathcal{F}}

\newcommand{\eps}{\epsilon}
\newcommand{\rd}{d}
\newcommand{\bw}{\bar{w}}

\newmuskip\pFqmuskip

\newcommand*{\pFq}[7][8]{ \begingroup % only local assignments
\pFqmuskip=#1mu\relax \mathchardef\normalcomma=\mathcode`,
\mathcode`\,=\string"8000
\begingroup\lccode`\~=`\, \lowercase{\endgroup\let~}\pFqcomma
{}_{#2}{#3}_{#4}{\left[\left.\genfrac..{0pt}{}{#5}{#6}\right|#7\right]} \endgroup
}
\newcommand{\pFqcomma}{{\normalcomma}\mskip\pFqmuskip}

\allowdisplaybreaks[2]

\begin{document}
\vspace*{1cm}
    \begin{center}
        {\LARGE Resonance and Differential Reduction\\[.3cm] of Feynman Integrals} 

        \vspace{.6cm}
    \end{center}

    \vspace{0.35cm}
    \begin{center}
        Ruth Britto,$^a$ Thomas W.~Grimm,$^{b,c,d}$ Arno~Hoefnagels\,$^b$
    \end{center}

\vspace{.5cm}
\begin{center} 
\vspace{0.25cm} 
\emph{
$^a$ School of Mathematics and Hamilton Mathematics Institute,\\
Trinity College, Dublin 2, Ireland\\
\vspace{0.25cm} 
}
$^b$\,
\emph{Institute for Theoretical Physics, Utrecht University,
\\
Princetonplein 5, 3584 CC Utrecht, 
The Netherlands } \\
\vspace{0.25cm} 

$^c$\,
\emph{Center of Mathematical Sciences and Applications %\& Jefferson Physical Laboratory
Harvard University, Cambridge, MA 02138, USA } \\
\vspace{0.25cm} 

$^d$\,
\emph{Jefferson Physical Laboratory,\\
Harvard University, Cambridge, MA 02138, USA}

\end{center}

    %%%%%%%%%%%%%%%%%%%%%%%%%%%%%%%%%%%%%%%%%%%%%%%
    %%%%%%%%%%%%%%%%%%%%%%%%%%%%%%%%%%%%%%%%%%%%%%%
    %%%%%%%%%%%%%%%%%%%%%%%%%%%%%%%%%%%%%%%%%%%%%%%
    %%%%%%%%%%%%%%%%%%%%%%%%%%%%%%%%%%%%%%%%%%%%%%%
    %%%%%%%%%%%%%%%%%%%%%%%%%%%%%%%%%%%%%%%%%%%%%%%
    %%%%%%%%%%%%%%%%%%%%%%%%%%%%%%%%%%%%%%%%%%%%%%%
    %%%%%%%%%%%%%%%%%%%%%%%%%%%%%%%%%%%%%%%%%%%%%%%
    %%%%%%%%%%%%%%%%%%%%%%%%%%%%%%%%%%%%%%%%%%%%%%%

\vspace*{2cm}
    \begin{abstract}
        \noindent
        Feynman integrals may be viewed as generalized hypergeometric functions, and specifically as solutions of GKZ systems of partial differential equations that typically exhibit resonance. 
        Resonance is a type of non-genericity implying reducibility to subsystems.
        We use this resonance to construct reduction operators, which are differential operators that can contract edges of Feynman graphs. 
        Correspondingly, their action is naturally compatible with cuts of Feynman graphs.
        Reduction operators may be used to close the system of differential equations for a given integral. The remaining GKZ data lead to algebraic relations identifying a smaller system that is fully reduced to master integrals. We develop the construction for one-loop, sunrise and banana graphs and discuss restrictions to physical kinematics. 
        While reduction operators can generally shift both propagator powers and spacetime dimension, certain combinations isolate a pure dimension shift together with contraction of a chosen edge.

    \end{abstract}

    \clearpage

    \tableofcontents

    %%%%%%%%%%%%%%%%%%%%%%%%%%%%%%%%%%%%%%%%%%%%%%%
    %%%%%%%%%%%%%%%%%%%%%%%%%%%%%%%%%%%%%%%%%%%%%%%
    %%%%%%%%%%%%%%%%%%%%%%%%%%%%%%%%%%%%%%%%%%%%%%%
    %%%%%%%%%%%%%%%%%%%%%%%%%%%%%%%%%%%%%%%%%%%%%%%
    %%%%%%%%%%%%%%%%%%%%%%%%%%%%%%%%%%%%%%%%%%%%%%%
    %%%%%%%%%%%%%%%%%%%%%%%%%%%%%%%%%%%%%%%%%%%%%%%
    %%%%%%%%%%%%%%%%%%%%%%%%%%%%%%%%%%%%%%%%%%%%%%%
    %%%%%%%%%%%%%%%%%%%%%%%%%%%%%%%%%%%%%%%%%%%%%%%

    \newpage

    \parskip=.2cm

\section{Introduction}

Feynman integrals are central objects in perturbative quantum field theory.  They encode scattering amplitudes and correlation functions as multidimensional integrals that depend on external kinematics, masses, propagator powers, and the dimension of spacetime.
A large part of modern multi-loop technology is built on integration-by-parts (IBP) identities and on differential equations for master integrals (see  e.g.~\cite{Vanhove:2021osr,Weinzierl:2022eaz,abreu_sagex_2022,Blumlein:2022qci,Badger:2023eqz}).  These methods have made many high-precision computations possible, but they also expose a persistent structural problem: the systems of differential equations obtained for integrals can be large, and the function spaces needed to describe their solutions can be difficult to organize.  This motivates looking for mechanisms that explain when the system of differential equations governing a physical integral is smaller than a generic system with the same combinatorial data.

A useful starting point is the Lee-Pomeransky representation \cite{Lee:2013hzt}.  It rewrites a scalar Feynman integral as an Euler-Mellin integral over Feynman parameters, with the integrand determined by the graph polynomial $\mathcal G=\mathcal U+\mathcal F$ built from the Symanzik polynomials \cite{Bogner:2010kv,Weinzierl:2022eaz}.  This representation makes graph operations particularly transparent: for instance, contracting an edge corresponds to setting the associated Feynman parameter to zero in the graph polynomial.  
It also places Feynman integrals in the context of GKZ, or $A$-hypergeometric, systems. 

Building on the foundational work of Gelfand, Kapranov and Zelevinski\u\i~\cite{Gelfand1989,gelfand_generalized_1990,gelfand_discriminants_1994,saito2000grobner}, the coefficients of $\mathcal G$ can be treated as independent variables, so that a Feynman integral is realized as a solution of a GKZ system   \cite{Vanhove:2018mto,delaCruz:2019skx,klausen_hypergeometric_2020,ananthanarayan_feyngkz_2023}.    
In the generic case, GKZ systems form regular holonomic $D$-modules whose rank and monodromy are controlled by a matrix $A$ together with a parameter vector $\nu$ \cite{Gelfand1989,gelfand_generalized_1990,gelfand_discriminants_1994,saito2000grobner}.  For Feynman integrals, most entries of the parameter vector are associated to the powers of the propagators. Because these powers often take integer values in physics, we leave the realm of the generic case and typically find some degree of \emph{resonance}.

The central point of this work is that since physically relevant parameter values are frequently resonant,  
the GKZ system can become reducible: its total holonomic rank need not drop,\footnote{Conditions for the rank to drop have been classified in \cite{tellander_cohenmacaulay_2023,Walther:2022pli,Michaelsen:2025fzz}.} but it contains subsystems of differential equations of smaller rank.  Reducibility and parameter shifts in $A$-hypergeometric systems are well studied mathematically \cite{Saito2001,beukers_irreducibility_2011,reichelt_algebraic_2021,dwork_generalized_1990,caloro_ahypergeometric_2023}.  In recent work on cosmological correlators, this resonant structure was used to construct differential operators that isolate smaller systems and relate different contributions \cite{Grimm:2024tbg,Grimm:2025zhv}.  
Here we adapt this resonance mechanism to Feynman integrals, where certain resonant face subsystems admit a direct interpretation in terms of simpler graphs.

Concretely, we construct differential operators from the Euler and toric operators of the resonant GKZ system. We call them \emph{reduction operators}, following \cite{Grimm:2024tbg,Grimm:2025zhv}, 
because they realize a reduction of the GKZ system to a subsystem.
Reduction operators act within the system of differential equations and relate solutions associated with different graph sectors. 
Resonance is a property of faces of the convex hull of the columns of $A$ in relation to the parameter vector $\nu$. For Feynman integrals, some of the faces correspond to edges of the graph. We show that if one of these faces is resonant, then we can construct a reduction operator that implements edge contraction: the action of a reduction operator maps the integral to the GKZ subsystem of the contracted graph, possibly with shifted parameters.  The fact that differential operators realizing edge contractions exist for Feynman graphs appeared as early as 1977 \cite{Kashiwara:1977yy}. We now offer a route for constructing them directly.

Our approach can be placed into the broader setting of periods, twisted cohomology, and $D$-modules, where linear differential equations and their solution spaces can be studied algebraically. On the period side, Feynman integrals, and in particular their cut or maximal-cut restrictions, can be regarded as solutions to Picard--Fuchs/Gauss--Manin differential equations. These can often be obtained from GKZ systems \cite{hosono_mirror_1995,hosono_gkzCY_1996,hosono_gkzapp_1996} which have been used to study Feynman integrals using geometric spaces in~\cite{Bloch:2014qca,bloch_local_2017,muller-stach_picardfuchs_2014,klemm_lloop_2020,bonisch_analytic_2021,bonisch_feynman_2022,lairez_algorithms_2023}. In the present setting, we focus on exploring the resonance structure in the GKZ data, keeping boundary terms, and avoid restricting to cut integrals. A particularly close predecessor to our approach is the differential reduction program for hypergeometric representations of Feynman integrals  \cite{Bytev:2009kb,Kalmykov:2011yy,Kalmykov:2012rr,Bytev:2022tav,Feng:2024xio}.
While close in spirit, we work directly in the GKZ coefficient space and attach reduction operators to resonant faces.
Similar ideas are found in certain works on
parametric annihilators~\cite{bitoun_feynman_2018,Bertolini:2025zud}, Pfaffian systems and their restrictions~\cite{chestnov_macaulay_2022,chestnov_restrictions_2023}, and recent $D$-module and Griffiths--Dwork/Macaulay constructions of Feynman differential equations and twisted Picard--Fuchs operators~\cite{agostini_vector_2024,chestnov_differential_2025,vanhove_picardfuchs_2026}.

The framework suggested in this work has several advantages.  Reduction operators are often of lower order than the toric GKZ operators and lead to simpler differential equations. Their application yields parameter shifts and edge contractions and organizes the integrals into a \textit{systematic reduction ladder} with a graphical interpretation. This ladder has a natural compatibility with cuts: on cut contours the relevant boundary terms are absent \cite{Britto_2025}, so cutting and contraction appear as compatible projections of the same differential system. Furthermore, the Euler and toric equations that are not needed in constructing a closing first-order system still carry information and give algebraic relations among the elements of the reduction ladder.  In addition, resonances involving the dimensional parameter lead to natural dimension-shift operators.  The main technical challenge that remains is the restriction from the enlarged GKZ coefficient space to the physical locus, where the coefficients of the graph polynomial are expressed in terms of masses and kinematic invariants.  We perform this restriction explicitly in the one-loop examples, but a systematic treatment belongs to the general problem of restricting Pfaffian and $D$-module systems \cite{FernandezFernandezWalther2011,chestnov_macaulay_2022,chestnov_restrictions_2023,Fevola:2023fzn}.

We develop the general framework and then illustrate it through explicit examples.
In section~\ref{sec:Feyn_GKZ_review}, we review Feynman integrals in the Lee-Pomeransky representation and discuss the associated GKZ systems. 
The conceptual core of the paper is section~\ref{GKZresonance-reduction}. There, we explain how resonance is related to reducibility, construct the corresponding reduction operators, and analyze their action on Feynman integrals associated with a graph $G$.
Sections~\ref{sec:bubble} and~\ref{sec:triang_example} show how the bubble and triangle diagrams realize this construction explicitly. For generic kinematics, their resonant edge faces give rise to first-order reduction operators, which reproduce the expected contractions to tadpole and bubble sectors. In section~\ref{sec:generic1loop}, we treat arbitrary massive one-loop graphs at generic kinematics. We derive a uniform formula for the edge reduction operators and describe its restriction to physical variables. We also discuss how the same operators act on cut integrals, where contraction and cutting operations become compatible ways of projecting the system of differential equations.  Beyond one loop, in section~\ref{sec:higherloops} we examine sunrise and banana integrals as first tests of the method.  In section~\ref{sec:dimension}, we discuss additional resonances that occur in integer dimensions and show how some of the associated operators can be interpreted through dimension shifts.
Finally, section~\ref{sec:conclusions} collects further directions and possible extensions to more general graph families.
Appendices A, B, and C respectively contain a short proof describing a condition for facets to be resonant, explicit formulas for tadpole and bubble integrals, and an illustration of resonance and reduction operators in the GKZ system for the hypergeometric function ${}_2F_1$.

\section{Feynman integrals and GKZ systems} \label{sec:Feyn_GKZ_review}

This section first provides a brief overview of Feynman integrals and graph polynomials in section~\ref{intro_Feyn_graph}, and then summarizes several basic facts about GKZ systems in section~\ref{sec:GKZ_basics}. The purpose of this review is to also establish our notation and conventions; readers already familiar with these topics may skim the following material quickly.

\subsection{Feynman integrals and graph polynomials} \label{intro_Feyn_graph}

In this section we first recap some basics on Feynman integrals. Much of this background can be found in various reviews \cite{Bogner:2010kv,Weinzierl:2022eaz}. 
Scalar Feynman integrals are relevant in the study of scattering amplitudes in momentum spaces when expanding in a small coupling constant. Such an expansion leads to the well-known Feynman graphs, and we will focus on a single such graph $G$ to present the associated Feynman integral. We denote the spacetime dimension by $D$ and consider a graph 
with $L$ loops and $N_e$ internal edges. In general, there can be an exchange of a mass $m_e$ along each edge, and the external legs carry momenta $p_k$.
The standard Feynman integral then takes the form 
\beq  \label{Feynman1}
   I^D_G(p,m;\nu) = \int \prod_{r=1}^L \frac{d^D k_r}{i\pi^{D/2}} \prod_{e=1}^{N_e} \frac{1}{(-q_e^2+ m_e^2)^{\nu_e}}\ ,
\eeq
where $q_e$ are linear combinations of the external momenta $p_k$ and internal momenta $k_r$ associated to the edge with label $e$ (see e.g.~\cite{Bogner:2010kv} for details). Note that we have parameters $\nu_e$ associated to the edges, which allow for non-unit powers of propagators.  Without subscripts, the expressions $p,m,\nu$ represent the full list of kinematic quantities and propagator powers. The dimension $D$ and the parameters $\nu$ play  an important role in the following and we will not fix them to any particular value. 
However, for physical scalar Feynman integrals, each $\nu_e$ will take a positive integer value.

A convenient way to rewrite the Feynman integrals \eqref{Feynman1} is the Lee-Pomeransky form~\cite{Lee:2013hzt}. This involves integrating over the edge parameters $x_e,\ e=1,...,N_e$ as 
\begin{equation}\label{eq:LPrep}
  I^D_{G}(p,m;\nu) = \frac{  \Gamma \left( \frac{D}{2} \right) }{ 
    \Gamma\left( \frac{(L+1)D}{2}-\sum_{e} \nu_e\right)
    \prod_{e = 1}^{ N_e } \Gamma( \nu_{e} ) }
    \int_{\R_+^{N_e}} d^{N_e}x_e\frac{\prod_{e=1}^{ N_e } x_{e}^{\nu_e-1  }}{\left[ \G(p,m,x) \right]^{  \frac{D}{2} }}\ . 
\end{equation}
where $\R_+$ is the positive orthant $x_e \geq 0$, and $\G(p,m,x)$ is the Lee-Pomeransky polynomial.
The goal of section \ref{sec:GKZ_basics} is to view the parametric integral as being associated to a GKZ system.

The Lee-Pomeransky polynomial $\G(p,m,x)$ is given by 
\beq \label{cGdef}
  \G(p,m,x)=\mathcal{U}(x)+\F(p,m,x)\ ,
\eeq 
where $\mathcal{U}$ and $\F$ are the usual Symanzik polynomials associated to the Feynman graph and take the form
\begin{align} \label{def-UF} \begin{split}
\mathcal{U}(x)
 & = 
 \sum_{T\in {\mathcal T}_1} 
     \prod_{e \notin T} x_e \,,
 \\
\F(p,m,x)
 & = \mathcal{U}(x) \sum_{e=1}^{N_e} x_e m_e^2 
 - \sum\limits_{(T_1,T_2)\in {\mathcal T}_2} 
     \left( \prod_{e \notin (T_1,T_2)} x_e \right) 
p_{(T_1,T_2)}^2 \,,
\end{split}\end{align}
where 
${\mathcal T}_1$ is the set of spanning trees of the Feynman graph, and ${\mathcal T}_2$ is the set of spanning 2-forests with connected components $T_1$ and $T_2$, in which $p_{(T_1,T_2)}$ is the momentum flowing from one connected component to the other. To make the dependence on the external momenta explicit, we denote by $P_{T_i}$ the set of external momenta attached to $T_i$, and  $p^2_{(T_1,T_2)} = - \sum_{p_i \in P_{T_1}} \sum_{p_j \in P_{T_2}}  p_i\cdot p_j$. Taken together, we can systematically read off  the Lee-Pomeransky polynomial $\cG(p,m,x)$ for any connected graph $G$.

\paragraph{Contracting an edge.} The Symanzik polynomials $\mathcal{U}$, $\cF$ have a number of important properties. First, they are homogeneous in the parameters $x_e$ of degree $L$ and $L+1$, respectively. Moreover, $\mathcal{U}$ is of degree 1 in each individual parameter $x_e$, while $\cF$ is of degree 2 in $x_e$ if $m_e \neq 0$, and degree 1 otherwise.
Second, the coefficients of $\cF$ are linear combinations of the kinematic invariants. This dependence will label the physical locus in a larger parameter space in section \ref{sec:GKZ_basics}.   
Third, they follow simple rules when two Feynman integrals are related to each other by edge contractions of their associated graphs. Let us denote by $G/e$ the graph obtained from $G$ by contracting the edge $e$. That is, $G/e$ is obtained from $G$ by deleting $e$ and identifying its two endpoints.
Contracting an edge $e$ of a graph $G$ has a natural effect on the graph polynomials, simply by setting $x_e$ to zero as  \cite{Bogner:2010kv,brown_feynman_2017} 
\begin{equation}\label{eq:UFG-edge-contraction}
    {\mathcal U}_{G/e} = \left.{\mathcal U}_G\right|_{x_e=0}\,, \qquad   {\mathcal F}_{G/e} = \left.{\mathcal F}_G\right|_{x_e=0}\,,\qquad \cG_{G/e} = \cG_G|_{x_e=0}\ .
\end{equation}

\paragraph{Cuts of Feynman integrals.}  
Feynman integrals can be generalized to allow a subset $J$ of its edges to be {\em cut}. Various definitions of cut integrals have been introduced for different purposes, such as computing discontinuities or monodromies, or for solving systems of differential equations.\footnote{See \cite{Britto_2025} for a review.} Their common element is that the Feynman integral is modified so that the momenta flowing through the cut edges are on-shell, i.e.\ that $q_e^2 = m_e^2$ for $e \in J$. In the integral (\ref{Feynman1}), the on-shell condition can be implemented by replacing the factor involving $e$ with the delta function $\delta(q_e^2-m_e^2)$ \cite{Cutkosky:1960sp}. In the parametric form of equation (\ref{eq:LPrep}), where there is no direct way to implement on-shell conditions,  we write 
\begin{equation}\label{eq:parametric-cuts}
 \mathcal{C}_J I^D_{G}(p,m;\nu) = \frac{  \Gamma \left(D/2\right) }{ 
    \Gamma\left( (L+1)D/2-\sum_{e} \nu_e\right)
    \prod_{e = 1}^{ N_e } \Gamma(\nu_{e}) }
    \int_{\Gamma_J} d^{N_e}x_e\frac{\prod_{e=1}^{ N_e } x_{e}^{\nu_e-1  }}{\left[ \G(p,m,x) \right]^{  D/2 }}\ , 
\end{equation}
where the integrand is the same as in equation (\ref{eq:LPrep}), but 
the integration contour is replaced with a new $\Gamma_J$ whose boundary $\partial \Gamma_J$ is constrained as follows \cite{britto_generalized_2023}:
\begin{equation}\label{eq:Gamma-contours}
    \partial \Gamma_J \subset \{ \mathcal{G}=0\} \cup \bigcup_{e \notin J} \{x_e=0\}  \,.
\end{equation} 

\subsection{Basics of GKZ systems for Feynman integrals   \label{sec:GKZ_basics}}

In this section we give a brief introduction to GKZ systems. We will recall the integrals that can be treated within this framework and give the
associated GKZ differential equations. We will not consider the most generic GKZ systems, instead adapting the general framework to the needs of Feynman integrals.

\paragraph{GKZ integrals.}
GKZ systems can be defined for more general integrals than are considered here,\footnote{Two common generalizations are increasing the number of polynomials in the integrand, as well as allowing for the exponential of a polynomial \cite{Gelfand1989,gelfand_generalized_1990,gelfand_discriminants_1994,saito2000grobner}.} but to lighten the exposition, we will only consider an integral of the form 
\begin{equation} \label{eq:gengkzint}
        I_{\cA}(z;\nu)= \kappa(\nu)\,\int_{\Gamma} \rd^{N_e}x\;   
  \frac{\prod_{e=1}^{ N_e } x_{e}^{\nu_e-1  }}{\cG(z,x)^{\nu_0}}\, ,
\end{equation}
where $\Gamma$ is an integration chain, $x_{e}$ are  integration
variables, and $\kappa(\nu)$ is a prefactor depending only on $\nu$ that is included for later convenience.\footnote{\label{footnote:gammas}Integrals of the form \eqref{eq:gengkzint} are referred to as Euler-Mellin integrals. In order for an Euler-Mellin integral to be an exact solution to the GKZ system shown below, it should be regularized such that $\kappa(\nu)$ contains certain gamma functions \cite{berkesch2013eulermellinintegralsahypergeometricfunctions}. The Lee-Pomeransky representation of \eqref{eq:LPrep} is suitably regularized.} 
Note that we allow complex values of $\nu_e$ and $\nu_0$.
The function $\cG(z,x)$ is a polynomial in $x_e$ with powers listed in  a matrix $\cA$ and coefficients being complex variables $z_i$. Any further condition on the $z_i$ is called a {\em restriction}.
To set up the matrix $\A$, let us write 
\beq\label{eq:poly} \cG(z,x)=
\sum_{i=1}^N z_i \prod_{e=1}^{N_e} x_e^{(b_i)_e}\, , 
\eeq 
where the $b_i$ are $N_e$-dimensional integer vectors determining the exponents of $x_e$ in each term of $\cG$. We will lift these to $N_e+1$ dimensional vectors $a_i$ and combine them into a matrix $\cA$ as
\begin{equation} \label{cA-a-form}
    \cA =\begin{pmatrix}
        1 & 1 & \cdots & 1\\
        b_1 & b_2& \cdots &b_N
    \end{pmatrix} := (a_1,\ldots,a_N)\ .
\end{equation}
Note that we index these vectors \emph{starting from zero}, such that $(a_i)_0=1$ and $(a_i)_e$ corresponds to a power of $x_e$ in $p$.
The property that all entries of the  first row of $\mathcal{A}$ are equal to  $1$ is usually referred to as $\mathcal{A}$ being homogenized. 

\paragraph{GKZ differential equations -- toric equations.}     A GKZ system is formulated in terms of the matrix $\A$ and the parameter vector $\nu$. It is a system of  differential equations that are split into two classes. First, there are the \textit{toric equations}, which are given by 
\beq \label{toric_GKZ} \L_{u,v}f(z;\nu)=0\ ,\qquad
    \L_{u,v}\coloneqq \partial^u-\partial^v\ ,
    \eeq 
where $f(z;\nu)$ is a $z_i$-dependent solution at parameter $\nu$, and we have introduced 
the notation 
\beq \label{partial-u}
  \partial^u = \prod_{i=1}^N \partial_i^{u_i}\,.
\eeq
To
define the toric operators $\L_{u,v}$ we have to find vectors $u$ and $v$ in $\N
    ^{N}$ satisfying $\A u=\A v$, which can be done systematically.
It is useful to interpret \eqref{toric_GKZ} as giving rise
    to equivalence relations on differential operators
    on some solution space. We introduce the symbol $\simeq$ if we \textit{only} want to impose the relations for toric operators $\L_{u,v}$, i.e.~we write $\L_{u,v}\simeq 0$.

\paragraph{GKZ differential equations -- Euler equations and inhomogeneity.}  
Second, there are \textit{Euler
    equations} expressed in terms of the Euler operators which are defined as follows: 
\begin{equation} \label{def-Euler}
    \cE_{\underline{e}} = \sum_{i=1}^N (a_i)_{\underline{e}}\theta_i\, , \qquad \underline{e} = 0,1,...,N_e\ ,
\end{equation}
where $\theta_i \coloneqq z_i \partial_i$. Note that $(a_i)_{\underline{e}}$ corresponds to the element in the $(e+1)$-st row and $i$-th column of $\mathcal{A}$. There is an Euler operator $\cE_e$ for each $e = 1,\ldots ,N_e$, as well as an additional one, $\cE_0 = \sum_i \theta_i$. To describe the Euler equations,
we first observe that they can be verified explicitly on the integral in \eqref{eq:gengkzint}, but the verification involves integration by parts and relies on the assumption that boundary terms vanish.\footnote{It is possible for the boundary contribution to be suppressed by the prefactor $\kappa(\nu)$ to preserve the Euler relations, as mentioned in footnote~\ref{footnote:gammas}. Since these boundary terms are essential to our results, we set aside the prefactors in this discussion  and restore them after performing the reduction, which is fully consistent with the mathematical foundation of \cite{Grimm:2024tbg, hoefnagels_2025}.} Hence we 
distinguish two fundamentally different situations: (1) choices of integration chain $\Gamma$ which have no boundary, or combinations of $(\Gamma,\nu)$ for which the boundary term vanishes; (2) integrals with $(\Gamma,\nu)$ for which the boundary term does not vanish. 

Let us first consider case (1), where the Euler equations are homogeneous and take the form    
    \beq \label{Euler_GKZ} 
        (\cE_{\underline{e}}+\nu_{\underline{e}})f(z;\nu)=0\, .
    \eeq 
Inspecting \eqref{eq:gengkzint}, we see that for nonvanishing $\partial \Gamma$, this condition is satisfied if the parameters are set so that $\cG(z;x)^{-\nu_0}\prod_{e=1}^{ N_e } x_{e}^{\nu_e-1}$ vanishes fast enough 
to suppress boundary terms along $\partial \Gamma$.
However, in case (2), 
the differentiation in $\cE_{\underline{e}}$ does yield 
a boundary contribution. This turns \eqref{Euler_GKZ} into  
\beq \label{inhomog-euler1}
  (\cE_{\underline{e}}+\nu_{\underline{e}}) f(z,\nu) = g_{\underline{e}}(z,\nu) \ ,
\eeq
where $g_{\underline{e}}$ is a boundary term that makes the Euler equation inhomogeneous. Such GKZ systems have been considered, for example, in references \cite{doi:10.1142/9789814383462_0012,Li:2009dz}. We will discuss the  conditions for having a boundary term of the form $g_{\underline{e}}$ and its precise form for Feynman integrals in section \ref{app_GKZ_integrals}.

\paragraph{$\cA$-degree of a differential operator.} The Euler operators $\cE_i$ introduce a natural grading on the space of differential operators built from the coordinates $z_i$ and their derivatives $\partial_i$. We introduce the $\cA$-degree $\text{deg}_{\underline{e}}(\cO)$ for operators $\cO$ satisfying 
\beq \label{def-A-degree}
   [\cE_{\underline{e}}, \cO ] = \text{deg}_{\underline{e}}(\cO)\, \cO\ . 
\eeq
This degree is only well-defined on operators $\cO$ that are sums of these building blocks with the same degree. 
Consider the operator $z^{v} \partial^u$, where we again used the multi-index notation $z^v= z_1^{v_1}...z^{v_N}_{N}$ and \eqref{partial-u}. A straightforward computation yields 
\beq \label{deg-zdz}
   \text{deg}_{\underline{e}} (z^v \partial^u) = (\cA v - \cA u)_{\underline{e}}\ .
\eeq

\paragraph{Contiguity shifts.} Applied to a solution $f(\nu)$ of a GKZ system $(\cA,\nu)$, the equation \eqref{deg-zdz} implies that $\partial^u f(\nu)$ is a solution $\hat f$ of a GKZ system $(\cA,\nu')$ at shifted parameter $\nu' =\nu + \cA u$, i.e.~we have 
\beq \label{contiguity-map}
   \hat f(\nu + \cA u) = \partial^u f(\nu)\ .
\eeq
These relations are known as contiguity morphisms and play an important role in the study of GKZ systems \cite{Saito2001,walther_duality_2005,reichelt_laurent_2014,
Beukers2007,reichelt_hypergeometric_2020}, see \cite{reichelt_algebraic_2021} for a review.

\paragraph{Convex polytope and faces.}
Given an
$(N_{e}+1) \times N$-matrix $\A$, of the form \eqref{cA-a-form}, we can define the $N_e$-dimensional polytope $\rm{Conv}(\A)$
    to be the convex hull of the column vectors $\{a_i\}$ for $i=1,\ldots,N$. The faces $F$ of the polytope $\rm{Conv}(\A)$ are certain subsets of the indices $\{
    1,\cdots,N\}$. Specifically, the subset $F$ is a face of $\A$ if there exists a linear
    functional $L_{F}:\bbZ^{N_e+1}\rightarrow \bbZ$ such that \beq \label{eq:LF}
    \begin{array}{rl}
        L_F(a_i) =0 & \text{ for } i\in F \, ,     \\
        L_F(a_i)>0  & \text{ for } i\not\in F \, .
    \end{array}
    \eeq
    There can be many different linear functionals associated to a face $F$. This changes for a face of dimension $N_e-1$, which is called a \textit{facet}. Such facets admit linear functionals $L_F$ that are unique up to rescaling.\footnote{The scaling freedom can be fixed by using the primitive integral support function satisfying the conditions listed in appendix \ref{app:Lnutest}.}

    \paragraph{GKZ subsystems.} Given a face $F$ of $\rm{Conv}(\A)$, we can construct a subsystem of differential operators associated to a smaller matrix $\A_F$ built from some of the columns of $\A$ by setting
    \beq \label{def-AF}
       \cA_F = (a_i)_{i\in F}\ ,
    \eeq
    where we keep the order of columns used for $\cA$, i.e.~simply delete the ones that are not indexed by $F$. If $F$ is a face and $\bar\nu$ is in the complex span of the columns of $F$, then   $(\cA_F,\bar\nu)$ is a GKZ system with toric relations obtained by restricting the relations from $\cA$. It is therefore a true subsystem, in the sense that solutions of the system $(\cA_F,\bar\nu)$ are also solutions of the system $(\cA,\bar\nu)$.

    \paragraph{Faces for edges.} For a one-particle-irreducible Feynman graph, the polynomial $\mathcal{U}$ must contain a term in which any given $x_e$ is absent. Since all entries of the vectors $a_i$ are nonnegative, $\rm{Conv}(\A)$ has a face $F_e$ associated to the edge $e$, corresponding to the linear functional 
    \begin{equation}\label{eq:LeFunctional}
        L_{F_e}(a_i)=(a_i)_e\,,
    \end{equation}
    where $(a_i)_{e}$ is the $e$-th entry of the vector $a_i$. 
  In other words, $L_{F_e}$ reads the $e$-th row of the matrix $\A$ (recalling that the row indices start from 0), and $F_e$ is the set of columns of $\A$ with a zero entry in position $e$, corresponding to terms of $\mathcal{G}$ in which $x_e$ is absent. Note that $F_e$ is not necessarily a facet.
    The contracted GKZ system for the face $F_e$ is equivalent to deriving the matrix $\A_{F_e}$ in which we set $x_e \to 0$ in the graph polynomials.
    Thus the contraction of an edge $e$ of the graph $G$ is identified with the contraction of the GKZ system to the face $F_e$, corresponding to  the GKZ subsystem with the matrix $\cA_{F_e}$.

\paragraph{Fixing the solution.} The integral~\eqref{eq:gengkzint} is a specific solution to the differential equations \eqref{toric_GKZ} and \eqref{Euler_GKZ} that depends on the choice of the integration chain $\Gamma$.
GKZ systems have finite-dimensional solution spaces.
A general way to find the solution of interest is to construct a complete basis of solutions $f_{r}(z;\nu)$, $r=1,...,R$, to the GKZ system, which then ensures that we can write  
    \begin{equation}
        I_{\cA}(z;\nu)=\sum_{r=1}^{R} c_{r}(\Gamma;\nu)f
        _{r}(z;\nu)\, .
    \end{equation}
The particular coefficients can either be fixed numerically or by evaluating the integral in specific limits for the $z_i$. There are various ways of determining the asymptotic solutions, e.g.~using complete basis of convergent series expansions~\cite{saito2000grobner,cattani_three_2006}. In this work, we analyze the solution space restricting to physically relevant $\Gamma$. The application of the preceding abstract ideas to the concrete GKZ integrals will be presented in section \ref{app_GKZ_integrals}.

\section{GKZ resonance and reduction operators} \label{GKZresonance-reduction}

This section contains the main conceptual contribution of our work. In section \ref{review:Red-op}, we first review the conditions under which GKZ systems become resonant, leading to a reducible system of differential equations. We then introduce the associated reduction operators, emphasizing how they are constructed and how they act on the GKZ solution space. In section \ref{app_GKZ_integrals}, we employ these reduction operators for edge faces in graphs corresponding to general $L$-loop Feynman integrals, demonstrating that they relate different integrals through edge contractions or parameter shifts. Finally, in section \ref{diff-equations_cuts} we comment on the resulting differential equation system and clarify how reduction operators act on cut integrals.  
The abstract findings of this section are illustrated in the rest of this work through several Feynman integral examples. Appendix \ref{ap:2f1} additionally shows reduction operators in the simple example of Gauss's hypergeometric function ${}_2F_1$.

\subsection{Reduction operators for resonant GKZ systems
    \label{review:Red-op}}
Feynman integrals are solutions to GKZ systems that are not generic, but rather exhibit a feature known as resonance.
For GKZ systems, resonance is known to be equivalent to the well-known property of \emph {reducibility }\cite{saito_irreducible_2011,beukers_irreducibility_2011,schulze_resonance_2012}. When a GKZ system is reducible, there exist additional differential operators, the reduction operators \cite{Grimm:2024tbg,Grimm:2025zhv}, that annihilate \textit{some} but not all of the solutions. These operators can be systematically constructed starting from the GKZ data. We now review this construction following \cite{Grimm:2025zhv}.

    \paragraph{Resonance and reducibility of GKZ systems.}
    A face $F$ of $\rm{Conv}(\A)$ is  a
    \textit{resonant face} for $\nu$ if there exist complex numbers $c_j \in {\mathbb C}$ and integers
    $n_i \in {\mathbb Z}$ such that
    \begin{equation}
        \label{check_resonance}
        \nu=\sum_{j\in F}c_j\, a_j+\sum_{i \notin F} n_i
        \,a_i\ .
    \end{equation} 
     In other words,
    the vector $\nu$ lies in the span of the face $F$, up to shifts given by
    integer multiples of the columns of $\A$. If we focus on facets and require that $\A \bbZ^N= \Z^{N_e+1}$,  
    a facet $F$ is resonant for a parameter $\nu$ if and only if 
\begin{equation}\label{eq:facet-resonance}
    L_F(\nu)\in \Z\,.
\end{equation} 
The proof is given in appendix \ref{app:Lnutest}.

According to Theorem 3.1 of~\cite{schulze_resonance_2012} (see also \cite{Grimm:2024tbg}), the GKZ system with data $\A,~\nu$ is reducible if 
there is a resonant face $F$ for $\nu$ such that  $\A$ is not a {\em pyramid} over $F$. 
The non-pyramid condition can be 
stated as the condition that the simplicial volumes of $F$ and $\cA$ are not equal, and it is always satisfied for the faces and GKZ integrals considered in this work.\footnote{See \cite{schulze_resonance_2012} for a formal definition of pyramid in Def.~3.4 and equivalent conditions in Lemma~3.5.} 

\paragraph{Euler operators for resonant faces.} Let us denote by $L_{F}$ the linear functional defining $F$ as in \eqref{eq:LF}. This functional can be used to define an Euler operator for $F$, \begin{equation}\label{eq:EF}
        \cE_F=L_F(\cE)\,,
\end{equation} 
where $\cE$ is the matrix whose rows are the Euler operators $\cE_j$ defined in \eqref{def-Euler}. The Euler equations \eqref{inhomog-euler1}  and the linearity of $L_F$ imply the relation
\beq \label{eval_Euler}
   \cE_F f(z;\nu) = - L_F(\nu) f(z;\nu) + g_F(z;\nu)\ , 
\eeq
where $g_F = L_F(g)$ is a boundary contribution associated to the face $F$ by acting with $L_F$ on the vector $g_{\underline{e}}$ introduced in \eqref{inhomog-euler1}.

\paragraph{Constructing reduction operators.} 
We now define a \textit{reduction operator} for a face $F$  
over which $\A$ is not a pyramid. 
We consider integer vectors $u,v \in \bbZ^N_{\geq 0}$ such that
\beq \label{uv-support}
    \cA u\equiv  \sum_{i} u_i a_i \in \text{span}_{\bbZ} \cA_F\,, \qquad  \cA v \equiv \sum_{i} v_i a_i
       \notin \text{span}_{\bbZ} \cA_F\, .
\eeq 
Nontrivial pairs $u,v$ exist, because $\A_F=\A$ is trivially excluded by the pyramid condition. 
A reduction operator $\cQ^F_{u,v}$
associated to $F$ and two vectors $u,v$ is obtained if one can find a nontrivial solution to the relation
\begin{equation}
        \label{eq:PEFsimQd}
      \boxed{ \quad \partial^u \cE_F \simeq \cQ^{F}_{u,v} \partial^v\ ,\quad }
\end{equation}
where we recall that $\simeq$ indicates equality up to using the toric relations \eqref{toric_GKZ} and we have used the notation \eqref{partial-u} for $\partial^u$ and $\partial^v$.\footnote{There exists an algorithm to obtain a suitable vector $u$ \cite{dwork_generalized_1990,beukers_irreducibility_2011,Grimm:2025zhv}, but for the purposes of this paper it is enough to solve this condition by inspection.} We 
additionally require that 
\beq  \label{Q-commutes-with-toric}
   [\cQ^F_{u,v}, \mathcal{L}_{u',v'}] \simeq 0 \ ,
\eeq
for all toric operators $\mathcal{L}_{u',v'}$ defined in \eqref{toric_GKZ}. We stress that our assumption \eqref{uv-support} is that $\cA u$ is supported on the face $F$, which implies that $[\cE_F,\partial^u]=0$. 
Roughly speaking, the operators $\cQ^{F}_{u,v}$ translate derivatives $\partial^v$ in directions supported not on the face, into derivatives $\partial^u$ supported on the face $F$. 

The defining conditions imply that $\cQ_{u,v}^F$ has order and $\cA$-degree \eqref{def-A-degree} given by 
\beq  \label{order-degr-Q}
   \text{ord}(\cQ_{u,v}^F) = \sum_{i \in F} u_i - \sum_{i \notin F} v_i +1 \ , \qquad \text{deg}_j(\cQ_{u,v}^F) =  (\cA v -\cA u)_j \ .
\eeq
This can be inferred by matching the order and degree of the differential operators appearing on both sides of \eqref{eq:PEFsimQd}. It is important to stress that while introduced generally, the operators $\cQ_{u,v}^F$ are only nontrivial, e.g.~genuinely different from the Euler operators, if $F$ is a resonant face for some parameter $\nu$. 

\paragraph{Acting with reduction operators.} One of the crucial properties of reduction operators is that they provide a link between resonance of a face and reducibility of the system of differential equations. We present this link in general terms here, with concrete applications to be demonstrated in the following sections. 

To explore the properties of the reduction operators $\cQ_{u,v}^F$ for resonant faces $F$, we evaluate \eqref{eq:PEFsimQd} 
on a GKZ solution $f(z;\nu)$ and use \eqref{eval_Euler} and the contiguity relations \eqref{contiguity-map} to write
\beq \label{red-on-solutions}
    \cQ^{F}_{u,v} \hat f(z;\nu +\cA v) = - L_F(\nu) \hat g(z;\nu + \cA u) + \partial^u g_F(z;\nu) \ ,
\eeq
where $\hat f = \partial^v f,\ \hat g= \partial^u f$ are solutions of the GKZ system with $\cA$ but at the different parameter values indicated in their arguments. 
In the absence of the boundary term $g_F$, the reduction operators relate GKZ solutions at different parameters consistent with \eqref{order-degr-Q}. Now consider a parameter 
$\bar \nu \in \text{span}_{\bbC} \cA_F$.
Then  $L_F(\bar\nu)=0$, and hence the  first term on the right-hand side of \eqref{red-on-solutions} vanishes. We conclude that 
\beq \label{QF-action}
   L_F(\bar \nu)=0: \qquad \cQ^{F}_{u,v} \hat f(z;\bar \nu + \cA v) =  \partial^u g_F(z;\bar \nu) \ .
\eeq
If additionally $g_F(z,\bar \nu)=0$,\footnote{This vanishing can occur if the contour of integration is chosen such that the  Euler equation (\ref{inhomog-euler1}) remains homogeneous.}  then we may  find special solutions $h$ annihilated by 
the reduction operators 
\beq
    \cQ^F_{u,v} h(z,\bar \nu + \cA v)  = 0\,,
\eeq
which are therefore seen as solutions to homogeneous  differential equations.

\paragraph{Reducing to a GKZ subsystem.} An interesting special case, which will be central in our study of the Feynman integral examples, is when 
$g_F(z;\nu)$ is a solution to the GKZ subsystem $(\cA_F,\nu)$. 
This implies that $g_F$ only depends on $\hat z_i = z_i$ with $i\in F$, while it is independent of the variables $z_i$ with $i \notin F$. 
An explicit realization arises when we can find a lower-dimensional integration cycle $\Gamma'$, such that 
\beq \label{gf-asI}
    g_F(z;\nu) = C(\nu) I_{\cA_F}(\hat z;\nu)\ ,
\eeq
where $ I_{\cA_F}(\hat z;\nu)$ is of the form \eqref{eq:gengkzint}, integrated over $\Gamma'$, and $C$ is an overall constant.
Such reductions arise only when additional assumptions are imposed, since one has to guarantee that it is valid to express the effect of a quotient by the operator $\partial_i$ with $i\notin F$ in terms of the subsystem with $\A_F$ together with a potential parameter shift.\footnote{Conditions for this reinterpretation are given generally in \cite{arno-felix} in the framework of D-modules.} In section \ref{app_GKZ_integrals}, we will explicitly check \eqref{gf-asI}.

\subsection{GKZ integrals, Feynman integrals, and reductions} \label{app_GKZ_integrals}

In this section, we apply the general results of sections~\ref{sec:GKZ_basics} and~\ref{review:Red-op} to GKZ integrals obtained from Feynman integrals, with the relevant integration chains. This will allow us to make many of the abstract formulas explicit and examine the action of the reduction operators in detail. 

\paragraph{GKZ integrals from Feynman integrals.} To begin with, we compare \eqref{eq:gengkzint}, which we repeat for the reader's convenience, 
\beq
    \label{eq:gengkzint1}
        I_{\cA}(z;\nu)= \kappa_G(\nu)\,\int_{\Gamma} \rd^{N_e}x\;   
  \frac{\prod_{e=1}^{ N_e } x_{e}^{\nu_e-1  }}{\cG(z,x)^{\nu_0}}\, ,
\eeq
to the Feynman integral $I_G$ in Lee-Pomeransky form  \eqref{eq:LPrep}. We see that the parameter vector $\nu$ is given by 
\beq \label{nu-LP}
   \nu = (D/2,\nu_1,...,\nu_{N_e})\ , 
\eeq
while the prefactor is 
\beq \label{def-kappa}
   \kappa_G(\nu) = \frac{  \Gamma \left( \frac{D}{2} \right) }{ 
    \Gamma\left(  \frac{(L+1)D}{2}-\sum_{e} \nu_e\right)
    \prod_{\substack{e=1 \\ \nu_e \neq 0}}^{N_e} \Gamma( \nu_{e} ) }\ . 
\eeq
While the parameters $\nu_e$ take positive values for physical scalar Feynman integrals, the effect of reduction will be to shift these parameters by integer values. A shift resulting in some $\nu_e=0$ is of particular interest, as it pertains to contracting edge $e$.
When we relate $G$ to an edge-contracted graph $G/e$, we should not naively set $\nu_e=0$ in the integral \eqref{eq:gengkzint1}, but instead eliminate the parameter entirely by computing the prefactor directly for the contracted graph, ensuring that
\beq \label{restriction_kappa}
   \kappa_G(\nu)|_{\nu_e=0} = \kappa_{G/e}(\nu_{F_e})\ , \qquad \nu_{F_e} \equiv (\nu_0,...,\nu_{e-1},\nu_{e+1},....,\nu_{N_{e}})
\eeq
The integration chain $\Gamma$ is taken such that
\beq \label{integration-chain}
   \Gamma = \bbR_+^{N_e}\ ,\qquad \partial \Gamma = \bigcup_{e=1}^{N_e} \Big\{x_e = 0, \bbR_{+}^{N_e-1} \Big\} \cup \Gamma_\infty\ ,
\eeq
where $\Gamma_\infty$ is the boundary at $x_e \rightarrow \infty$.

To complete the match of the GKZ integral and Lee-Pomeransky integral, we need to determine the matrix $\cA$ by comparing the polynomial $\cG(z,x)$, of the form \eqref{eq:poly}, with $\cG(p,m,x)$ given in \eqref{cGdef}, \eqref{def-UF}. 
Note that $(b_i)_e \geq 0$ for all $i$ and $e$, and that all $x_e$-coefficients in $\cG$ have been promoted to complex variables $z_i$, $i=1,...,N$. This implies that the higher-dimensional complex parameter space spanned by $z_i$ restricts to the physical variables $p,m$ on a  lower-dimensional locus. This \textit{restriction to the physical locus} is an essential and nontrivial step in relating GKZ systems and Feynman integrals. 

\paragraph{Boundary terms.} The fact that $\Gamma$ in \eqref{integration-chain} has a nontrivial boundary implies that the Euler equations obeyed by $I_\cA(z;\nu)$ can be inhomogeneous as in \eqref{inhomog-euler1}. To determine the boundary contribution $g_i$, we note that the integral is 
performed over the differential form $\omega =  \prod_{e=1}^{ N_e } x_{e}^{\nu_e-1  } \cG(z,x)^{-\nu_0} dx_1 \wedge ... \wedge dx_{N_e}$. We define  
\beq \label{explicit-eta}
    \eta_f = (-1)^{f-1} x_f \prod_{e=1}^{ N_e }  x_{e}^{\nu_e-1  } \cG(z,x)^{-\nu_0} dx_1 \wedge ...\wedge \widehat{dx_f} \wedge ...\wedge dx_{N_e}\ ,
\eeq
where the hat indicates that this differential is omitted. We now use \eqref{eq:poly} to derive 
\beq 
  \cE_e \cG(z,x)= x_e \frac{\partial \cG(z,x)}{\partial x_e}\ , \qquad \cE_0 \cG(z,x) = \cG(z,x)\ . 
\eeq
where we recall that $\cE_e$ contains the $z_i$-derivatives $\theta_i = z_i \partial_i$. This relation allows us to show that $(\cE_e+\nu_e) \omega = d_x \eta_e$ and $(\cE_0 + \nu_0)\omega =0$. Integrating  these conditions, we obtain
\begin{eqnarray} \label{inhomog-euler1-fi}
  &&(\cE_e+\nu_e) I_{\cA}(z,\nu) = g_e(z,\nu) \ ,\qquad g_e(z,\nu) = \kappa_G(\nu) \int_{\partial \Gamma} \eta_e\ ,\\
  &&(\cE_0+\nu_0) I_{\cA}(z,\nu) = 0 \ .\nonumber
\end{eqnarray}
The boundary $\partial \Gamma$ as identified in \eqref{integration-chain} contains the $N_e$ components with $x_e=0$. Inspecting \eqref{explicit-eta}, we realize that the boundary integral for $\eta_e$ has support only on the $x_e = 0$ component of $\partial \Gamma$. Moreover, the 
corresponding integral contains positive powers of $x_e$ for $\nu_e >0$, and therefore vanish on each $x_e=0$ boundary component unless $\nu_e = 0$. In addition, we find that the boundary integral vanishes at $x_e \rightarrow \infty$ for a suitable noninteger choice of $\nu_0 =D/2$, which we will assume in the following. Together we obtain
\beq \label{explicitge-boundary}
   g_e(z,\nu) =  \begin{cases} \displaystyle - \kappa_{G/e}(\nu_{F_e})  \int_{\bbR_+^{N_e-1}} d^{N_e-1}x \frac{\prod_{f\neq e}  x_{f}^{\nu_f-1  } }{\cG(z,x)^{\nu_0}|_{x_e = 0}} & \text{if\ \ }  \nu_e =0\ , \\[.2cm]
   \displaystyle 0 & \text{if\ \ }  \nu_e >0\ , 
                         \end{cases} 
\eeq
where $ d^{N_e-1}x$ is the integration measure omitting the $x_e$ variable. The factor $(-1)^{e-1}$ in \eqref{explicit-eta} has been absorbed by the orientation of the domain, and an additional minus sign arises from $x_e=0$ being the lower boundary in the integral. We have also changed the prefactor $\kappa_G$ using \eqref{restriction_kappa}.

\paragraph{Contiguity shifts.} As discussed in section~\ref{sec:GKZ_basics}, a derivative on a GKZ solution shifts 
it to another solution valid at different parameters. 
Applied to the integral representation \eqref{eq:gengkzint1}, we can explicitly evaluate \eqref{contiguity-map} by performing the $\partial^u$-derivative in the integral and reading off the new integrand. We first evaluate the shift for $\partial^u = \partial_i$ and find 
\beq \label{shifts1}
   \partial_i I_\cA(z;\nu) = C_i(\nu) I_\cA(z;\nu + a_i)\ , \qquad C_i(\nu) = -\frac{\kappa_G(\nu) \nu_0}{\kappa_G(\nu + a_i)}\ , 
\eeq
and iteratively generalize the result to 
\beq \label{shifts2}
   \partial^u I_\cA(z;\nu) = C_u(\nu) I_\cA(z;\nu + \cA u)\ , \quad C_u(\nu) =  \frac{\kappa_G(\nu) (-\nu_0)_{\sum_i u_i}}{\kappa_G(\nu + \cA u)}
  \,,
\eeq
where in the last equation we use the Pochhammer symbol to denote the falling factorial, $(x)_n = x(x-1)\cdots(x-n+1)$.
In this evaluation we have ignored convergence issues, i.e.~assumed sufficiently generic parameters.

\paragraph{Resonance for Feynman integrals.}

The resonance of the Feynman GKZ system follows from the existence of resonant faces of the matrix $\cA$ associated to the Feynman graph $G$ with the given kinematics (nonvanishing masses). While determining the resonant faces is in principle algorithmic, and can be carried out for any given example, it is a challenging task to make general statements about \textit{all} reductions for \textit{all} $G$. Since resonance is related to the properties of the vector $\nu$, we first recall 
that for scalar Feynman integrals $\nu$ is of the form
\beq \label{nu-integer}
  \nu=(D/2,\nu_1,\cdots,\nu_{N_e})^T\  \ \text{with}\ \ D\in \bbC, \ \nu_e \in \bbZ_{\geq 0}\ ,
\eeq
where we stress that we work in dimensional regularization, so that  $D$ is not necessarily an integer,\footnote{If $D$ is an even integer, we expect more resonance to arise, as we discuss in section \ref{sec:dimension}. It is well known that certain Feynman integrals admit further simplifications in specific dimensions.} but we do not employ analytic regularization, so the parameters $\nu_e$ are integers.
Since all entries of $\cA$ are integral and resonance arises when \eqref{check_resonance} is satisfied, we expect that many resonant faces can arise for these integral $\nu_e$. In the following, we will focus on the analysis of faces $F_e$ associated 
to the edges of a graph $G$. 
Since $L_{F_e}(\nu) \in \bbZ$ for the  parameter $\nu$ of \eqref{nu-integer}, it follows from the arguments above that in the case where the face
$F_e$ is moreover a \textit{facet}, it is resonant, i.e.
\beq \label{resonance_facet}
  F_e\ \text{is an edge facet}\ \Rightarrow \ F_e\ \text{is resonant for }\nu \text{ with } \nu_e \in \bbZ_{\geq 0}\ . 
\eeq

\paragraph{Reduction operators and boundary terms.} As described in section \ref{review:Red-op}, we can associate both an Euler operator $\cE_F$ and a set of reduction operators $\cQ^F_{u,v}$ to a face $F$. If $F$ is resonant, both have special properties when acting on solutions on certain parameter values. In the following we focus on resonant 
edge facets $F_e$ for a Feynman graph $G$ satisfying \eqref{resonance_facet}.
Let us consider $I_\cA(z;\nu)$ and use \eqref{eval_Euler} to evaluate
\beq \label{Euler-on-solutions}
 \cE_{F_e} I_\cA (z;  \nu) = \cE_e I_\cA (z; \nu) = \begin{cases}   g_e (z; \nu)\ & \text{if} \ \ \nu_e = 0 \ ,\\
 - \nu_e I_\cA(z;\nu)\ & \text{if} \ \ \nu_e > 0 \ ,
 \end{cases}
\eeq
where we have used \eqref{eq:LeFunctional}, i.e.~that $L_{F_e}$ is the projection to the $e$-th entry. Importantly, we see that either the $L_F(\nu)$-term or the boundary term in \eqref{eval_Euler} survives for any given value of $\nu$. The actions of the reduction operators 
$\cQ^{F_e}_{u,v}$ follow a similar split. Using \eqref{red-on-solutions} we find
\beq \label{red-on-solutions1}
    \cQ^{F_e}_{u,v} \partial^v I_\cA (z;\nu)  = 
    \begin{cases}   \partial^u g_e (z; \nu)\ & \text{if} \ \ \nu_e = 0 \ ,\\
 - \nu_e  \partial^u I_\cA(z;\nu)\ & \text{if} \ \ \nu_e > 0 \ ,
 \end{cases}
\eeq
where we recall that $\partial^v$ are derivatives whose indices are not on $F_e$, while the indices of $\partial^u$ are on $F_e$.

\paragraph{Edge contraction from reduction operators.}
Let us now have a closer look at $g_e$ appearing in \eqref{Euler-on-solutions}, \eqref{red-on-solutions2} and show that it can be viewed as 
a solution of a GKZ subsystem as in \eqref{gf-asI}. The full GKZ integral corresponds to the Feynman integral $I_G$ and we henceforth write 
\beq
   I_{G}(z;\nu)= I_\cA(z;\nu)\ , 
\eeq
where we recall that $\cA$ is determined from $G$ with the appropriate kinematic input. We realize from \eqref{explicitge-boundary} that $g_e$ contains the restriction $\cG(z,x)|_{x_e=0}$. From \eqref{eq:UFG-edge-contraction},  we infer that $g_e$ is associated to the graph $G/e$, i.e.~the graph obtained by contracting the edge $e$. This leads us to identify 
\beq
    g_e(z;\nu) = - I_{G/e}(z_{F_e};\nu_{F_e}) = - I_{F_e}(z_{F_e};\nu_{F_e})\ ,
\eeq
with 
\beq \label{def-ZFe}
   z_{F_e} = (z_i)_{i\in F_e}\ ,\qquad \nu_{F_e} = (\nu_0,...,\widehat \nu_e,...,\nu_{N_e})\ ,
\eeq
where the hatted entry is omitted. 
By construction of the edge face $F_e$, as discussed after \eqref{eq:LeFunctional}, the terms independent of $x_e$ are labeled by $z_i$, $i\in F$. Inserted into \eqref{Euler-on-solutions} the Euler relations read 
\beq  \label{Euler-onsol2}
\boxed{\quad \cE_0 I_G(z;\nu) = -\nu_0 I_G(z;\nu)\ , \quad  \cE_e I_G (z; \nu) = \begin{cases}  - I_{G/e} (z_{F_e};\nu_{F_e}) & \text{if} \ \ \nu_e = 0 \ ,\\
 - \nu_e I_G(z;\nu) & \text{if} \ \ \nu_e > 0 \ .
 \end{cases} \quad }
\eeq

Having identified $g_e$ as a solution to the GKZ subsystem 
$(\cA_{F_e},\bar \nu)$, we can now use the contiguity shifts \eqref{contiguity-map}. 
After shifting $\nu \rightarrow \nu - \cA v$ in \eqref{red-on-solutions1}, this leads us to the final expression 
\beq \label{red-on-solutions2}
  \boxed{\   \cQ^{F_e}_{u,v} I_G (z;\nu)  = 
    \begin{cases}  -  C_{u,v}(\nu)\, I_{G/e} (z_{F_e};(\nu - \cA v +\cA u)_{F_e})\ & \text{if} \ \ (\nu-\cA v)_e = 0 \ ,\\
 - (\nu-\cA v)_e C_{u,v}(\nu)\,  I_G(z;\nu-\cA v +\cA u)\ & \text{if} \ \ (\nu-\cA v)_e > 0 \ .
 \end{cases}\ }
\eeq
 The prefactors are evaluated using \eqref{shifts2} and give 
\beq \label{Cuv-evaluated}
  C_{u,v}(\nu) \equiv \frac{C_u(\nu - \cA v)}{C_v(\nu - \cA v)}=  \frac{\kappa_G(\nu)}{\kappa_G(\nu - \cA v+ \cA u)} \frac{(-\nu_0+(\cA v)_0)_{\sum_i u_i}}{(-\nu_0+(\cA v)_0)_{\sum_i v_i}} \, .
\eeq

The identities \eqref{Euler-onsol2}, \eqref{red-on-solutions2} are the key results that allow us to derive differential equation systems for Feynman integrals. Given a $I_G(z;\nu)$ and a set of reduction operators $\cQ^{F_e}_{u,v}$, they can be used iteratively to stepwise lower $\nu$ until it hits zero and the edge gets contracted. Here it is important to realize that all entries in $\cA$ and in $u,v$ are positive. Hence, the $v$-shift in \eqref{red-on-solutions2} actually lowers the values of $\nu$ in directions outside the face $F_e$.

\subsection{Differential equation systems and cut simplifications} \label{diff-equations_cuts}

Starting from \eqref{Euler-onsol2} and \eqref{red-on-solutions2} we can try to systematically construct a differential equation system that closes on a finite set of $I_{G'}(z;\nu')$ associated to an integral $I_{G}(z;\nu)$ of interest. Although providing this construction in full generality lies outside the scope of this work, we record several observations that hold for all of the examples examined below. 

\paragraph{Shifting and reducing.} First, we observe that the set $I_{G'}(z;\nu')$ contains a finite set of graphs $G' \subset G$, since the reduction operators act by removing edges and yielding subgraphs. Second, to each graph there will also be a number of $\nu'$-values that need to be included. Consider a $\cQ_{u,v}^e$ acting as in \eqref{red-on-solutions2} without removing an edge. This implies that $\nu$ decreases in the $v$-direction and increases in the $u$-direction, when considering the basis vectors $a_i$. All reduction operators that we will construct have  
\beq
    v=\delta_e\ , \qquad \partial^v = \partial_e\ , 
\eeq
where $\delta_e = (0,...,1,...,0)^T$ is the unit normal vector with a $1$ at the $e$-th position. Henceforth we will introduce the shorthand notation
\beq \label{short-not}
   \cQ_{u,e} \equiv \cQ^{F_e}_{u,\delta_e}\ . 
\eeq
The reduction operator shifts 
\beq
 \cQ_{u,e}: \ \nu' \mapsto  \nu' - a_e+ \cA u\ . 
\eeq
We note that the vectors $a_e$ associated to $\cU(z,x)$ contain $L+1$ times the number $1$, where the $+1$ comes from the homogenization in \eqref{cA-a-form}, with all remaining entries being $0$, since $\cU(z,x)$ contains no $x_e^2$-terms. The highest entries in $\nu'$ arise from adding $\cA u$ when many shifts are needed to send $\nu_e' =0$. We will see an explicit pattern in figure~\ref{Fig-reduction-triangle}. Clearly, the story becomes particularly simple at one-loop level, since in this case $a_e = (1,\delta_e)$, i.e.~only contains one entry $1$ in addition to the homogenization.

\paragraph{Differential equation system.} In case one has identified the finite set of integrals $I_{G'}(z;\nu')$ reached from $I_{G}(z;\nu)$ using the reduction operators, one can now obtain a differential equation system $d \vec I = A\cdot \vec I$, where $A$ is a one-form-valued matrix. For first-order reduction operators, the construction of $\vec I$ and the matrix $A$ is straightforward and amounts to solving a linear system. This is precisely the situation we encounter for one-loop diagrams with all edges massive in sections~\ref{sec:bubble}--\ref{sec:generic1loop}. With more loops or in partially massless cases, we find higher-order reduction operators. In these situations, one can translate the higher-order system into a first-order system in the usual fashion by introducing additional entries for $\vec I$. 

It is important to stress that the differential equation systems obtained from reduction operators alone are not yet minimal. This has a simple reason: we have not yet imposed the toric relations $\cL_{u,v} I^D_G(z;\nu) = 0$. For first-order systems for one-loop integrals, these additional differential equations yield algebraic relations and thus reduce the rank of the first-order system $d \vec I = A\cdot \vec I$. We will see this in a simple example below, but leave an exploration of the full differential equation systems for more involved situations to future work.    

\paragraph{Reduction operators on cut integrals.} 

Cuts of Feynman integrals were defined in (\ref{eq:parametric-cuts}), with the integration
contours specified in (\ref{eq:Gamma-contours}), such that if an edge $e$ is cut, there is no boundary contribution from the coordinate hyperplane at $x_e=0$. Therefore, there is no way to pick up the face subsystem $g_{F_e}$ under resonance conditions. 
The boundary of the cut integration region does include part of $\mathcal{G}=0$, but there is no contribution from there, by the same reasoning by which we ignore the boundary terms at infinity (convergence of the integral is assumed for noninteger $D$.)
As a result, we have a system of linear differential equations for cuts that are a more homogeneous version of the ones for the uncut integral $I$, because the choice of contour eliminates potential boundary terms.
\begin{equation}
    \cQ^e_{u,v} \mathcal{C}_J I = \mathcal{C}_J \cQ^e_{u,v} I \qquad \textrm{for any $e$ and $J$},
\end{equation}
\begin{equation}
    \cQ^e_{u,v} \mathcal{C}_J I = \mathcal{C}_J \cQ^e_{u,v} I = 0 \qquad \textrm{if $e \in J$ and $\nu_e=1$.}
\end{equation}

\section{GKZ reduction for the bubble diagram} \label{sec:bubble}

To illustrate the use of reduction operators associated GKZ systems and Feynman integrals, we present two detailed examples: the one-loop bubble and the one-loop triangle diagram. This allows us to set the stage for a more general study of one-loop diagrams in section~\ref{sec:generic1loop}. We determine the GKZ system and present reduction operators that realize edge contractions. 
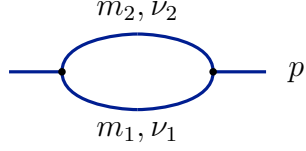
\begin{figure}[h!]
\begin{center}
\begingroup
\definecolor{rlLineBlue}{RGB}{0,30,145}
\begin{tikzpicture}[
    diagram/.style={draw=rlLineBlue, very thick},
    text=black
]
\draw[diagram] (-1.7,0) -- (-1,0);
\draw[diagram] (1,0) -- (1.7,0);
\draw[diagram] (-1,0) .. controls (-1,0.4) and (-0.3,0.5) .. (0,0.5);
\draw[diagram] (-1,0) .. controls (-1,-0.4) and (-0.3,-0.5) .. (0,-0.5);
\draw[diagram] (1,0) .. controls (1,0.4) and (0.3,0.5) .. (0,0.5);
\draw[diagram] (1,0) .. controls (1,-0.4) and (0.3,-0.5) .. (0,-0.5);

\node at (-1,0) [circle,fill=black,inner sep=1pt] {};
\node at (1,0)  [circle,fill=black,inner sep=1pt] {};

\node at (0,-0.8) {$m_1,\nu_1$};
\node at (0,0.8) {$m_2,\nu_2$};
\node at (2.1,0) {$p$};

\end{tikzpicture}
\endgroup
\end{center}
\caption{The 1-loop bubble diagram.}
\label{fig:bubble}
\end{figure}

\subsection{GKZ system for the bubble diagram and resonance}

The Feynman integral for the bubble diagram, shown in figure~\ref{fig:bubble}, takes the following form in the Lee-Pomeransky representation:
\begin{equation}\label{eq:bubblephysint}
  I^D_{\textrm{bub}}(p,m;\nu)=
        \kappa_{\textrm{bub}}(\nu) 
    \int_{\mathbb{R}_+^2}\!\! d^{2}x\, 
        \frac{x_{1}^{\nu_1-1} x_{2}^{\nu_2-1}}{
        \big(x_{1}+x_{2}+ \left(m_1^2+m_2^2-p^2\right)x_{1} x_2+m_1^2 x_1^2+m_2^2 x_2^2\big){}^{D/2 }}\,,
\end{equation}
where $\kappa_{\rm bub}(\nu)$ is given by \eqref{def-kappa} with $L=1$. 
To obtain the GKZ system, we turn the coefficients of the Lee-Pomeransky polynomial into variables $z=\{z_i\}=\{z_1,z_2,z_{1,2},z_{1,1},z_{2,2}\}$ and obtain the integral 
\begin{equation}\label{eq:bubblegkzint}
 I^D_{\textrm{bub}}(z;\nu)=\kappa_{\textrm{bub}}(\nu) 
\int_{\R_+^2}d^2x\,        \frac{x_1^{\nu_1-1} x_2^{\nu_2-1}}{
\big( z_1x_1+z_2 x_2+ z_{1,2} x_1 x_2 +z_{1,1} x_1^2+z_{2,2} x_2^2\big){}^{\frac{D}{2 }}}\,.
\end{equation}
In this expression we have labeled the variables according to the appearance of $x_1$ and $x_2$ in their corresponding monomials.
In particular, for any pair of edges $e$ and $e'$, we write $z_{e,e'}$ for the coefficient of  $x_e x_{e'}$. In this notation, we allow for permutations of the edges, such that $z_{e,e'}=z_{e',e}$.

\begin{figure}[t]
\begin{center}
   \includegraphics[width=0.5\linewidth]{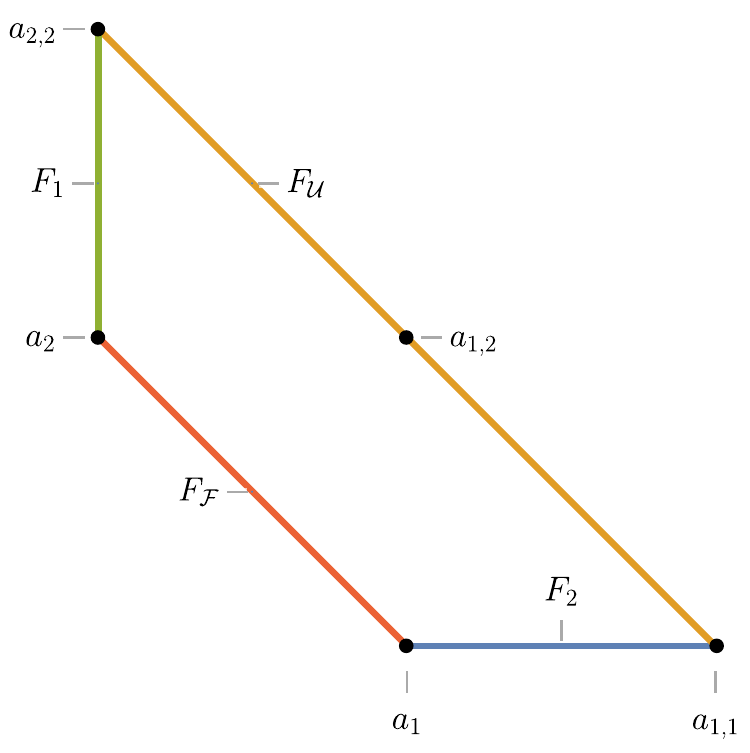}
   \caption{The polytope ${\rm Conv}({\mathcal A})$ for the bubble with generic kinematics, with the matrix $\cA$ given in \eqref{eq:bubbleAandV}.  }\label{fig:bubble-polytope}
\end{center}
\end{figure}

Comparing the integral~\eqref{eq:bubblegkzint} with the more general GKZ expression \eqref{eq:gengkzint}, \eqref{eq:poly}, allows us to read off the GKZ data,
\begin{equation}\label{eq:bubbleAandV}
   \A= \left(
\begin{array}{ccccc}
 1 & 1 & 1 & 1 & 1 \\
 1 & 0 & 1 & 2 & 0 \\
 0 & 1 & 1 & 0 & 2 \\
\end{array}
\right)\, , \quad \nu
=\begin{pmatrix}
    D/2\\
    \nu_1\\
    \nu_2
\end{pmatrix}\,.
\end{equation}
We assume $\nu_1, \nu_2 \in {\mathbb N}$ and apply the labeling of the variables by $z_e$ and $z_{e,e'}$ to the column vectors of $\A$ as well.
The polytope $\rm{Conv}(\A)$ is depicted in figure \ref{fig:bubble-polytope}.
Using equations \eqref{toric_GKZ}, \eqref{def-Euler} we determine the toric operators
\begin{equation}\label{eq:bubbletoric}
    \L_{1}=\partial_1 \partial_{2,2}-\partial_2 \partial_{1,2}\, , \quad \L_2=\partial_1 \partial_{1,2}-\partial_2\partial_{1,1} \,, \quad \L_3=\partial_{1,2}^2-\partial_{1,1}\partial_{2,2}\, ,
\end{equation}
as well as the Euler operators
\begin{equation}\label{eq:bubbleeuler}
    \cE\equiv \left(\begin{array}{l}
        \cE_0\\
        \cE_1\\
        \cE_2
    \end{array}\right)=\A\Theta=\left(\begin{array}{l}
        \theta_1+\theta_2+\theta_{1,2}+\theta_{1,1}+\theta_{2,2}\\
        \theta_1+\theta_{1,2}+2\theta_{1,1}\\
        \theta_2+\theta_{1,2}+2\theta_{2,2}
    \end{array}\right)\, .
\end{equation}
The integral in equation~\eqref{eq:bubblegkzint} then satisfies 
\begin{equation}\label{eq:bubbletorandeuleqs}
\begin{split}
    \L_n I^D_{\mathrm{bub}}(z;\nu)=0 & \text{ for } n\in\{1,2,3\}\, , \\
     (\cE_{\underline{e}} + \nu_{\underline{e}})I^D_{\mathrm{bub}}(z;\nu)=0& \text{ for } {\underline{e}}\in\{0,1,2\}\, .
\end{split}
\end{equation}

\paragraph{Resonant facets.}

We will now show that in the case of generic kinematics, the edge faces are resonant.
We have established that there is a face $F_e$ associated to each edge $e$ of the Feynman graph, with the linear functional $L_{e}$ given in equation (\ref{eq:LeFunctional}). In the case of the bubble with generic kinematics, these faces are facets. It is clear from figure \ref{fig:bubble-polytope} that there are two more facets. We denote them by $F_{\mathcal U}$ and $F_{\mathcal F}$, as they vanish for the columns of the terms in the corresponding Symanzik polynomials.
Explicitly, we have the following expressions for the facets,
\beq
F_1 =\{a_2,a_{2,2}\}\, ,\ \ F_2 = \{a_1,a_{1,1}\}\,,\ \ F_{\cF} = \{a_1,a_2\}\,,\ \ F_{\cU} = \{a_{1,1},a_{1,2},a_{2,2}\}\,,  
\eeq
with linear functionals 
\begin{equation}
    L_1 = (0,1,0)\,, \quad
    L_2 = (0,0,1)\,, \quad
    L_{\mathcal F} = (-1,1,1)\,, \quad
    L_{\mathcal U} = (2,-1,-1)\,,
\end{equation}
where the vectors are understood as multiplying  the columns of $\A$  on the left.
Taking the  expression for $\nu$ in \eqref{eq:bubbleAandV}, and using the condition \eqref{eq:facet-resonance} that a {\em facet} $F$ is resonant for a parameter $\nu$ if and only if
 $L_F(\nu)\in \Z$,
 we see that the facet $F_{\mathcal F}$ is resonant only if $D/2 \in \mathbb {Z}$, and the facet $F_{\mathcal U}$ is resonant only if $D \in \mathbb {Z}$, while the edge facets $F_1$ and $F_2$ are resonant for any value of $D \in \mathbb{C}$.

\subsection{Reduction operators and physical locus}

We now construct the reduction operators and take the restriction to the physical locus, arriving at a system of  inhomogeneous differential equations for the bubble.

\paragraph{Constructing reduction operators in GKZ coordinates.}

We begin with the edge $e=1$ and the corresponding face $F_1$ with $\cA_{F_1}=(a_2,a_{2,2})$.
The Euler operator defined for faces in \eqref{eq:EF} is
\begin{equation}
    \cE_{F_1}=\cE_1=\theta_1+\theta_{1,2}+2\theta_{1,1}\, .
\end{equation}
The terms of $\cE_{F_1}$ correspond to the columns $i\not \in F$.
According to \eqref{eq:PEFsimQd} we now have to look for operators $\cQ^{F_1}_{u,v}$ and vectors $u,v$ as in \eqref{uv-support} such that we can write
\begin{equation}
    \partial^u \cE_{F_1}\simeq \cQ^{F_1}_{u,v}\partial^v\, ,
\end{equation}
where we recall that $\simeq$ denotes equality modulo the toric relations. 
We choose  $v$ to pick out the direction $a_1$, while $u$ picks out the direction $a_2$, i.e.~we consider\footnote{The other possibilities for $v$ are such that one gets the directions $a_{1,2}$ or $a_{1,1}$. We find that all three choices lead to the same reduction operator.}
\beq
    \partial^v = \partial_1\ , \qquad \partial^u = \partial_2\ . 
\eeq
With these choices, 
we can use the toric relations $
    \partial_2 \partial_{1,2} \simeq \partial_{2,2} \partial_{1}$ and $\partial_2 \partial_{1,1} \simeq \partial_{1,2} \partial_{1}$
to see that 
\begin{equation}\label{eq:u2EF1}
        \partial_2 \cE_{F_1} = \partial_2 (z_1 \partial_1+z_{1,2}\partial_{1,2}+2 z_{1,1}\partial_{1,1}) \\
      \simeq ( z_1  \partial_2 + z_{1,2} \partial_{2,2}+2 z_{1,1}\partial_{1,2})\partial_1\, ,
\end{equation}
We can then read off the reduction operator $\cQ^{F_1}_{2,1} \equiv \cQ_{2,1}$, which is given by
\begin{equation}\label{eq:bubbleQ1}
    \cQ_{2,1} =  z_1  \partial_2 + z_{1,2} \partial_{2,2} +2 z_{1,1}\partial_{1,2}\,.
\end{equation}
Similarly for $e=2$, $\partial^v=\partial_2$, $\partial^u=\partial_1$, we obtain a second reduction operator
\begin{equation}
    \cQ_{1,2} = z_2 \partial_1+z_{1,2} \partial_{1,1}+2 z_{2,2 
}\partial_{1,2}\, .
\end{equation}

\paragraph{Physical coordinates.}
The GKZ bubble integral of equation~\eqref{eq:bubblegkzint} is related to the  physical integral~\eqref{eq:bubblephysint} by the following restriction,
\begin{equation}\label{eq:bubblechangeofvars}
    z_{1,2}=m_1^2+m_2^2-p^2\, , \quad z_{1,1}=m_1^2\,,\quad z_{2,2}=m_2^2\, ,\quad
    z_1=z_2=1\,.
\end{equation}
The change of variables~\eqref{eq:bubblechangeofvars} can be implemented in the differential equation through simple applications of the chain rule. To perform the restriction, we first eliminate partial derivatives with respect to $z_1$ and $z_2$ with the Euler operators of equation (\ref{eq:bubbleeuler}) with the parameter $\nu$ of equation (\ref{eq:bubbleAandV}). On a GKZ solution $f(z;\nu)$ we find
\begin{equation}\label{eq:bubblederivrestr}
\begin{split}
    \partial_2 f(z;\nu) = \,-\tfrac{1}{z_2}(z_{1,2} \partial_{1,2}+2 z_{2,2} \partial_{2,2}+\nu_2)f(z;\nu)\, ,\\
    \partial_1 f(z;\nu)= \,-\tfrac{1}{z_1}(z_{1,2} \partial_{1,2}+2 z_{1,1} \partial_{1,1}+\nu_1) f(z;\nu)\, .
\end{split}
\end{equation}
 Applying these replacements to the reduction operators, we find that they can be written as 
\begin{equation}\label{eq:bubbleGredphys}
    \begin{array}{rl}
        \cQ_{2,1} f(p,m;\nu)&= \big((m_1^2 - m_2^2-p^2)\partial_{m_2^2}-2p^2 \partial_{p^2}-\nu_2 \big) f(p,m;\nu) \,,  \\
        \cQ_{1,2} f(p,m;\nu) &= \big((m_2^2 - m_1^2-p^2)\partial_{m_1^2}-2p^2 \partial_{p^2}-\nu_1 \big)f(p,m;\nu) \,.
    \end{array}
\end{equation}
The Euler operator $\cE_0$, corresponding to the first row of the matrix in equation~\eqref{eq:bubbleeuler}, allows us to eliminate any single remaining physical variable, and enforces that the integral itself only depends on ratios of the kinematic variables, up to an overall dimensional factor.
For example, if we choose to eliminate $p^2$, the reduction operators can be rewritten as 
\begin{equation}\label{eq:bubbleQm2}
    \begin{array}{rl}
        \cQ_{2,1} f(p,m;\nu)&= \big(2\nu_1 + \nu_2 - D +2 m_1^2 \partial_{m_1^2}+(m_1^2+m_2^2-p^2)\partial_{m_2^2} \big) f(p,m;\nu) \,,\\
        \cQ_{1,2} f(p,m;\nu)&= \big( 2\nu_2 + \nu_1 - D +2 m_2^2 \partial_{m_2^2}+(m_1^2+m_2^2-p^2)\partial_{m_1^2}\big) f(p,m;\nu)\,.
    \end{array}
\end{equation}
The physical restriction of $\cE_0$ itself is
\begin{equation}\label{eq:bubbleE0}
    \cE_0 f(p,m;\nu) =-\big(\nu_1+\nu_2  + p^2 \partial_{p^2} + m_1^2 \partial_{m_1^2} +  m_2^2 \partial_{m_2^2}\big) f(p,m;\nu)\,.
\end{equation}

\subsection{Contractions from reduction operators}

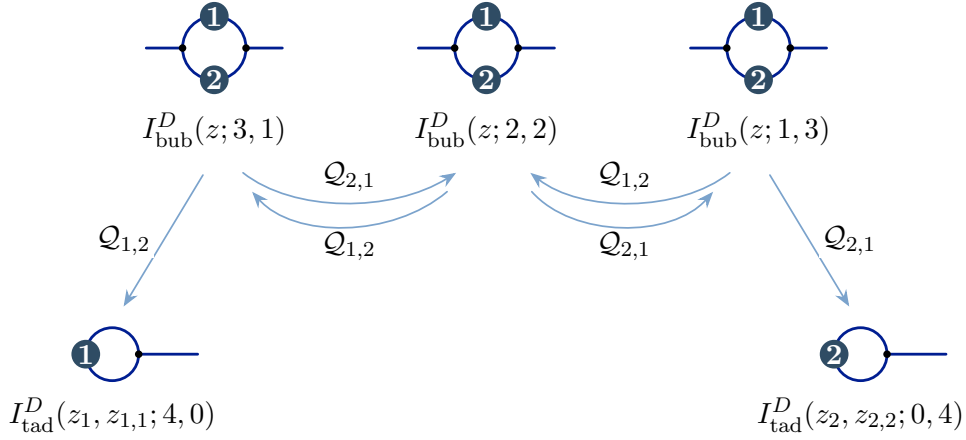
\begin{figure}[t]
\begin{center}
\begingroup
\definecolor{rlLineBlue}{RGB}{0,30,145}
\definecolor{rlArrowBlue}{RGB}{123,163,204}
\definecolor{rlNode}{RGB}{47,76,100}

\begin{tikzpicture}[
  x=1cm,y=-1cm,
  line cap=round,line join=round,
  >=Stealth,
  diagram/.style={draw=rlLineBlue,line width=1.05pt},
  thinarrow/.style={draw=rlArrowBlue,line width=0.65pt,
                    -{Stealth[length=2.2mm,width=1.7mm]}},
  nodecircle/.style={circle,fill=rlNode,text=white,inner sep=0pt,
                     minimum size=0.38cm,
                     font=\fontsize{10.5}{10.5}\selectfont\bfseries},
  vertexdot/.style={circle,fill=black,inner sep=1pt},
  mathlabel/.style={font=\fontsize{11.0}{12.5}\selectfont},
  qlabel/.style={font=\fontsize{10.5}{12}\selectfont,
                 fill=white,inner sep=1pt}
]

\path[use as bounding box] (0,0) rectangle (15.6,8.3);

% Reusable pieces -------------------------------------------------------------
\newcommand{\rlbubbletwo}[4]{%
  \draw[diagram] (#1,#2) ++(-0.90,0) -- ++(0.48,0);
  \draw[diagram] (#1,#2) ++( 0.42,0) -- ++(0.48,0);
  \draw[diagram] (#1,#2) circle[radius=0.42];
  \node[vertexdot] at ($(#1,#2)+(-0.42,0)$) {};
  \node[vertexdot] at ($(#1,#2)+( 0.42,0)$) {};
  \node[nodecircle] at ($(#1,#2)+(0,-0.42)$) {#3};
  \node[nodecircle] at ($(#1,#2)+(0, 0.42)$) {#4};
}

\newcommand{\rltadpoletwo}[3]{%
  \draw[diagram] (#1,#2) circle[radius=0.35];
  \draw[diagram] (#1,#2) ++(0.35,0) -- ++(0.78,0);
  \node[vertexdot] at ($(#1,#2)+(0.35,0)$) {};
  \node[nodecircle] at ($(#1,#2)+(-0.35,0)$) {#3};
}

% Shift the whole drawn diagram slightly to the left --------------------------
\begin{scope}[xshift=-0.7cm]

% Top row: bubble integrals ---------------------------------------------------
\rlbubbletwo{4.20}{2.40}{1}{2}
\rlbubbletwo{7.80}{2.40}{1}{2}
\rlbubbletwo{11.40}{2.40}{1}{2}

\node[mathlabel] at (4.20,3.48)
  {\small $I_{\mathrm{bub}}^{D}(z;3,1)$};
\node[mathlabel] at (7.80,3.48)
  {\small $I_{\mathrm{bub}}^{D}(z;2,2)$};
\node[mathlabel] at (11.40,3.48)
  {\small $I_{\mathrm{bub}}^{D}(z;1,3)$};

% Back-and-forth arrows between left and center -------------------------------
\draw[thinarrow] (4.58,4.05) .. controls (5.25,4.58) and (6.55,4.58) .. (7.40,4.05);
\node[qlabel] at (5.98,4.1) {\small $\cQ_{2,1}$};

\draw[thinarrow] (7.28,4.30) .. controls (6.55,4.92) and (5.25,4.92) .. (4.70,4.30);
\node[qlabel] at (5.98,5.02) {\small $\cQ_{1,2}$};

% Back-and-forth arrows between center and right ------------------------------
\draw[thinarrow] (8.4,4.30) .. controls (8.95,4.92) and (10.25,4.92) .. (10.80,4.30);
\node[qlabel] at (9.62,4.1) {\small $\cQ_{1,2}$};

\draw[thinarrow] (11.02,4.05) .. controls (10.35,4.58) and (9.05,4.58) .. (8.38,4.05);
\node[qlabel] at (9.62,5.02) {\small $\cQ_{2,1}$};

% Arrows down to the two tadpoles ---------------------------------------------
\draw[thinarrow] (4.05,4.08) -- (3.00,5.82);
\node[qlabel] at (3.,4.95) {\small $\cQ_{1,2}$};

\draw[thinarrow] (11.55,4.08) -- (12.60,5.82);
\node[qlabel] at (12.6,4.95) {\small $\cQ_{2,1}$};

% Bottom row: tadpoles --------------------------------------------------------
\rltadpoletwo{2.85}{6.45}{1}
\rltadpoletwo{12.75}{6.45}{2}

\node[mathlabel] at (2.85,7.25)
  {\small $I_{\mathrm{tad}}^{D}(z_{1},z_{1,1};4,0)$};
\node[mathlabel] at (12.75,7.25)
  {\small $I_{\mathrm{tad}}^{D}(z_{2},z_{2,2};0,4)$};

\end{scope}

\end{tikzpicture}
\endgroup
\end{center}
\caption{Reduction ladder for the one-loop bubble integral $I^D_{\rm bub}(z;2,2)$. The two reduction operators $\cQ_{1,2}$ and $\cQ_{2,1}$ act by shifting the bubble's $\nu$-vectors or contraction to the twisted tadpole integral.}
\label{fig:bubblecontracted}
\end{figure}

We now study the action of the reduction operators on the bubble integral, which will illustrate the general identity \eqref{red-on-solutions2} for this special case. It is instructive to return to the GKZ integral rather than the restriction to physical variables. We observe that, when acting on $\cG(z,x)$, the differential operator in GKZ variables can be replaced by a differential operator in the integration variables,
\begin{equation}
    \cQ_{2,1} \cG(z,x) = z_1 x_2 + z_{1,2} x_2^2 + 2 z_{1,1} x_1 x_2 =  x_2 \partial_{x_1}\cG(z,x)\,,
\end{equation}
Proceeding similarly to section \ref{app_GKZ_integrals} by integrating by parts over $x_1$, we obtain 
\begin{align}\label{eq:bubbleQ1action}
    \cQ_{2,1} I^D_{\textrm{bub}}(z;\nu) = \begin{cases} - \kappa_{\textrm{bub}}(\nu) \int_{\R_+} dx_2 \, x_2^{\nu_2} \big( z_2 x_2+z_{2,2} x_2^2\big)^{-D/2} & \text{if}\ \ \nu_1 =1\,, \\
    -(\nu_1-1)\kappa_{\textrm{bub}}(\nu) \int_{\R_+^2} d^2x \, x_1^{\nu_1-2} x_2^{\nu_2}  \cG(z,x)^{-D/2} & \text{if} \ \ \nu_1>1\ . 
    \end{cases}
\end{align}
The case $\nu_1=1$ is related to the GKZ integral for a tadpole, defined as
\begin{equation}\label{eq:tadpolegkzint}
\begin{split}
&    I^D_{\textrm{tad}}(z_2,z_{2,2};\nu_2)=\frac{  \Gamma \left( D/2 \right) }{ 
    \Gamma\left( D- \nu_2 \right)
     \Gamma( \nu_2 ) } \int_{\R_+}dx_2\,          x_2^{\nu_2-1} 
\big( z_2 x_2+ z_{2,2} x_2^2\big){}^{-D/2 }\,.
\end{split}
\end{equation}
In terms of the full integrals, we have 
\begin{equation} \label{Q-bubble}
     \cQ_{2,1} I^D_{\textrm{bub}}(z;\nu)= 
     \begin{cases}
          - \nu_2 I^D_{\textrm{tad}}(z_2,z_{2,2};\nu_2+1) &  \textrm{if}\ \  \nu_1=1\,, \\
          - \nu_2 I^D_{\textrm{bub}}(z;\nu_1-1,\nu_2+1) &  \textrm{if}\ \  \nu_1 > 1 \,.
     \end{cases}
\end{equation}
where $z= (z_1,z_2,z_{1,2},z_{1,1},z_{2,2})$ is the full set of bubble GKZ coordinates. Noting that $\cA v= a_1$ and $\cA u=a_2$, we find $(\nu -\cA v)_1 = \nu_1-1$ and see that \eqref{Q-bubble} exactly matches the general result \eqref{red-on-solutions2}. Indeed, one evaluates \eqref{Cuv-evaluated} to give $C_{u,v}(\nu)=\nu_2/(\nu_1-1)$ for $\nu_1>1$ and $C_{u,v}(\nu)=\nu_2$ for $\nu_1=1$, which reproduces the correct prefactor when used in \eqref{red-on-solutions2}.
After $\nu_1$ iterations we arrive at the result
\begin{equation} \label{Q-bubble-it}
     \left(\cQ_{2,1}\right)^{\nu_1} I^D_{\textrm{bub}}(z;\nu_1,\nu_2) = 
    (-\nu_2)_{\nu_1}
     I^D_{\textrm{tad}}(z_2,z_{2,2};\nu_2+\nu_1) \,.
\end{equation}
In the Feynman graph, the original edge 1 has effectively been contracted, while the power of the other propagator has increased by $\nu_1$. Equations \eqref{Q-bubble} and \eqref{Q-bubble-it} are readily translated into equations for $\cQ_{1,2}$ under relabeling $1 \leftrightarrow 2$. The full set of reduction relations involving $\cQ_{2,1}$ and $\cQ_{1,2}$ can be used on any integral $I_{\rm bub}^D(z;\nu_1,\nu_2)$ and leads to a finite reduction ladder. We have depicted the result for $I_{\rm bub}^D(z;2,2)$ in figure~\ref{fig:bubblecontracted}.

\paragraph{Reduction on the physical locus.} 
Restricting to the physical locus, we can verify that the known solutions $I_{\rm bub}^D(p,m;\nu_1,\nu_2)$, $I_{\rm tad}^D(m;\nu)$ satisfy \eqref{Q-bubble}, and its $\cQ_{1,2}$-equivalent, in both the general case and the cut case.
In terms of the first-order differential operators given in (\ref{eq:bubbleQm2})-(\ref{eq:bubbleE0}), we now have a system of differential equations for the bubble, 
\begin{align} \label{Q21red}
             &\cQ_{2,1} I^D_{\textrm{bub}}(p,m_1,m_2;\nu_1,\nu_2)= 
     \begin{cases}
          - \nu_2 I^D_{\textrm{tad}}(m_2;\nu_2+1) &  \textrm{if}\ \  \nu_1=1\,, \\
          - \nu_2 I^D_{\textrm{bub}}(p,m_1,m_2;\nu_1-1,\nu_2+1) &  \textrm{if}\ \  \nu_1 > 1 \,,
     \end{cases}\\ \label{Q12red}
                   & \cQ_{1,2} I^D_{\textrm{bub}}(p,m_1,m_2;\nu_1,\nu_2)= 
     \begin{cases}
          - \nu_1 I^D_{\textrm{tad}}(m_1;\nu_1+1) &  \textrm{if}\ \  \nu_2=1\,, \\
          - \nu_1 I^D_{\textrm{bub}}(p,m_1,m_2;\nu_1+1,\nu_2-1) &  \textrm{if}\ \  \nu_2 > 1 \,,
     \end{cases} \\
&     \big(\mathcal{E}_0 + \tfrac{D}{2} \big) I^D_{\textrm{bub}}(p,m_1,m_2;\nu_1,\nu_2)= 0\,. \label{E0bub}
\end{align}
 Explicit expressions for the bubble and tadpole functions are given in appendix~\ref{ap:tadbub}. One readily checks that these solutions indeed satisfy these differential relations. 

\paragraph{Using rationalized coordinates.} There is a natural change of coordinates on the physical locus that `rationalizes' the poles of bubble integrals, which otherwise are only  algebraic functions of $m_1^2$, $m_2^2$, and $p^2$. The relevant coordinates are $w,\bw,p$ defined via
$w\bw = m_1^2/p^2$ and $(1-w)(1-\bw)= m_2^2/p^2$.
In terms of these coordinates, the reduction operators (\ref{eq:bubbleQm2}) take the form 
\begin{equation}\label{eq:bubble-w}
   \begin{array}{rl}
       \cQ_{2,1} f(p,w,\bw;\nu)&= \big(2\nu_1 + \nu_2  - D + w \partial_{w}+\bw\partial_{\bw} \big) f(p,w,\bw;\nu) \,,\\
       \cQ_{1,2} f(p,w,\bw;\nu)&= \big( 2\nu_2 + \nu_1 -D -(1-w) \partial_{w} - (1-\bw)\partial_{\bw}\big) f(p,w,\bw;\nu)\,.
   \end{array}
\end{equation}
In the canonical basis ($\epsilon$-form) \cite{henn_multiloop_2013}, the bubble family is represented by the integral $J_2$, with single powers of the propagators $\nu_1=\nu_2=1$, evaluated in $D=2-2\eps$ dimensions and normalized by its leading singularity,
\begin{equation}
    J_2(p,w,\bw) = -\frac{1}{2}p^2 (w-\bw)I_{\rm bub}^{2-2\epsilon}(p,w,\bw;1,1)\,.
\end{equation}
The tadpole family is represented by the integral $J_1$, likewise evaluated in $D=2-2\eps$ dimensions,
\begin{equation}
    J_1(m) = I_{\rm tad}^{2-2\epsilon}(m;1)\,.
\end{equation}
Translating the reduction operator relations \eqref{Q21red}, \eqref{Q12red} into 
differential relations for $J_2$ and $J_1$, we find
\begin{align}
    \big(2-D+w \partial_{w}+\bw\partial_{\bw} \big) J_2(p,w,\bw)
    &= \epsilon \frac{w-\bw}{2(1-w)(1-\bw)} J_1(m_2)\,, \\
        \big(2-D-(1-w) \partial_{w}-(1-\bw)\partial_{\bw} \big) J_2(p,w,\bw)
    &= \epsilon \frac{w-\bw}{2 w \bw} J_1(m_1)\,.
\end{align}
We observe that the appropriate reduction operators for $J_2$ are given by an integer shift of those used for the original integrals.

\paragraph{Reduction for cut integrals.} Note that the differential equations \eqref{Q21red}, \eqref{Q12red}, \eqref{E0bub} also  hold for the cuts of the bubble, whose expressions are also given in appendix \ref{ap:tadbub}. The integrals on the right-hand side must be cut in the same way, so they vanish whenever a cut is applied to a contracted edge.
In the special case that $\nu_1=\nu_2=1$, we have for a one-line cut,
\begin{align}
             \cQ_{2,1} \mathcal{C}_1 I^D_{\textrm{bub}}(p,m_1,m_2;1,1) &= 
        0\,,
\\
                    \cQ_{1,2} \mathcal{C}_1I^D_{\textrm{bub}}(p,m_1,m_2;1,1) &= 
          -  \mathcal{C}_1 I^D_{\textrm{tad}}(m_1;2)\,,  \\
     \big(\mathcal{E}_0 + \tfrac{D}{2} \big) \mathcal{C}_1I^D_{\textrm{bub}}(p,m_1,m_2;1,1) &= 0\,,
\end{align}
and for the maximal cut,
\begin{align}
             \cQ_{2,1} \mathcal{C}_{12} I^D_{\textrm{bub}}(p,m_1,m_2;1,1) &= 
        0\,,
\\
                    \cQ_{1,2} \mathcal{C}_{12} I^D_{\textrm{bub}}(p,m_1,m_2;1,1) &= 
         0 \,, \\
     \big(\mathcal{E}_0 + \tfrac{D}{2} \big) \mathcal{C}_{12}I^D_{\textrm{bub}}(p,m_1,m_2;1,1) &= 0\,,
\end{align}
These equations can be verified for the expressions given in appendix \ref{ap:tadbub}.

\subsection{Differential equations}

The reduction operators can be used to determine a first order system of differential equations satisfied by the integrals $I_{\rm bub}^D(p,m;\nu)$ and $I_{\rm tad}^D(p,m;\nu)$.

\paragraph{First order system.} Let us stress that the differential relations \eqref{Q21red}, \eqref{Q12red} close on a finite set of integrals $I_{\rm bub}^D(p,m;\nu)$ and $I_{\rm tad}^D(p,m;\nu)$.
For example, if we are interested in the system of differential equations satisfied by $I_{\rm bub}^D(p,m;2,2)$, the action of the reduction operators $\cQ_{1,2}$ and $\cQ_{2,1}$ forces us also to consider the bubble integrals $I_{\rm bub}^D(p,m;3,1)$, $I_{\rm bub}^D(p,m;1,3)$ and the tadpole integrals $I_{\rm tad}^D(p,m;4,0)$, $I_{\rm tad}^D(p,m;0,4)$. The complete reduction ladder is shown in figure~\ref{fig:bubblecontracted}. 

The reduction ladder can also be turned into a first-order differential equation system taking the Pfaffian form 
$d \vec I = A \cdot \vec I$, where $A$ is a one-form-valued matrix. For the bubble $I_{\rm bub}^D(p,m;2,2)$, we have to introduce a vector 
\beq \label{vecI}
  \vec I(p,m) = (I_{\rm bub}^D(2,2),I_{\rm bub}^D(3,1),I_{\rm bub}^D(1,3),I_{\rm tad}^D(4,0),I_{\rm tad}^D(0,4) )^T\ . 
\eeq
The action of the reduction operators $\cQ_{1,2}$, $\cQ_{2,1}$ and the Euler operator $\cE_0$ on $I_{\rm bub}^D(2,2),I_{\rm bub}^D(3,1),I_{\rm bub}^D(1,3)$ yields nine first-order differential equations that can be solved for 
$\partial_{p^2} I_{\rm bub}^D(\nu )$, $\partial_{m_1^2}I_{\rm bub}^D(\nu )$, and $\partial_{m_2^2}I_{\rm bub}^D(\nu )$ with $\nu \in \{(2,2),(3,1),(1,3)\}$. The derivatives of the tadpoles $I_{\rm tad}^D(m_1;4,0)$, $I_{\rm tad}^D(m_2;0,4)$ are fixed by the Euler relation alone. Performing these steps, we find 
\beq \label{explicit-dI}
   d \vec I = A \cdot \vec I\ , \qquad A = \left(\begin{array}{ccccc} \Omega & 2 \omega_{12}& 2 \omega_{21}& 0& 0\\ \omega_{21} & \Omega_{+} & 0 & 3 \omega_{12} &0 \\ \omega_{12}& 0 & \Omega_{-} & 0 & 3\omega_{21} \\
   0& 0 & 0 &  \omega_{1} &0\\ 0& 0& 0 &0 &\omega_{2}\end{array} \right)\ ,
\eeq
where $\omega_{i}= \frac{D-8}{2} d\log m_i^2$, $\Omega_{\pm} = \Omega \pm (\omega_{21} - \omega_{12})$, and 
\begin{equation}
    \omega_{ij} = \frac{m_i^2}{\sqrt{\lambda}}d \log \frac{p^2+m_i^2-m_j^2-\sqrt\lambda}{p^2+m_i^2-m_j^2+\sqrt\lambda}\,,
    \quad \Omega= -\tfrac{D-4}{2} d \log p^2 + \tfrac{D-6}{2} d \log \lambda\, ,
\end{equation}
with $\lambda=\big(p^2 - (m_1+m_2)^2 \big)\big( p^2-(m_1-m_2)^2\big)$. Although it is not essential for the subsequent discussion, we have included the explicit result as a proof of principle. Evidently, the same procedure can be applied to any bubble diagram for arbitrary values of $\nu_1,\nu_2$.

\paragraph{Algebraic relations from toric equations.} The above system of differential equations has rank 5. However, it is known that there are only 3 master integrals for the bubble $I^D_{\rm bub} (2,2)$, which implies that there must be algebraic relations among the entries of \eqref{vecI}. These relations follow directly from the toric equations
\beq
   \big(\partial_1 \partial_{2,2}-\partial_2 \partial_{1,2}\big)I_{\rm bub}^D(\nu_1,\nu_2) = 0\, , \quad \big(\partial_1 \partial_{1,2}-\partial_2\partial_{1,1}\big) I_{\rm bub}^D(\nu_1,\nu_2) =0  
    \ . 
\eeq
To find the corresponding algebraic relations one first reduces the derivatives in $z$ to derivatives in physical variables as above. Then we apply the fact that the derivatives of $I_{\rm bub}^D(\nu_1,\nu_2)$ can be expressed in terms of the entries of $\vec I$ using \eqref{explicit-dI}. Doing this twice we find two algebraic relations. The first is given by 
\beq \label{algebraic-relation1}
(D-6)I_{\rm bub}^D(3,1)
= \frac{1}{\mathcal{N}}\Big(
(D-6)\gamma_{22}\,I_{\rm bub}^D(2,2)
+ \gamma_{40}\,I_{\rm tad}^D(4,0)-
12m_2^4\,I_{\rm tad}^D(0,4)\Big)\, ,
\eeq
with polynomial factors
\begin{eqnarray}
   \gamma_{22} &=& \frac{1}{2}(D-4)(p^2-m_1^2+m_2^2)^2-2 p^2m_2^2\, , \nonumber
 \\
   \gamma_{40}&=& 
6m_1^2\left[(D-6)(p^2-m_1^2)+(D-4)m_2^2\right]
\, ,\\
   \mathcal N &=&
(D-6)(p^2)^2
+2p^2(m_1^2+m_2^2)
-(D-4)(m_1^2-m_2^2)^2\ , \nonumber
\end{eqnarray}
while the second is obtained by exchanging $I_{\rm bub}^D(1,3) \rightarrow I_{\rm bub}^D(3,1)$, $I_{\rm tad}^D(4,0) \leftrightarrow I_{\rm tad}^D(0,4)$, $m_1 \leftrightarrow m_2$. These relations 
can also be found without going to the physical variables. Using the fact that both the Euler equations and the reduction equations \eqref{Q-bubble} are first order, one readily solves for the $\partial_z$-derivatives in terms of the entries of $\vec I$ and eliminates all partial derivatives in the toric equations. This strategy can be used for all one-loop diagrams to find the missing algebraic relations.

\subsection{Some cases of special kinematics}

We close this section with comments on reduction operators for bubbles with one massless propagator or with equal masses.

\paragraph{Bubble with one massless propagator.} Consider the bubble depicted in figure~\ref{fig:bubble}, now with one of its internal masses set to zero, $m_2=0$. The Lee-Pomeransky polynomial now lacks the $x_2^2$ term, and the $\A$ matrix for the GKZ system is the same as before except that the column $a_{2,2}$ is absent:
\begin{equation}\label{eq:bubbleA1m}
   \A= \left(
\begin{array}{cccc}
 1 & 1 & 1 & 1  \\
 1 & 0 & 1 & 2  \\
 0 & 1 & 1 & 0  \\
\end{array}
\right)\, , \quad \nu=\begin{pmatrix}
    D/2\\
    \nu_1\\
    \nu_2
\end{pmatrix}\,.
\end{equation}

\begin{figure}[t]
\begin{center}
   \includegraphics[width=0.5\linewidth]{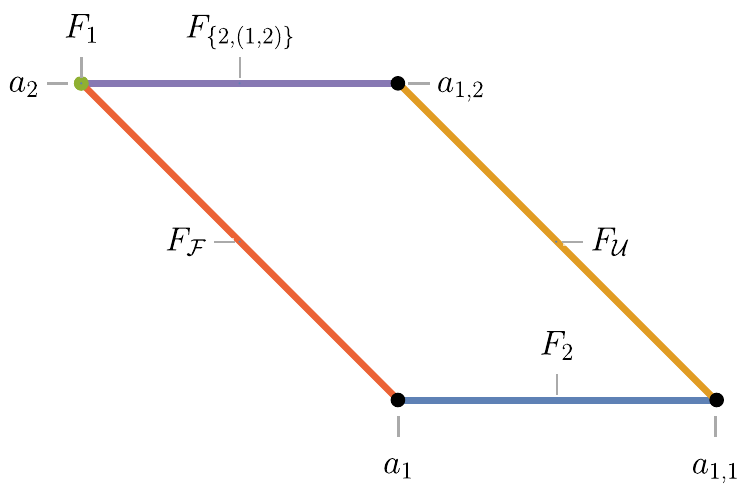}
   \caption{The polytope ${\rm Conv}({\mathcal A})$ for the bubble with $m_2=0$, with the matrix $\cA$ given in \eqref{eq:bubbleA1m}.
   }\label{fig:ConvBub1m}
\end{center}
\end{figure}

We have depicted the polytope $\rm{Conv}(\A)$ in figure \ref{fig:ConvBub1m}. Notice that the edge face $F_1$ still exists, but it is no longer a facet, and it is no longer resonant for $D/2 \notin \mathbb{Z}$. There is a new facet $F_{2,(1,2)}$, which is not resonant either.
However, $F_2$ is still a resonant facet. Its reduction operator can be constructed as before, and is found to be 
\begin{equation}
    \cQ_{1,2} = z_2 \partial_1+z_{1,2} \partial_{1,1}\, ,
\end{equation}
which is the same as the operator $\cQ_{1,2}$ in the generic case after setting $z_{2,2} \to 0$. Its restriction to physical kinematics is
\begin{equation}
    \cQ_{1,2} f(p,m;\nu)= \big((m_1^2-p^2) \partial_{m_1^2} + 2 \nu_2+\nu_1 - 2 \nu_0\big) f(p,m;\nu)\,.
\end{equation}

The absence of a reduction operator for edge 1 has two implications. First, the contraction of edge 1 is impossible: this would result in a tadpole with a massless propagator, which is zero in dimensional regularization. Second, the single-propagator cut of edge 2 is not sensible. Such a cut integral is likewise interpreted to give zero, as the limit of a mass-discontinuity in the case of zero mass \cite{Abreu:2017ptx}.
With the restricted Euler operator, 
\begin{equation}\label{eq:1mbubE0}
    \cE_0 f(z;\nu)= \big(\nu_1+\nu_2  + p^2 \partial_{p^2} + m_1^2 \partial_{m_1^2}\big) f(p,m;\nu) \,,
\end{equation}
we can write the system of differential equations
\begin{align}
                   & \cQ_{1,2} I^D_{\textrm{bub}}(p,m_1,0;\nu_1,\nu_2)= 
     \begin{cases}
          - \nu_1 I^D_{\textrm{tad}}(m_1;\nu_1+1) &  \textrm{if}\ \  \nu_2=1\,, \\
          - \nu_1 I^D_{\textrm{bub}}(p,m_1,0;\nu_1+1,\nu_2-1) &  \textrm{if}\ \  \nu_2 > 1 \,,
     \end{cases} \\
&     \big(\mathcal{E}_0 + \tfrac{D}{2} \big) I^D_{\textrm{bub}}(p,m_1,0;\nu_1,\nu_2)= 0\,.
\end{align}
Explicit expressions are given in appendix \ref{ap:tadbub}. The cut $\mathcal{C}_2$ of the massless edge gives zero on all terms. Noting that the cut $\mathcal{C}_1$ does not change the tadpole at all, we observe that $\mathcal{C}_1 I^D_{\textrm{tad}}(m_1^2;\nu_1+1)$ satisfies exactly the same differential equations as the uncut bubble. Only the boundary conditions differ. Both are expressed in terms of ${}_2F_1$ hypergeometric functions. This is the consequence of the absence of resonance of the face $F_1$. We do, however, find a simpler system for the maximal cut in the case $\nu_1=1$, which then reduces to  a power function.
\begin{align}
                   & \cQ_{1,2} \mathcal{C}_{12} I^D_{\textrm{bub}}(p,m_1,0,1,\nu_2)= 
0
     \\
&     \big(\mathcal{E}_0 + \tfrac{D}{2} \big) \mathcal{C}_{12} I^D_{\textrm{bub}}(p,m_1,0;1,\nu_2)= 0\,.
\end{align}

\paragraph{Bubble with equal nonzero masses.}

If $m_1=m_2$, then there is no straightforward restriction from the GKZ system to physical kinematics, so we cannot follow the procedure described above to obtain a reduction operator. However, there still exists a first-order operator that acts upon the bubble to result in the tadpole. We use the rationalized variables introduced above, where now with equal masses, we have $\bw=1-w$. For unit powers of propagators, the reduction operator is found to be
\begin{equation}
    \cQ_w  = 3 - D + \left(w-\tfrac{1}{2}\right) \partial_w \,,
\end{equation}
which is not a straightforward limit or symmetrization of the reduction operators with $m_1 \neq m_2$, although the same terms appear with different relative coefficients.
It acts on the bubble integral to give
\begin{equation}
    \cQ_w I_{\rm bub}^{2-2\epsilon}(p,w,1-w;1,1) = 
    -\frac{\epsilon}{m^2} I_{\rm tad}^{2-2\epsilon}(m;1)\,.
\end{equation}

\section{GKZ reductions for the triangle diagram \label{sec:triang_example}}

We now study the triangle diagram with generic kinematics, and construct its GKZ system and reduction operators. Here, large parts will be similar to what we have done so far for the bubble diagram. 
Again, we will see that these reduction operators will act on the triangle integral by contracting an edge, while simultaneously performing a shift. A new feature here, compared to the bubble, is that we find multiple reduction operators for each edge.

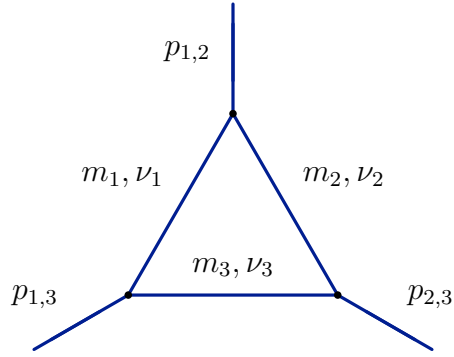
\begin{figure}[h!]
\centering
\begingroup
\definecolor{rlLineBlue}{RGB}{0,30,145}
\begin{tikzpicture}[
line cap=round,
line join=round,
very thick,
diagram/.style={draw=rlLineBlue},
text=black
]
% Symmetric triangle vertices
\coordinate (v12) at (0,1.6);
\coordinate (v13) at (-1.3856,-0.8);
\coordinate (v23) at ( 1.3856,-0.8);

    % External-leg endpoints
    \coordinate (e12) at (0,3.05);
    \coordinate (e13) at (-2.6326,-1.52);
    \coordinate (e23) at ( 2.6326,-1.52);

    % Internal triangle
    \draw[diagram] (v13) -- node[midway, above left=2pt] {$m_1, \nu_1$} (v12);
    \draw[diagram] (v12) -- node[midway, above right=2pt] {$m_2, \nu_2$} (v23);
    \draw[diagram] (v23) -- node[midway, above=2pt] {$m_3, \nu_3$} (v13);

    % External legs
    \draw[diagram] (e12) -- (v12);
    \draw[diagram] (e13) -- (v13);
    \draw[diagram] (e23) -- (v23);

    % Momentum arrows, all incoming
    \draw[diagram] ($(e12)!0.18!(v12)$) -- ($(e12)!0.70!(v12)$)
        node[midway, left=4pt] {$p_{1,2}$};

    \draw[diagram] ($(e13)!0.18!(v13)$) -- ($(e13)!0.70!(v13)$)
        node[midway, above left=2pt] {$p_{1,3}$};

    \draw[diagram] ($(e23)!0.18!(v23)$) -- ($(e23)!0.70!(v23)$)
        node[midway, above right=2pt] {$p_{2,3}$};

    \node at (0,1.6)[circle,fill=black,inner sep=1pt]{};
    \node at (-1.3856,-0.8)[circle,fill=black,inner sep=1pt]{};
    \node at ( 1.3856,-0.8)[circle,fill=black,inner sep=1pt]{};
\end{tikzpicture}
\endgroup
\caption{The 1-loop triangle diagram.}
\label{fig:triangle}

\end{figure}

\subsection{GKZ system and reduction operators}\label{sec:trianggkz}

The Lee-Pomeransky polynomial for the scalar triangle is
\begin{equation} \label{LeeP-tri}
   \cG= \left(x_1+x_2+x_3 \right)(1+m_{1}^{2}x_{1}+m_{2}^{2}x_{2}+m_{3}^{2}x_{3})- p_{1,2}^2 x_1 x_2 -p_{2,3}^2 x_2x_3 - p_{1,3}^2 x_1x_3 \,,
\end{equation}
which determines the Feynman integral $I^D_{\mathrm{tri}}(p,m;\nu)$ when using 
\eqref{eq:LPrep}.
 In GKZ variables, the integral takes the form
    \begin{equation}
        \label{eq:triangint-gkz}
        I^D_{\mathrm{tri}}(z;\nu)=\kappa_{\rm tri}(\nu)\int_{\R_+^3}d^{3}x \;  x_1^{\nu_1-1} x_2^{\nu_2-1} x_3^{\nu_3-1} \cG(z,x)^{-D/2}\,.
    \end{equation}
    where $\kappa_{\rm tri}(\nu)$ is obtained from \eqref{def-kappa}, and 
      \begin{equation}
\label{eq:triangpoly}\cG(z,x)=\sum_{e=1}^{3}z_{e} x_{e}+\sum_{e=1}^{3}\sum_{e'=e}
        ^{3} z_{e,e'}x_{e} x_{e'}\,.
    \end{equation}
    Comparing \eqref{eq:triangpoly} with \eqref{LeeP-tri}, we see that the restriction of the GKZ variables to physical variables is implemented by taking
    \begin{equation}
        \begin{array}{rl}
            z_e|_{\rm phys} =         & 1\ ,                                              \\
            z_{e,e'}|_{\rm phys}=     & m_e^2+m_{e'}^2-p_{e,e'}^2\; \text{ for } e<e' \, , \\
            z_{e,e}\vert_{\rm phys} = & m_e^2\,.                                          
        \end{array}
    \end{equation}
Using \eqref{eq:poly} and  \eqref{cA-a-form}, we thus read off GKZ data for the triangle integral, 
    \begin{equation}
        \label{eq:triangAandv}\A =
        \begin{pmatrix}
            1 & 1 & 1 & 1 & 1 & 1 & 1 & 1 & 1    \\
            1 & 0 & 0 & 1 & 1 & 0 & 2 & 0 & 0    \\
            0 & 1 & 0 & 1 & 0 & 1 & 0 & 2 & 0    \\
            0 & 0 & 1 & 0 & 1 & 1 & 0 & 0 & 2
        \end{pmatrix}
        \qquad \nu=
        \begin{pmatrix}
            D/2 \\
            \nu_1   \\
            \nu_2   \\
            \nu_3
        \end{pmatrix}\ .
    \end{equation}
Recall that each column vector $a_i$ corresponds to a variable $z_i$. Here we have chosen the ordering of the columns associated to $z = (z_{e},z_{1,2},z_{1,3},z_{2,3},z_{e,e})$.
As in \eqref{nu-integer}, we assume that $\nu_1, \nu_2, \nu_3$ are positive integers. Using the general expressions \eqref{toric_GKZ} and \eqref{def-Euler}, the data \eqref{eq:triangAandv} uniquely fix the toric and Euler operators. Instead of presenting their explicit forms here, we refer to the general one-loop results \eqref{toric-general}, \eqref{Euler-oneloop}, which specialize directly to the triangle diagram.

\paragraph{Resonant faces.}

\begin{figure}
\begin{center}
   \hspace*{2cm}\includegraphics[width=0.55\linewidth]{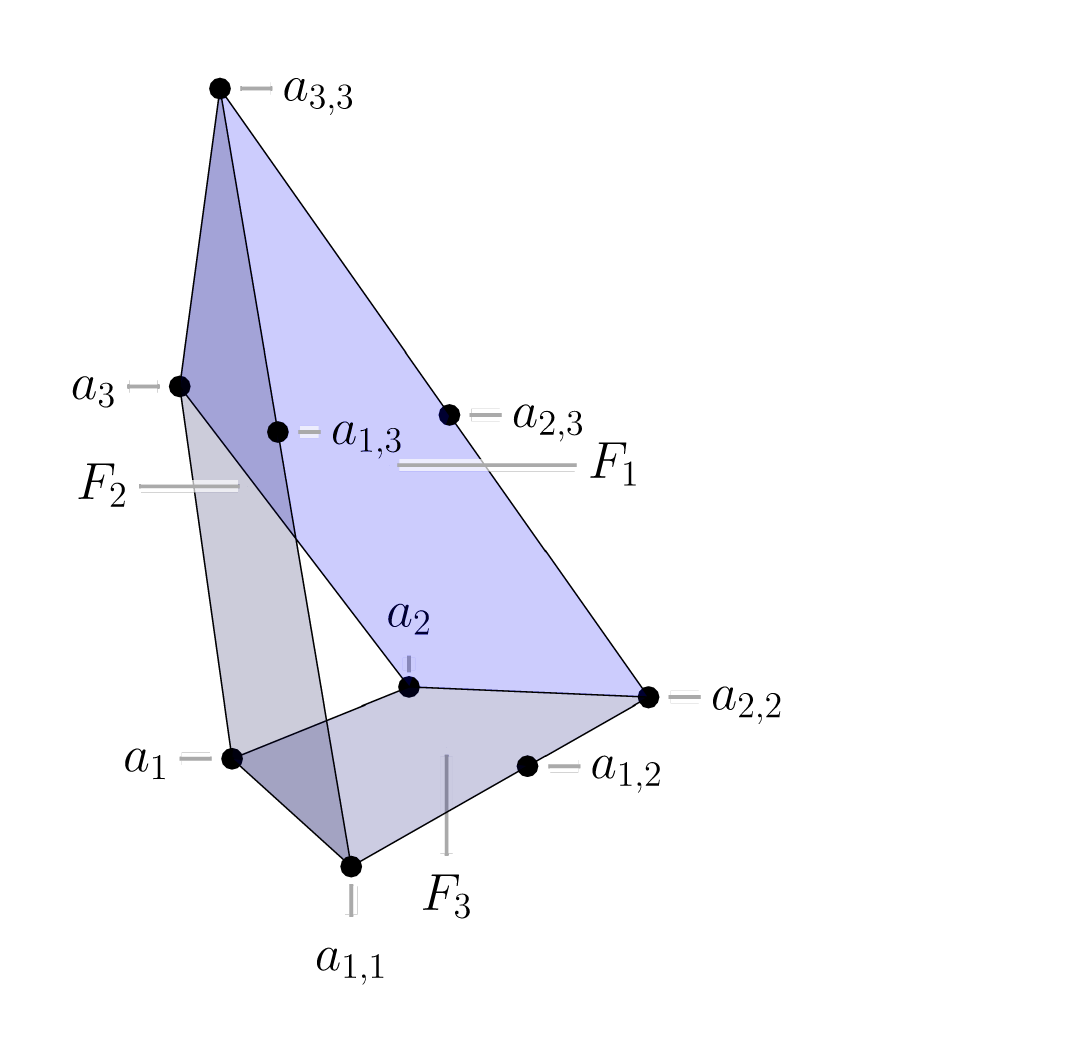}
   \vspace*{-.5cm}
\end{center}
    \caption{The polytope ${\rm Conv}({\mathcal A})$ for the triangle with three massive propagators. The column vectors contained in the face $F_1$  are given by $a_2, a_3, a_{2,3}, a_{2,2}, a_{3,3}$, and similarly for the other edge faces. Notice that each edge face has the geometry of the polytope of a bubble. }
\label{fig:triangle-polytope}
\end{figure}

As in the case of the bubble diagram, the edge faces are facets which are resonant, and there are two more facets corresponding to the Symanzik polynomials. The linear functionals associated to edges were given in equation (\ref{eq:LeFunctional}).
The explicit expressions of their linear functionals are 
\begin{equation}
\begin{split}
&    L_1 = (0,1,0,0)\,, \quad
    L_2 = (0,0,1,0)\,, \quad
    L_3 = (0,0,0,1)\,, \quad \\
&    L_{\mathcal F} = (-1,1,1,1)\,, \quad
    L_{\mathcal U} = (2,-1,-1,-1)\,,
\end{split}
\end{equation}
where the vectors are understood as multiplying  the columns of $\A$  on the left.
Again, assuming integer $\nu_e$,
  the facet $F_{\mathcal F}$ is resonant only if $D/2 \in \mathbb {Z}$, and the facet $F_{\mathcal U}$ is resonant only if $D \in \mathbb {Z}$, while the three edge facets are resonant for any value of $D \in \mathbb{C}$. 
We proceed under the assumption that $D$ is generic, and hence the only resonant facets are the edge facets.
To each edge facet $F_i$ we associate a GKZ subsystem $\cA_{F_i}$ via \eqref{def-AF}. For $F_1$ this takes the form
\begin{equation}\label{eq:triangF!matrix}
    \A_{F_1}=\begin{pmatrix}
        1 & 1 & 1 &1 &1\\
        0 & 0& 0& 0& 0\\
        1 & 0 & 1 &2 &0\\
        0 & 1 & 1 &0 &2
    \end{pmatrix}\ .
\end{equation}
Observe that this matrix is the $\cA$-matrix of the bubble diagram \eqref{eq:bubbleAandV} studied in the previous section, with an additional row of zeros, as also seen in figure~\ref{fig:triangle-polytope}.

\paragraph{Reduction operators.}

For each of the resonant faces, we obtain reduction operators. Here, we will see the first essential difference from the simpler case of the bubble: each edge contraction gives rise to \emph{multiple} reduction operators. For the triangle, we obtain an independent reduction operator for every ordered pair of edges. 
To see this, let us consider the reduction operators for the face $F_1$ as above.
From its linear functional $L_1$, we construct the operator
\begin{equation}
    \cE_{F_1} = L_{F_1}(\cE_1)= \cE_1 =\theta_1+\theta_{1,2}+\theta_{1,3}+2 \theta_{1,1}\, .
\end{equation}
The next step is to find vectors $u,v$ that admit an associated reduction operator $\cQ^{F_1}_{u,v}$. Let us pick $\partial^v = \partial_1$.
Again we find that we can  satisfy the relation (\ref{eq:PEFsimQd}) by choosing $\partial^u = \partial_2$. Concretely, we use the notation \eqref{short-not} and solve the condition 
\begin{equation}\label{eq:triangleQ12deriv}
 \partial_2 \cE_{F_1}   \simeq \cQ_{2,1} \partial_1 \,,
\end{equation}
with the reduction operator given by
\begin{equation}\label{eq:triangleQ12}
    \cQ_{2,1} =  z_1  \partial_2 + z_{1,2} \partial_{2,2}+z_{1,3}\partial_{2,3}+2 z_{1,1}\partial_{1,2}\,, \qquad \cA u = a_2, \quad \cA v =a_1\ . 
\end{equation}
Here, the subscripts correspond to the index of $a_i$ selected by $u,v$, respectively.

By the symmetry of the diagram, we are able to freely exchange the edges $2$ and $3$ and thus obtain another reduction operator. In particular, the relation
\begin{equation}\label{eq:triangleQ13deriv}
 \partial_3 \cE_{F_1}   \simeq \cQ_{3,1} \partial_1 \,,
\end{equation}
is satisfied for the following reduction operator:
\begin{equation}\label{eq:triangleQ13}
    \cQ_{3,1}= z_1  \partial_3+ z_{1,2} \partial_{2,3}+z_{1,3}\partial_{3,3}+2 z_{1,1}\partial_{1,3}\,, \qquad \cA u = a_3, \quad \cA v =a_1\,.
\end{equation}
We now have two reduction operators, $\cQ_{2,1}$ and $\cQ_{3,1}$, both associated to the edge $x_1=0$. By permutation symmetry, each of the other two edges can again produce two reduction operators $\cQ_{e',e}$, $e\neq e'$, with suitable permutations of the indices in equations (\ref{eq:triangleQ12}) and (\ref{eq:triangleQ13}). It is not hard to check that all such reduction operators satisfy \eqref{Q-commutes-with-toric}, i.e.~commute with the toric operators yielding linear combinations of toric operators.
Furthermore, modulo the Euler relations, not all the reduction operators obtained for the triangle are independent, and it is straightforward to derive a linear relation among them.

\subsection{Contractions, physical locus, and differential equations}\label{ssec:triangcontract}

We can explicitly verify equation~\eqref{red-on-solutions2} for all reduction operators, thereby confirming this general result. This is simplified by the fact that, for example, $\cQ_{2,1} \cG(z,x) =   x_2 \partial_{x_1}\cG(z,x)$ and the partial integration can be performed without much effort. Concretely, the equation \eqref{red-on-solutions2} for the triangle takes the form 
\beq \label{red-on-triang}
 \cQ_{e',e} I_{\rm tri}^D (z;\nu)  = 
    \begin{cases}  - \nu_{e'} \, I_{\rm bub}^D (z_{F_e}; \nu_{F_e} +\delta_{e'})\ & \text{if} \ \ \nu_e = 1 \ ,\\
 - \nu_{e'}\,  I_{\rm tri}^D(z;\nu-\delta_e +\delta_{e'})\ & \text{if} \ \ \nu_e > 1 \ ,
 \end{cases}
\eeq
where $\delta_e$ are the unit vectors with a $1$ in the $e$-th entry, and $z_{F_e}$ are the coordinates associated with the face $F_e$, e.g.~for $F_1$ we find $z_{F_1} = (z_2,z_3,z_{2,3},z_{2,2},z_{3,3})$.
As for the bubble, we can iterate these conditions and, for example, obtain
\begin{equation}
  \left(\cQ_{2,1} \right)^{\nu_1} I^D_{\textrm{tri}}(z;\nu) \\
= (-\nu_2)_{\nu_1}
     I^D_{\textrm{bub}}(z_{F_1};\nu_2+\nu_1,\nu_3) \,.
\end{equation}

\paragraph{Reduction operators on the physical locus.}
Using the Euler relations $\cE_e + \nu_e$ to eliminate the differentials of the form $\partial_e$, we can impose the restriction to physical variables and find the reduction operator
\beq
    \cQ_{2,1} f(z;\nu) =\Big( (m_1^2-m_2^2-p_{1,2}^2)\partial_{m_2^2}+(p_{1,3}^2-p_{1,2}^2-p_{2,3}^2)\partial_{p_{2,3}^2}-2 p_{1,2}^2\partial_{p_{1,2}^2}-\nu_2\Big) f(p,m;\nu)\,,
\eeq
with the others following by permutation of the labels as displayed in \eqref{gen-one-loop-phys}. These six reduction operators are not linearly independent, as they are expressed in terms of differentials in six kinematic variables. They are constrained by the overall dimensional scaling imposed by the relation $(\cE_0+\nu_0)f(z;\nu) = 0$ on any GKZ solution, which translates to 
    \begin{equation}\label{eq:triangphyseuler}
        \Big(\tfrac{D}{2}-\sum_{e} \big(  \nu_e + m_{e}^{2}\partial_{m_e^2}\big)-\sum
        _{e<e'}p_{e,e'}^{2}\partial_{p_{e,e'}^2}\Big)f(p,m;\nu) =
        0\,.
    \end{equation}

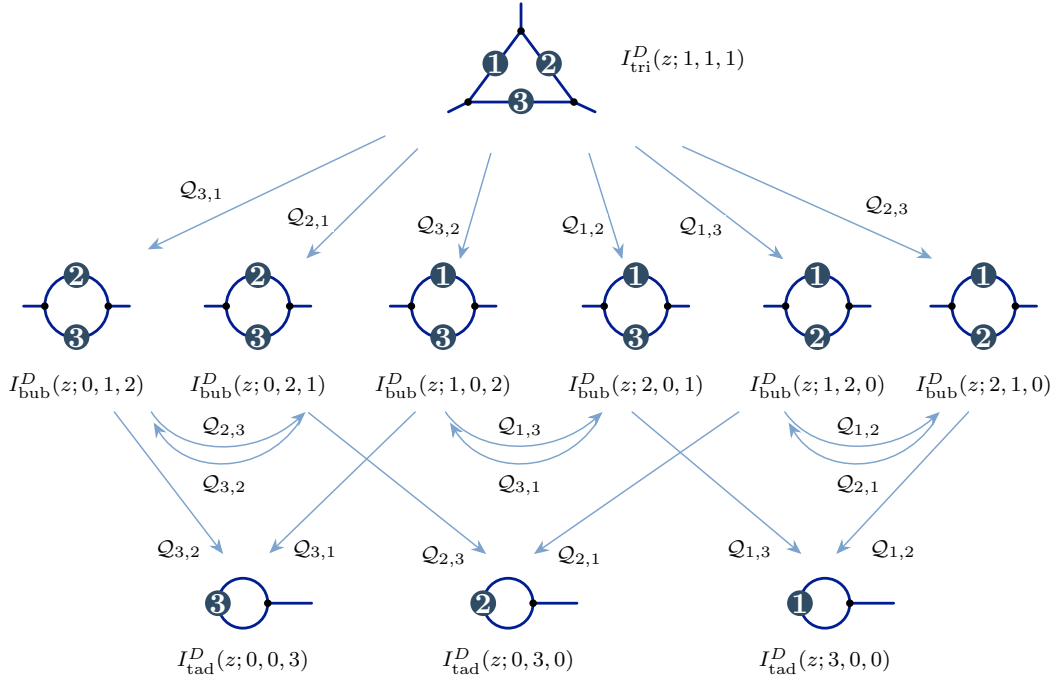
\begin{figure}[h!]
\begingroup
\definecolor{rlTitleBlue}{RGB}{0,42,210}
\definecolor{rlLineBlue}{RGB}{0,30,145}
\definecolor{rlArrowBlue}{RGB}{123,163,204}
\definecolor{rlNode}{RGB}{47,76,100}
\definecolor{rlPurple}{RGB}{145,31,150}
\begin{tikzpicture}[
  x=1cm,y=-1cm,
  line cap=round,line join=round,
  >=Stealth,
  diagram/.style={draw=rlLineBlue,line width=1.05pt},
  thinarrow/.style={draw=rlArrowBlue,line width=0.55pt,-{Stealth[length=2.0mm,width=1.6mm]}},
  nodecircle/.style={circle,fill=rlNode,text=white,inner sep=0pt,minimum size=0.34cm,
                     font=\fontsize{10.5}{10.5}\selectfont\bfseries},
  vertexdot/.style={circle,fill=black,inner sep=1pt},
  mainlabel/.style={text=rlPurple,anchor=west,font=\fontsize{15.5}{18}\selectfont},
  mathlabel/.style={font=\fontsize{9.6}{11.4}\selectfont},
  eqtop/.style={font=\fontsize{10.8}{12.8}\selectfont},
  qlabel/.style={font=\fontsize{10.0}{11.5}\selectfont,fill=white,inner sep=0.7pt},
  title/.style={fill=black!5,text=rlTitleBlue,anchor=west,inner xsep=0.22cm,inner ysep=0.13cm,
                font=\fontsize{16.5}{19}\selectfont}
]
\path[use as bounding box] (0,0) rectangle (14.4,10.8);

% Small reusable pieces --------------------------------------------------------
\newcommand{\rlnumber}[3]{\node[nodecircle] at (#1,#2) {#3};}
\newcommand{\rlbubble}[4]{%
  \draw[diagram] (#1,#2) ++(-0.70,0) -- ++(0.29,0);
  \draw[diagram] (#1,#2) ++( 0.41,0) -- ++(0.29,0);
  \draw[diagram] (#1,#2) circle[radius=0.42];
  % vertex dots at the two bubble vertices
  \node[vertexdot] at ($(#1,#2)+(-0.42,0)$) {};
  \node[vertexdot] at ($(#1,#2)+( 0.42,0)$) {};
  \node[nodecircle] at ($(#1,#2)+(0,-0.41)$) {#3};
  \node[nodecircle] at ($(#1,#2)+(0, 0.41)$) {#4};
}
\newcommand{\rltadpole}[3]{%
  \draw[diagram] (#1,#2) circle[radius=0.33];
  \draw[diagram] (#1,#2) ++(0.33,0) -- ++(0.58,0);
  % vertex dot at the tadpole vertex
  \node[vertexdot] at ($(#1,#2)+(0.33,0)$) {};
  \node[nodecircle] at ($(#1,#2)+(-0.33,0)$) {#3};
}

% Triangle ---------------------------------------------------------------------
\draw[diagram] (7.28,0.93) -- (7.28,1.29);
\draw[diagram] (7.28,1.29) -- (6.58,2.23) -- (7.98,2.23) -- cycle;
\draw[diagram] (6.58,2.23) -- (6.32,2.36);
\draw[diagram] (7.98,2.23) -- (8.26,2.36);

\node[vertexdot] at (7.28,1.29) {};
\node[vertexdot] at (6.58,2.23) {};
\node[vertexdot] at (7.98,2.23) {};

\rlnumber{6.94}{1.72}{1}
\rlnumber{7.65}{1.72}{2}
\rlnumber{7.28}{2.22}{3}
\node[eqtop,anchor=west] at (8.45,1.66) {\scriptsize $I_{\mathrm{tri}}^{D}(z;1,1,1)$};

% Arrows from triangle to bubble row -----------------------------------------
\draw[thinarrow] (5.48,2.68) -- (2.34,4.18);
\node[qlabel] at (3.05,3.42) {\scriptsize $\cQ_{3,1}$};
\draw[thinarrow] (5.91,2.85) -- (4.44,4.30);
\node[qlabel] at (4.47,3.78) {\scriptsize $\cQ_{2,1}$};
\draw[thinarrow] (6.88,2.91) -- (6.47,4.31);
\node[qlabel] at (6.19,3.88) {\scriptsize $\cQ_{3,2}$};
\draw[thinarrow] (8.18,2.91) -- (8.62,4.31);
\node[qlabel] at (8.1,3.88) {\scriptsize $\cQ_{1,2}$};
\draw[thinarrow] (8.80,2.82) -- (10.75,4.31);
\node[qlabel] at (9.66,3.89) {\scriptsize $\cQ_{1,3}$};
\draw[thinarrow] (9.42,2.82) -- (12.72,4.30);
\node[qlabel] at (12.12,3.62) {\scriptsize $\cQ_{2,3}$};

% Bubble row -------------------------------------------------------------------
\rlbubble{1.40}{4.93}{2}{3}
\rlbubble{3.8}{4.93}{2}{3}
\rlbubble{6.25}{4.93}{1}{3}
\rlbubble{8.8}{4.93}{1}{3}
\rlbubble{11.2}{4.93}{1}{2}
\rlbubble{13.4}{4.93}{1}{2}

\node[mathlabel] at (1.40,5.98) {\scriptsize $I_{\mathrm{bub}}^{D}(z;0,1,2)$};
\node[mathlabel] at (3.8,5.98) {\scriptsize $I_{\mathrm{bub}}^{D}(z;0,2,1)$};
\node[mathlabel] at (6.25,5.98) {\scriptsize $I_{\mathrm{bub}}^{D}(z;1,0,2)$};
\node[mathlabel] at (8.8,5.98) {\scriptsize $I_{\mathrm{bub}}^{D}(z;2,0,1)$};
\node[mathlabel] at (11.2,5.98) {\scriptsize $I_{\mathrm{bub}}^{D}(z;1,2,0)$};
\node[mathlabel] at (13.4,5.98) {\scriptsize $I_{\mathrm{bub}}^{D}(z;2,1,0)$};

% Horizontal paired bubble reductions ----------------------------------------
\draw[thinarrow] (2.39,6.33) .. controls (2.77,6.93) and (3.78,6.93) .. (4.43,6.33);
\node[qlabel] at (3.35,6.54) {\scriptsize $\cQ_{2,3}$};
\draw[thinarrow] (4.34,6.47) .. controls (3.85,7.18) and (2.84,7.18) .. (2.42,6.47);
\node[qlabel] at (3.35,7.3) {\scriptsize $\cQ_{3,2}$};

\draw[thinarrow] (6.28,6.33) .. controls (6.69,6.92) and (7.70,6.92) .. (8.36,6.33);
\node[qlabel] at (7.26,6.54) {\scriptsize $\cQ_{1,3}$};
\draw[thinarrow] (8.28,6.47) .. controls (7.80,7.18) and (6.79,7.18) .. (6.34,6.47);
\node[qlabel] at (7.26,7.3) {\scriptsize $\cQ_{3,1}$};

\draw[thinarrow] (10.77,6.33) .. controls (11.18,6.92) and (12.16,6.92) .. (12.82,6.33);
\node[qlabel] at (11.74,6.54) {\scriptsize $\cQ_{1,2}$};
\draw[thinarrow] (12.74,6.47) .. controls (12.25,7.18) and (11.24,7.18) .. (10.82,6.47);
\node[qlabel] at (11.74,7.3) {\scriptsize $\cQ_{2,1}$};

% Arrows from bubble row to tadpole row --------------------------------------
\draw[thinarrow] (1.9,6.33) -- (3.34,8.21);
\node[qlabel] at (2.75,8.16) {\scriptsize $\cQ_{3,2}$};
\draw[thinarrow] (5.88,6.33) -- (3.95,8.20);
\node[qlabel] at (4.55,8.16) {\scriptsize $\cQ_{3,1}$};
\draw[thinarrow] (4.47,6.34) -- (6.82,8.21);
\node[qlabel] at (6.25,8.25) {\scriptsize $\cQ_{2,3}$};
\draw[thinarrow] (10.16,6.33) -- (7.45,8.21);
\node[qlabel] at (8.05,8.25) {\scriptsize $\cQ_{2,1}$};
\draw[thinarrow] (8.75,6.33) -- (10.99,8.21);
\node[qlabel] at (10.3,8.16) {\scriptsize $\cQ_{1,3}$};
\draw[thinarrow] (13.2,6.33) -- (11.45,8.21);
\node[qlabel] at (12.2,8.16) {\scriptsize $\cQ_{1,2}$};

% Tadpole row ------------------------------------------------------------------
\rltadpole{3.60}{8.86}{3}
\rltadpole{7.11}{8.86}{2}
\rltadpole{11.30}{8.86}{1}

\node[mathlabel] at (3.60,9.62) {{\scriptsize $I_{\mathrm{tad}}^{D}(z;0,0,3)$}};
\node[mathlabel] at (7.11,9.62) {\scriptsize $I_{\mathrm{tad}}^{D}(z;0,3,0)$};
\node[mathlabel] at (11.30,9.62) {\scriptsize $I_{\mathrm{tad}}^{D}(z;3,0,0)$};
\end{tikzpicture}
\endgroup
\caption{Reduction ladder for the integral $I^{D}_{\rm tri}(z;1,1,1)$ associated to the triangle diagram with $\nu_1= \nu_2=\nu_3 = 1$.}
\label{Fig-reduction-triangle}
\end{figure}

\paragraph{Reduction ladder.} The condition \eqref{red-on-triang} implements shifts among the 
triangle diagrams and the reduction from the triangle to the bubble. To find the full reduction ladder that closes and determines a first-order system, we also need to apply a second reduction step to the bubble diagrams associated to the resonant edge facets. We have discussed the bubble reduction in section~\ref{sec:bubble}, but it is useful to realize that the reduction operators $\cQ_{e',e}$ of the triangle simply restrict to those of the bubble. For example, assume that we have reduced the triangle to the face $F_3$, the resulting $I^D_{\rm bub}(z_{F_3}, \nu_1,\nu_2)$ is independent of $z_3,z_{1,3},z_{2,3}$. Hence, acting with $\cQ_{2,1}$ given in \eqref{red-on-triang} on this integral yields 
\beq
   \cQ_{2,1} I^D_{\rm bub}(z_{F_3}, \nu_1,\nu_2) = \big( z_1 \partial_2 + z_{1,2} \partial_{2,2} + 2 z_{1,1} \partial_{1,2} \big) I^D_{\rm bub}(z_{F_3}, \nu_1,\nu_2)\ ,
\eeq
which we realize is precisely the reduction operator found for the bubble in \eqref{eq:bubbleQ1}. In contrast, we obtain $\cQ_{3,1 }I^D_{\rm bub}(z_{F_3}, \nu_1,\nu_2) =0$. More generally, we naturally restrict the operators to the face and observe that 
\beq
   \cQ_{e,e'}|_{F_e} = \cQ_{e',e}|_{F_e}  = 0 \ ,
\eeq
once the restriction to $F_e$ is imposed by setting $z_e=z_{e,f}=z_{f,e}=0$. In this situation, the derivatives $\partial_{z_e}$ and $\partial_{z_{e,f}},\partial_{z_{f,e}}$ vanish, since the integrals on this face do not depend on these variables. These observations show that the reduction operators of the triangle suffice to build the full reduction ladder: triangle $\rightarrow$ bubble $\rightarrow$ tadpole. 
To illustrate the connections between the integrals and diagrams, the complete reduction of $I_{\rm tri}^D(z;1,1,1)$ is shown in figure \ref{Fig-reduction-triangle}.

\paragraph{First order system and algebraic relations.}
As in the bubble example, the reduction ladder in figure~\ref{Fig-reduction-triangle} can be promoted to a system of first-order differential equations. One takes as a preliminary vector $\vec I_{\rm tri}$ all triangle, bubble and tadpole integrals appearing in the ladder, including the shifted values of the propagator powers. The first-order operators $\cQ_{e',e}$, together with the Euler operator $\cE_0$, then give a linear system for the derivatives of the entries of $\vec I_{\rm tri}$ and hence a Pfaffian equation
$d\vec I_{\rm tri}=A_{\rm tri}\cdot \vec I_{\rm tri}$.
This system is naturally overcomplete. Indeed, the construction of the ladder uses the Euler and reduction equations, but not yet the toric GKZ relations. Once the derivatives in the toric equations are eliminated using the first-order system, the toric operators become algebraic relations among the entries of $\vec I_{\rm tri}$. These relations reduce the number of independent functions in the preliminary vector and should be imposed before comparing the system with a minimal master integral basis.
 
\section{GKZ system for one-loop integrals with generic kinematics}\label{sec:generic1loop}

\begin{figure}[h!]
    \centering
    \begingroup
    \definecolor{rlLineBlue}{RGB}{0,30,145}
    \begin{tikzpicture}[
        scale=0.6667,
        transform shape,
        line cap=round,
        line join=round,
        thick,
        >=Latex,
        diagram/.style={draw=rlLineBlue, thick},
        vertexdot/.style={circle,fill=black,inner sep=1pt},
        text=black
    ]
        \def\R{3.25}
        \def\Rext{4.75}

        % Visible loop vertices from a regular octagon.
        % Hence all visible propagators between adjacent shown vertices have equal length.
        \coordinate (v34)    at (202.5:\R);
        \coordinate (v23)    at (157.5:\R);
        \coordinate (v12)    at (112.5:\R);
        \coordinate (v1n)    at ( 67.5:\R);
        \coordinate (vn1n)   at ( 22.5:\R);
        \coordinate (vn2n1)  at (-22.5:\R);

        % Hidden bottom guide points on the same regular-octagon radius.
        \coordinate (hL) at (247.5:\R);
        \coordinate (hR) at (292.5:\R);
        \coordinate (bC) at (270:\R);

        % External endpoints placed radially for symmetry.
        \coordinate (e34)    at (202.5:\Rext);
        \coordinate (e23)    at (157.5:\Rext);
        \coordinate (e12)    at (112.5:\Rext);
        \coordinate (e1n)    at ( 67.5:\Rext);
        \coordinate (en1n)   at ( 22.5:\Rext);
        \coordinate (en2n1)  at (-22.5:\Rext);

        % Internal loop edges and mass labels
        \draw[diagram] (v1n)   -- node[midway, above=3pt] {$m_1,\nu_1$} (v12);
        \draw[diagram] (v12)   -- node[midway, above left=2pt] {$m_2,\nu_2$} (v23);
        \draw[diagram] (v23)   -- node[midway, left=6pt] {$m_3,\nu_3$} (v34);
        \draw[diagram] (v34)   -- (hL);
        \draw[diagram] (hR)    -- (vn2n1);
        \draw[diagram] (vn2n1) -- node[midway, right=6pt] {$m_{n-1},\nu_{n-1}$} (vn1n);
        \draw[diagram] (vn1n)  -- node[midway, above right=2pt] {$m_n,\nu_n$} (v1n);

        % Omitted internal propagators
        \node at ($(bC)+(0,-0.05)$) {$\cdots$};

        % External legs
        \draw[diagram] (e34)   -- (v34);
        \draw[diagram] (e23)   -- (v23);
        \draw[diagram] (e12)   -- (v12);
        \draw[diagram] (e1n)   -- (v1n);
        \draw[diagram] (en1n)  -- (vn1n);
        \draw[diagram] (en2n1) -- (vn2n1);

        % Momentum arrows (incoming)
        \draw[diagram,-] ($(e34)!0.18!(v34)$) -- ($(e34)!0.72!(v34)$)
            node[midway, below left=10pt] {$p_{3,4}\ \ $};

        \draw[diagram,-] ($(e23)!0.18!(v23)$) -- ($(e23)!0.72!(v23)$)
            node[midway, left=10pt] {$p_{2,3}\ \ $};

        \draw[diagram,-] ($(e12)!0.18!(v12)$) -- ($(e12)!0.72!(v12)$)
            node[midway, above left=10pt] {$p_{1,2}\ $};

        \draw[diagram,-] ($(e1n)!0.18!(v1n)$) -- ($(e1n)!0.72!(v1n)$)
            node[midway, above right=10pt] {$\ \, p_{1,n}$};

        \draw[diagram,-] ($(en1n)!0.18!(vn1n)$) -- ($(en1n)!0.72!(vn1n)$)
            node[midway, right=10pt] {$\ \ p_{n-1,n}$};

        \draw[diagram,-] ($(en2n1)!0.18!(vn2n1)$) -- ($(en2n1)!0.72!(vn2n1)$)
            node[midway, below right=10pt] {$\ \ \ \ p_{n-2,n-1}$};

        % Vertex circles
        \node[vertexdot] at (v34) {};
        \node[vertexdot] at (v23) {};
        \node[vertexdot] at (v12) {};
        \node[vertexdot] at (v1n) {};
        \node[vertexdot] at (vn1n) {};
        \node[vertexdot] at (vn2n1) {};
    \end{tikzpicture}
    \endgroup
    \caption{A general 1-loop $n$-point diagram.}
    \label{fig:npoint-loop}
\end{figure}
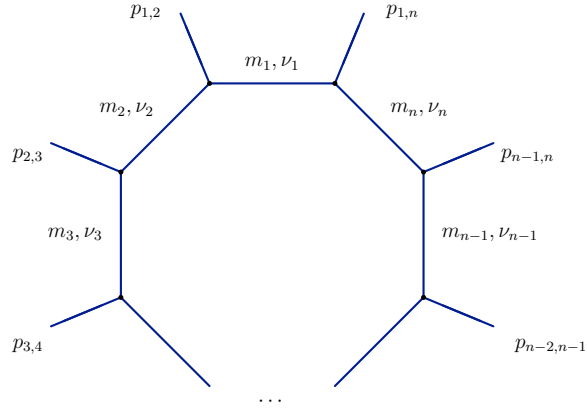

In this section we will consider general one-loop integrals, depicted in figure \ref{fig:npoint-loop}, and determine their associated reduction operators 
and system of differential equations. Our exposition will be concise, since the triangle example of section \ref{sec:triang_example} already illustrated  all the essential features appearing in  one-loop integrals with generic values of kinematic parameters. 

\subsection{GKZ data of one-loop integrals}

To arrive at the GKZ data, we consider the Lee-Pomeransky polynomials at generic kinematics and with nonvanishing 
masses $m_e$ associated to each edge in an $n$-point diagram. Complexifying the coefficients, the polynomial $\cG(z,x)$ in the GKZ integral \eqref{eq:gengkzint} is given by 
\begin{equation} \label{G-gen}
    \cG_{\bf{1\cdots n}}(z,x) =  \sum_{e=1}^n z_ex_e+\sum_{e=1}^n\sum_{e'=e}^n z_{e,e'}x_ex_{e'}
     \ ,
\end{equation}
where we use boldface subscripts to label the $n$ edges in the diagram. 
To return to the physical locus $\cG(p,m,x)$ we have to restrict the parameters $z_{e},z_{e,e'}$ as
\begin{equation} \label{phys-one-loop}
    z_e|_{\rm phys} = 1\,,\quad z_{e,e'}|_{\rm phys} = m_e^2+m_{e'}^2 - p_{e,e'}^2\, , \quad z_{e,e}|_{\rm phys}=m_e^2\, ,
\end{equation}
where $p_{e,e'}$ are the external momenta associated to the leg between the $e$-th and $e'$-th edge as depicted in figure~\ref{fig:npoint-loop}.
Note that we formally identify $z_{e,e'}=z_{e',e}$.

\paragraph{$\cA$-matrix and $\nu$-vector.} The columns of the matrix $\A$ are denoted by $a_e$, $a_{e,e'}$ for $e<e'$, and $a_{e,e}$, and are given by
\beq \label{genA-oneloop}
   \cA = (a_e,a_{e,e'},a_{e,e}) =\begin{pmatrix}
        1 & 1 &  1\\
        \delta_e & \delta_e+\delta_{e'} & 2 \delta_e
    \end{pmatrix} \ ,
\eeq
where $(\delta_e)_f = \delta_{e f}$ are the unit normal vectors.
Here the coefficients $a_e$ are associated with the terms $z_e x_e$ in $\cG_{\bf 1\cdots n}$, whereas the coefficients $a_{e,e'}$ for $e \leq e'$ are associated with the terms $z_{e,e'} x_e x_{e'}$. The parameter vector is taken to be $ \nu = (D/2,\nu_1,\ldots\nu_n)^T$ as above, 
and we assume that $\nu_e \in \mathbb{N}$ while $D/2 \in \mathbb{C}$.

\paragraph{Toric and Euler operators.} Given $(\cA,\nu)$ it is straightforward to determine the GKZ system. We find toric operators of the form 
\begin{equation} \label{toric-general}
 \L_{e,f,g} =   \partial_{e}\partial_{f,g}-\partial_{e,f}\partial_{g}, \qquad
 \L_{e,e',f,f'} = \partial_{e,e'}\partial_{f,f'}-\partial_{e,f}\partial_{e',f'}\,
\end{equation}
where different edge labels are allowed to take equal values. The Euler operators are 
\begin{equation} \label{Euler-oneloop}
    \cE_0 = \sum_e \theta_e +\sum_{e=1}^n\sum_{e'=e}^n\theta_{e,e'}, \qquad
     \cE_e = \theta_e+2\theta_{e,e}+\sum_{e'\neq e} \theta_{e,e'}\,,
\end{equation}
with $\theta_e = z_e \partial_e$ and $\theta_{e,e'}=z_{e,e'}\partial_{e,e'}$.

\subsection{Resonance and reduction operators}

In order to examine the resonance condition \eqref{check_resonance}, we recall that we assume $\nu_e \in \mathbb{N}$, while generally $D/2 \in \mathbb{C}$.
It follows that the edge facets are resonant with the linear functionals $L_e = \delta_e$. The other two facets of ${\rm Conv}(\mathcal{A})$ are $F_{\mathcal F}$, associated with the linear functional $L_{\mathcal F} = (-1,1,\ldots,1)$, which becomes resonant when $D/2 \in \mathbb{Z}$; and $F_{\mathcal U}$, associated with the linear functional $L_{\mathcal U} = (2,-1,\ldots,-1)$, which is resonant when $D \in \mathbb{Z}$. We will discuss the resonances involving $D$ in section~\ref{sec:integer-dim} and focus on the edge faces in the following.

\paragraph{Reduction operators for edges.} In the following, we derive the reduction operators $\cQ^{F_e}_{u,v}$ associated with each edge $e$ of the diagram displayed in figure \ref{fig:npoint-loop}. As in the bubble and triangle examples, they are obtained by considering $\partial^v = \partial_e$ and $\partial^u = \partial_{e'}$, where $e' \notin F_e$, and can be denoted by $\cQ_{e',e}$. We thus solve the condition
\begin{equation}
        \partial_{e'}\cE_{e}\simeq \cQ_{e',e} \partial_{e}\,, 
    \end{equation}
and recall $\cA u = a_{e'}$, $\cA v = a_e$.  
Using the toric operators \eqref{toric-general} for the one-loop integrals we determine
\begin{equation} \label{eq:Qgen}
        \cQ_{e',e}= z_{e}\partial_{e'}+z_{e,e'}\partial_{e',e'}+2 z_{e,e}\partial
        _{e,e'}+\sum_{f \neq e,e'}
        z_{e,f}\partial_{f,e'}\,.
    \end{equation}
As in the triangle diagram, each edge is associated with several reduction operators. These play a key role in determining the full reduction ladder for a generic one-loop integral $I^D_{\rm 1-loop}(z;\nu)$. The operators $\cQ_{e',e}$ are not all independent when combined with the Euler equations; in 
fact, as $n$ increases, an increasing number of relations among them appears.

\paragraph{Restriction to the physical locus.} Using the Euler equations, one can express the reduction operators \eqref{eq:Qgen} using only physical variables \eqref{phys-one-loop}. 
Explicitly, when acting on a GKZ solution $f(p,m;\nu)$, they take the form
\begin{eqnarray} \label{gen-one-loop-phys}
    \cQ_{e',e}f(p,m;\nu) &=& \Big(-\nu_{e'} +\big(m_e^2-m_{e'}^2-p_{e,e'}^2\big)\partial_{m_{e'}^2} - 2 p_{e,e'}^2 \partial_{p_{e,e'}^2}
    +\\
    &&\qquad \qquad \qquad  +\sum_{f \neq e,e'} ( p_{e,f}^2 -p_{e',f}^2- p_{e,e'}^2)
    \partial_{p_{e',f}^2}\Big) f(p,m;\nu)\,.\nonumber
\end{eqnarray}
In addition, we evaluate the Euler operator $\cE_0$ as
\beq \label{E0phys}
  \cE_0 f(p,m;\nu) =- \Big(\sum_{e} \big(  \nu_e + m_{e}^{2}\partial_{m_e^2}\big)+\sum
        _{e<e'}p_{e,e'}^{2}\partial_{p_{e,e'}^2}\Big)f(p,m;\nu)\ .
\eeq
Together, the expressions \eqref{gen-one-loop-phys} and \eqref{E0phys} allow us to present the reduction ladder entirely in physical coordinates. 

\paragraph{First reduction -- removing edge $e$.} To show how the reduction operators act, we denote the $n$-point integral by $I^D_{\bf{1\cdots n}}(z;\nu)$.  
According to the general relation \eqref{red-on-solutions2}, the effect of the reduction operators is given by 
\beq \label{red-on-1-loop}
 \cQ_{e',e} I^D_{\bf{1\cdots n}}(z;\nu)  = 
    \begin{cases}  - \nu_{e'}\, I^D_{\bf{1\cdots \hat{e} \cdots n}} (z_{F_e}; \nu_{F_e} +\delta_{e'})\ & \text{if} \ \ \nu_e = 1 \ ,\\
 - \nu_{e'} \,  I^D_{\bf{1\cdots n}}(z;\nu-\delta_e +\delta_{e'})\ & \text{if} \ \ \nu_e > 1 \ ,
 \end{cases}
\eeq
and
\beq \label{gen-Euler}
   (\cE_e + \nu_e)I^D_{\bf{1\cdots n}}(z;\nu) = 0 \ , \quad \big(\cE_0 + \tfrac{D}{2}\big)I^D_{\bf{1\cdots n}}(z;\nu) = 0\quad \text{for}\quad \nu_e \geq 1\ .
\eeq
Here $I^D_{\bf{1\cdots \hat{e} \cdots n}}$ is the integral associated to the diagram with the edge $e$ removed and therefore depends only on the coordinates $z_{F_e}$ supported on $F_e$ and the shorter vector $\nu_{F_e}$, as introduced in \eqref{def-ZFe}. Iterating this equation 
we deduce 
\begin{equation}
     \left(\cQ_{e',e} \right)^{\nu_e} I^D_{\bf{1\cdots n}}(z;\nu) = (-\nu_{e'})_{\nu_e}
     I^D_{\bf{1\cdots \hat{e} \cdots n}}(z_{F_e};\nu + \nu_e \delta_{e'}) \,.
\end{equation}
The equations \eqref{red-on-1-loop} and \eqref{gen-Euler} suffice to obtain the first reduction from an $n$-point to an $(n-1)$-point one-loop diagram. As seen in \eqref{gen-one-loop-phys}, \eqref{E0phys}, the $\cE_e$-Euler equations can be used to restrict to physical variables. As for the triangle diagram, we can further reduce the number of independent integrals by using the toric equations. These are second order, so we need to iterate the reduction further to identify them as algebraic relations. 

\paragraph{Reduction ladder.} After reducing from $n$ to $n-1$ points, yielding the integrals $I^D_{\bf{1\cdots \hat{e} \cdots n}}(z_{F_e};\nu_{F_e})$, we can proceed further along the reduction ladder. To do so, we observe that
\beq
    \cQ_{e',f}|_{F_e} \equiv \cQ_{e',f}| ({z_g=z_{g,g'}=\partial_g=\partial_{g,g'}=0\ \text{for}\ g,g'\notin F_e})\ ,  
\eeq
yields precisely the reduction operators associated to the one-loop graph with edge $e$ removed. In fact, due to the form of \eqref{eq:Qgen} one has $\cQ_{e',e}|_{F_e}=\cQ_{e,e'}|_{F_e}=0$. Therefore, we can generate all the required integrals for subgraphs $G' \subset G$ purely by using the $\cQ_{e.e'}$ of the largest graph $G$. The resulting reduction ladder gives a preliminary vector of integrals $\vec I$ that is closed under the action of $\cQ_{e,e'}$.

\paragraph{Differential equations and algebraic relations.}
The reduction equations \eqref{red-on-1-loop}, together with the Euler equations, generate a finite first-order system for any one-loop integral. Let $\vec I$ denote the preliminary vector of all integrals appearing in the complete reduction ladder, including the lower-point subgraphs and the shifted values of $\nu$. For each entry $I_\alpha$ of this vector, we collect the first-order Euler and reduction equations in the form
\begin{equation}\label{eq:first-order-linear-system-gkz}
    \sum_{j=1}^N M^{(\alpha)}_{rj}(z)\,\partial_j I_\alpha
    =
    \sum_\beta R^{(\alpha)}_{r\beta}(z)\,I_\beta\, .
\end{equation}
Here the left-hand side sums over the $N=n+n(n+1)/2$ GKZ variables $(z_e,z_{e,e'})$ supported on the $n$-point graph. The index $r$ runs over a chosen full-rank subset of the $n+1$ available Euler operators $\cE_{\underline e}$ and the $n(n-1)$ reduction operators $\cQ_{e,e'}$. 
Where
$M^{(\alpha)}$ has maximal rank, 
\eqref{eq:first-order-linear-system-gkz} can be solved for the GKZ derivatives,
\begin{equation}\label{eq:partial-z-Pfaffian}
        \partial_j \vec I = B_j(z)\cdot \vec I\, .
\end{equation}
This gives the desired Pfaffian system in GKZ coordinates. Any of the $(n-1)(n-2)/2$ remaining first-order equations, not used in the inversion of $M^{(\alpha)}$, already give algebraic relations among the entries of $\vec I$.

The toric equations give further algebraic relations. Indeed, after \eqref{eq:partial-z-Pfaffian} has been obtained, every higher derivative can be eliminated recursively. 
Any toric operator $\partial_i\partial_j-\partial_k\partial_l$ from \eqref{toric-general} leads to
\begin{equation}\label{eq:toric-to-algebraic-basic}
    \Big(
        \partial_i B_j -\partial_k B_l +B_jB_i
       -B_lB_k
    \Big)\cdot \vec I=0\, ,
\end{equation}
which is an algebraic relation among the components of $\vec I$. Toric operators therefore reduce the preliminary reduction-ladder vector $\vec I$ to a smaller set of independent functions. The restriction to the physical locus \eqref{phys-one-loop} can be performed either before or after imposing these algebraic relations.

 \section{Sunrise and banana integrals}\label{sec:higherloops}

As initial examples going beyond one loop, we present reduction operators for the sunrise integral in section~\ref{sec:sunrise}. The generalization to the multi-loop banana family is then addressed in section \ref{sec:bananas}.

\subsection{Sunrise reduction operators} \label{sec:sunrise}

\begin{figure}[h!]
    \centering
    \begingroup
    \definecolor{rlLineBlue}{RGB}{0,30,145}
    \begin{tikzpicture}[
        scale=1,
        line cap=round,
        line join=round,
        diagram/.style={draw=rlLineBlue, very thick},
        vertexdot/.style={circle,fill=black,inner sep=1pt},
        text=black
    ]

        % Main horizontal propagator and external legs
        \draw[diagram] (-1.7,0) -- (-1.35,0);
        \draw[diagram] (-1.35,0) -- (0,0);
        \draw[diagram] (0,0) -- (1.35,0);
        \draw[diagram] (1.35,0) -- (1.7,0);

        % Labels
        \node at (0,1.05) {$m_1, \nu_1$};
        \node at (0,0.2) {$m_2, \nu_2$};
        \node at (0,-1.05) {$m_3, \nu_3$};
        \node at (2.1,-0.05) {$p$};

        % Upper and lower arcs on the left
        \draw[diagram] (-1,0) .. controls (-1,0.6) and (-0.3,0.8) .. (0,0.8);
        \draw[diagram] (-1,0) .. controls (-1,-0.6) and (-0.3,-0.8) .. (0,-0.8);

        % Upper and lower arcs on the right
        \draw[diagram] (1,0) .. controls (1,0.6) and (0.3,0.8) .. (0,0.8);
        \draw[diagram] (1,0) .. controls (1,-0.6) and (0.3,-0.8) .. (0,-0.8);

        % Vertex circles
        \node[vertexdot] at (-1,0) {};
        \node[vertexdot] at (1,0) {};
  %      \node[vertexdot] at (0,0.8) {};
  %      \node[vertexdot] at (0,-0.8) {};

    \end{tikzpicture}
    \endgroup
    \caption{The sunrise diagram.}
    \label{fig:sunrise}
\end{figure}
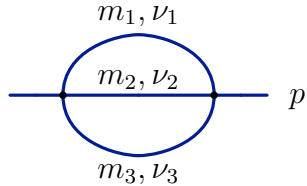

Our initial higher-loop example is the sunrise diagram depicted in figure \ref{fig:sunrise}. It is a two-loop diagram with external momentum $p$. We assume generic kinematics, i.e.~nonvanishing masses $m_1,m_2,m_3$ for the three internal propagators. 
 
\begin{figure}
\begin{center}
   \includegraphics[width=0.6\linewidth]{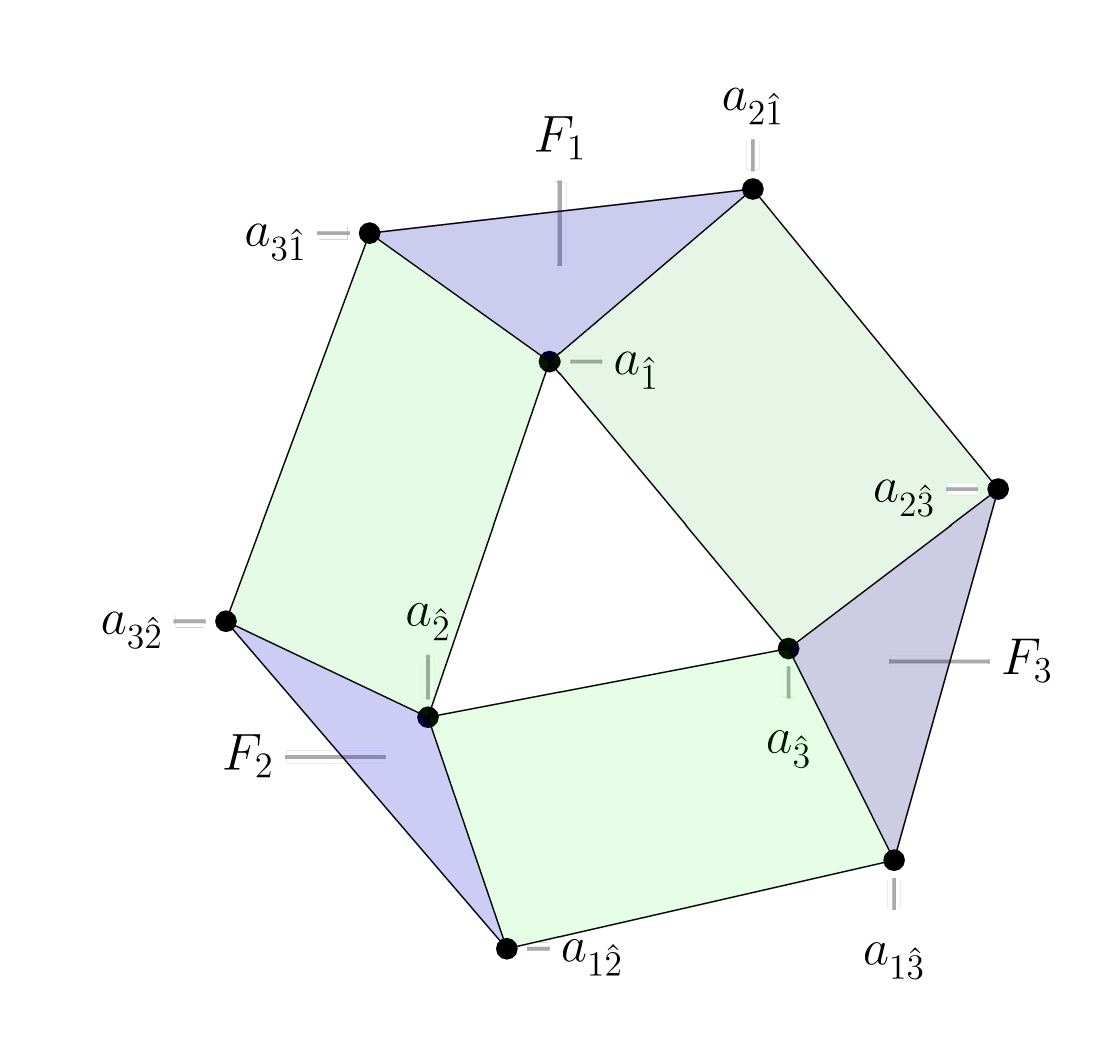}
   \caption{The polytope ${\rm Conv}({\mathcal A})$ for the sunrise graph.}\label{fig:3msunrise-polytope}
\end{center}
\end{figure}

\paragraph{GKZ data and physical locus}
The GKZ integral for the sunrise diagram is given by
    \begin{equation}
        \label{eq:sunriseint-gkz}
        I^D_{\mathrm{sun}}(z;\nu)= \kappa_{\rm sun}(\nu)
        \int_{\R_+^3}d^{3}x \;  x_1^{\nu_1-1} x_2^{\nu_2-1} x_3^{\nu_3-1} \cG(z,x)^{-\nu_0}\,,
    \end{equation}
    where
      \begin{align}\label{eq:sunrisepoly}
\cG(z,x) &=   
z_{\hat 1} x_2 x_3 +z_{\hat 2} x_1 x_3 +z_{\hat 3} x_1 x_2 
+ z_0 x_1 x_2 x_3 \nonumber \\ 
&
+ z_{1\hat 2} x_1^2 x_3 + z_{1\hat 3} x_1^2 x_2
+ z_{2\hat 1} x_2^2 x_3 + z_{2\hat 3}  x_1 x_2^2 
+ z_{3\hat 1} x_2 x_3^2  + z_{3\hat 2}  x_1 x_3^2
\,,
\end{align}
and $\kappa_{\rm sun}(\nu)$ is determined from \eqref{def-kappa} with $L=2$, $\nu=(D/2,\nu_1,\nu_2,\nu_3)$.
For brevity, we have now condensed the indices of the GKZ variables so that they represent the powers of monomials after subtracting one from each. 
Hats in the indices of the GKZ variables correspond to their absence in their respective monomials. 
The restriction of the GKZ variables to physical variables is implemented by taking
    \begin{equation}
        \begin{array}{rl}
            z_{\hat i}|_{\rm phys} =         & 1\ ,                                              \\
            z_{i \hat j}|_{\rm phys}=     &   m_i^2, \\
            z_{0}\vert_{\rm phys} = & m_1^2 + m_2^2 + m_3^2 - p^2\,.                            \end{array}
    \end{equation}
The full GKZ system for the sunrise has the matrix and parameter vector 
\begin{equation}
   \A_{{\rm sun}(1,2,3)}= \left(
\begin{array}{cccccccccc}
 1 & 1 & 1 & 1 & 1 & 1 & 1 & 1 & 1 & 1 \\
 0 & 1 & 1 & 1 & 2 & 2 & 0 & 1 & 0 & 1 \\
 1 & 0 & 1 & 1 & 0 & 1 & 2 & 2 & 1 & 0 \\
 1 & 1 & 0 & 1 & 1 & 0 & 1 & 0 & 2 & 2 \\
\end{array}
\right),
\qquad \nu=\begin{pmatrix}
        D/2\\
        \nu_1 \\
        \nu_2 \\
        \nu_3
    \end{pmatrix}\,  ,
\end{equation}
where we ordered the columns according to $z_{\hat 1},z_{\hat 2},z_{\hat 3},z_0,z_{1\hat 2},z_{1\hat 3},z_{2\hat 1},z_{2\hat 3},z_{3\hat 1},z_{3\hat 2}$ as in \eqref{eq:sunrisepoly}.
The polytope for the sunrise is depicted in figure~\ref{fig:3msunrise-polytope}.  

\paragraph{Reduction operators for edges.}
Assuming that $\nu_i \in \mathbb{N}$, we find that all edge faces are resonant. 
We find that there is a unique second-order reduction operator for each edge. Its expression may be obtained, for example, from the observation that $\partial_{\hat 1}\partial_{3\hat 1} \cE_1  \simeq \cQ^{F_1}_{\{\hat 1,3 \hat 1\},\hat 2} \partial_{\hat 2}$, with the expression for the reduction operator  
given by
\begin{align}\label{eq:Q-sunrise}
\begin{split}
    \cQ^1 \equiv \cQ^{F_1}_{\{\hat 1,3 \hat 1\},\hat 2} =& z_{0} \partial_{2 \hat 1}\partial_{3 \hat 1} + z_{\hat 3} \partial_{2 \hat 1}\partial_{\hat 1} + z_{\hat 2} \partial_{3 \hat 1}\partial_{\hat 1} 
    \\ & + 2 z_{1 \hat 3} \partial_{2 \hat 1} \partial_{0} + 2 z_{1 \hat 2} \partial_{3 \hat 1}\partial_{0}+z_{2 \hat 3}\partial_{2 \hat 1}^2 + z_{3 \hat 2}\partial_{3 \hat 1}^2\,.
    \end{split}
\end{align} 
The $\cA$-degree of $\cQ^1$ is $\text{deg}(\cQ^1)= \cA v - \cA u = (-1,1,-2,-2)$, which implies that $\cQ^1$ generally will shift the dimension $D$. Similarly, there are operators $\cQ^2$ and $\cQ^3$ corresponding to the remaining edge faces $F_2$ and $F_3$, obtained by cyclic permutations of the indices. In what follows, we will concentrate on the action of $\cQ^1$, as the generalization to the other cases is immediate.  

\paragraph{Reduction to double tadpole.} The reduction operator for edge 1 acts on the sunrise integral to give
\begin{align}\label{eq:sunriseQ1action}
    \cQ^1 I^D_{\textrm{sun}}(z;\nu) = \begin{cases} \frac{D}{2}\kappa_{\rm sun}(\nu) \int_{\R^2_+} d^2x \, x_2^{\nu_2+1} x_3^{\nu_3+1} \left[ \left.\cG(z,x)\right|_{x_1=0}\right]^{-D/2-1} & \text{if}\ \ \nu_1 =1\,, \\
    \frac{D}{2}(\nu_1-1)\kappa_{\rm sun}(\nu) \int_{\R_+^3} d^3x \, x_1^{\nu_1-2}  x_2^{\nu_2+1} x_3^{\nu_3+1} \cG(z,x)^{-D/2-1} & \text{if} \ \ \nu_1>1\ . 
    \end{cases}
\end{align}
These equations can be obtained by direct computation using \eqref{eq:Q-sunrise}, but can also be obtained from the general result \eqref{red-on-solutions2} after rewriting the prefactors.  It turns out to be convenient to include a factor of $z_{\hat 1}^2$, so that we obtain an operator $\widehat \cQ^1 = z_{\hat 1}^2 \cQ^1$ of $\cA$-degree $(1,1,0,0),$ which shifts the dimension, reduces the power of propagator 1, and leaves the other propagator powers unchanged.\footnote{Since the physical restriction of $z_{\hat 1}$ is simply 1, we can infer that in the sunrise family, parameter shifts of $(1,-1,2,2)$ and $(-1,-1,0,0)$ are equivalent.}
Let us specialize to the case $\nu_1=1$.
Because $z_{\hat 1} x_2 x_3 = \theta_{\hat 1} \cG(z,x)$, we have
\begin{align}\label{eq:sunriseQ1-dimshift}
\widehat \cQ^1 I^D_{\textrm{sun}}(z;1,\nu_2,\nu_3) &=  \frac{D}{2}\kappa_{\rm sun}(\nu) \int_{\R^2_+}\!\! d^2x \, x_2^{\nu_2-1} x_3^{\nu_3-1} (z_{\hat 1} x_2 x_3)^2 \left[ \left.\cG(z,x)\right|_{x_1=0}\right]^{-D/2-1} \\
&=  \frac{2}{D-2}\kappa_{\rm sun}(\nu) \int_{\R^2_+} d^2x \, x_2^{\nu_2-1} x_3^{\nu_3-1} \left(\theta_{\hat 1}^2-\theta_{\hat 1} \right) \cG(z,x)^{-D/2+1}\,. 
\end{align}
We can write $\theta_{\hat{1}} = 3\cE_0-\cE_2-\cE_3$ in terms of Euler operators on the contracted space where $x_1=0$, and then apply the Euler relations to exchange the differential operators for functions of $\nu$. 
Inserting the expressions for $\kappa_{\rm sun}(D/2,1,\nu_2,\nu_3)$ and the contracted double-tadpole integral, we find 
\begin{align}\label{eq:sunriseQ1contraction}
\widehat \cQ^1 I^D_{\textrm{sun}}(z;1,\nu_2,\nu_3) 
&= 
I^{D-2}_{\textrm{dtad}}(z;\nu_2,\nu_3)\,,
\end{align}
where
\begin{equation}
        \label{eq:dtadint-gkz}
        I^D_{\textrm{dtad}}(z;\nu_2,\nu_3)=  \kappa_{\textrm{dtad}}(\nu) 
        % \frac{\Gamma\left(\nu_0\right)}{\Gamma(\frac{3D}{2}-\nu_2-\nu_3)\Gamma(\nu_2) \Gamma(\nu_3) } \times \\
        \int_{\R_+^2} \!\! \rd x_2\, \rd x_3 \;  x_2^{\nu_2-1} x_3^{\nu_3-1}  \left( z_{\hat 1} x_2 x_3 + z_{2\hat 1} x_2^2 x_3 + z_{3 \hat 1} x_2 x_3^2\right)^{-D/2} ,
\end{equation}
and $\kappa_{\textrm{dtad}}(\nu)$ is determined from \eqref{def-kappa} with $\nu=(D/2,\nu_2,\nu_3)$. As expected, \eqref{eq:sunriseQ1contraction} generalizes to all reduction operators $\widehat \cQ^i = z_{\hat i}^2 \cQ^i$ by  permutation of the indices.  

\paragraph{Reduction ladder, first order system, and physical locus.} We now proceed to act systematically with $\widehat \cQ^i$ to find the set of integrals closing under these differential operators. Starting with $I^D_{\rm sun}(z;1,1,1)$, we obtain three double tadpoles in $D-2$ dimensions.
In contrast to the general one-loop construction of section \ref{sec:generic1loop}, this set does not suffice to build the first order system $d\vec I= A \cdot \vec I$ and perform the reduction to the physical locus. A central challenge is that $\widehat \cQ^i$ for the sunrise are second order operators, just as its toric operators. This implies that the vector $\vec I$ should include $z$-derivatives of the integrals connected by $\widehat \cQ^i$. This is, in fact, also needed in order to reduce to the physical locus, since the four Euler equations do not suffice to eliminate the six `nonphysical' derivatives $\partial_{\hat i}$ and $\partial_{i \hat j} -\partial_{i \hat k}$. For example, we can consider the preliminary vector 
\beq
   \vec I = (I^D_{\rm sun}(1,1,1), \partial_{i}I^D_{\rm sun}(1,1,1), I^{D-2}_{\rm dtad}(0,1,1), I^{D-2}_{\rm dtad}(1,0,1), I^{D-2}_{\rm dtad}(1,1,0))^T\ .
\eeq
Taking derivatives of this vector, we can use the reduction and toric operators when evaluating the second derivatives of $I^D_{\rm sun}$. Due to the second order of the operators, the matrix $A$ has a block structure that does not reduce to an upper-triangular system like in the one-loop case. This matches the fact that at dimension $D=2$ it is known that the solutions $I^D_{\rm sun}(1,1,1)$ are elliptic integrals that satisfy non-reducible second-order differential equations \cite{bloch_local_2017,klemm_lloop_2020,bonisch_analytic_2021}.\footnote{These differential equations are obtained from the toric equations of the GKZ system.} In our framework, the reduction operators enable the direct inclusion of the boundary terms, yet obtaining explicit solutions encounters obstacles analogous to those in the integer-$D$ case, where the solutions are expressed as (relative) period integrals of elliptic curves.

\paragraph{Special kinematics.}

In the case with one massless propagator, say $m_3^2=0$, the edge faces $F_1$ and $F_2$ are no longer facets, and are not resonant. The edge face $F_3$ is still a resonant facet and leads to one reduction operator, corresponding to the contraction of the massless propagator which results in a double tadpole. In terms of cuts, we recognize the property discussed in \cite{Abreu:2021vhb}, that the  maximal cuts are solutions to an Appell $F_1$ system of rank 3, within the larger space of all cuts, which is an Appell $F_4$ system of rank 4.

\begin{figure}
\begin{center}
\includegraphics[width=0.45\linewidth]{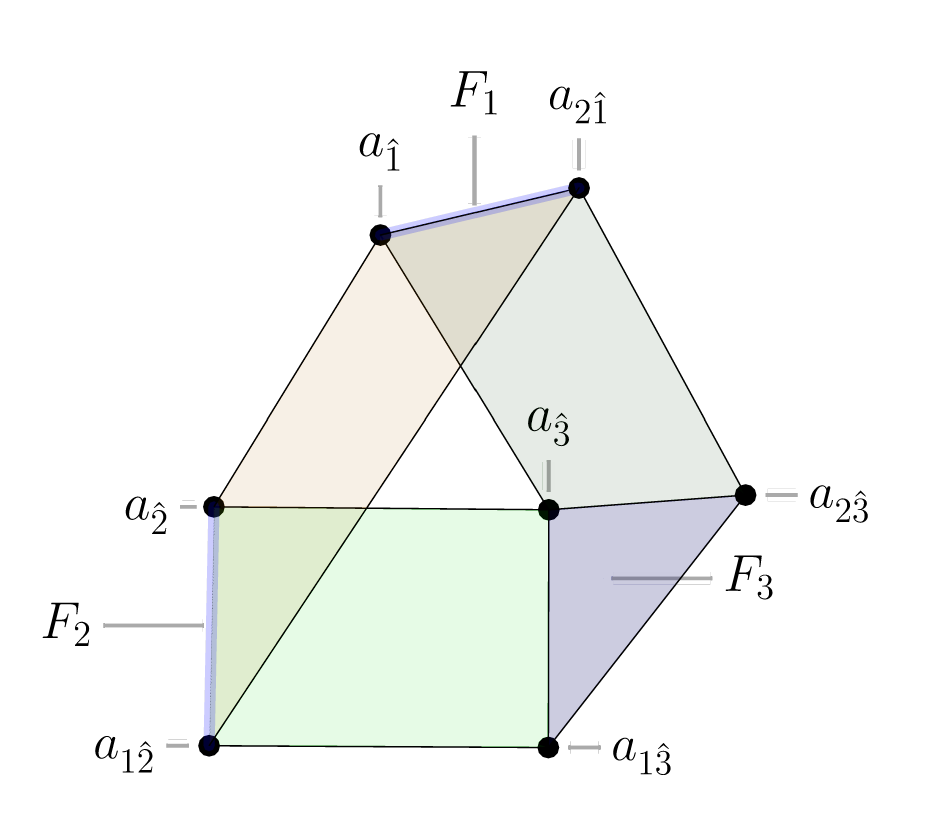}
\includegraphics[width=0.4\linewidth]{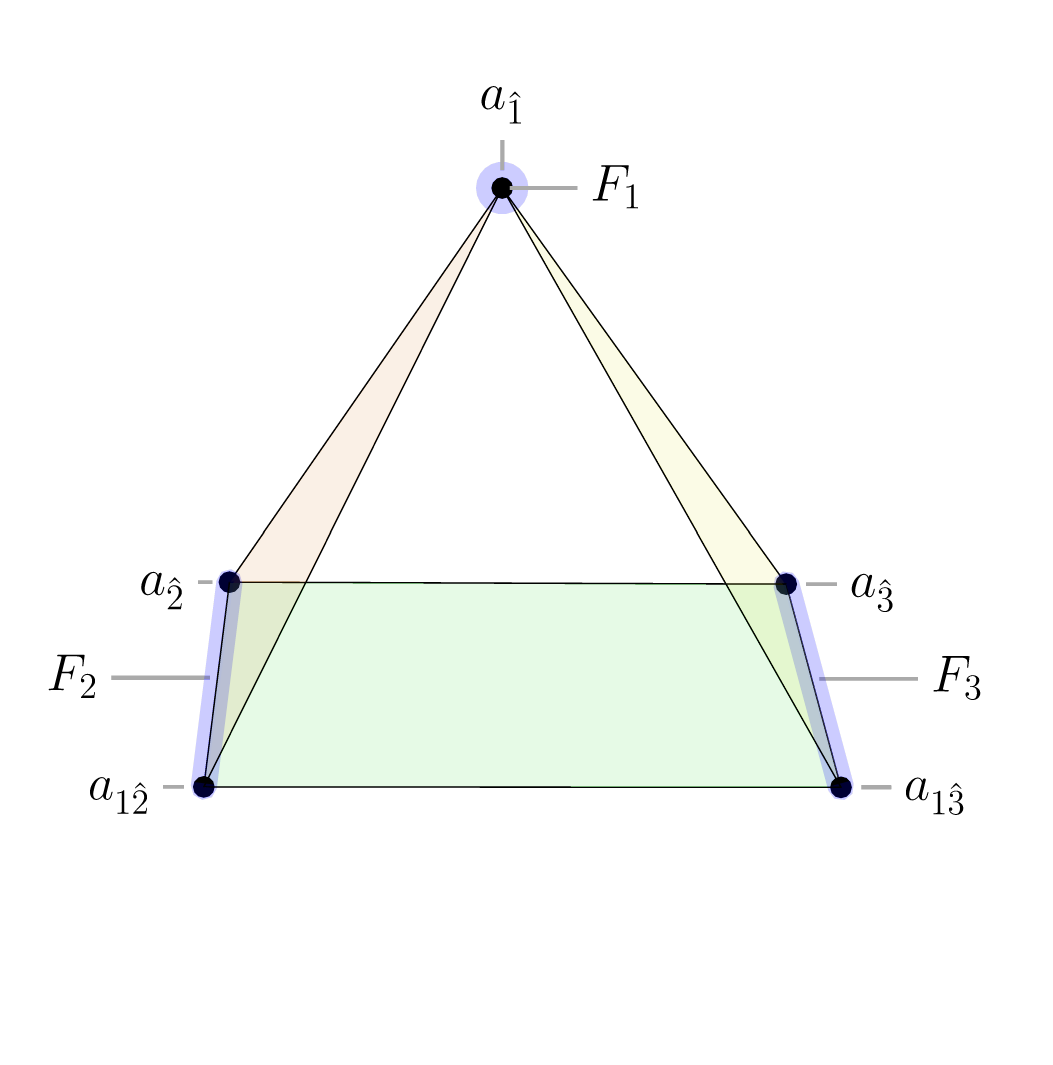}
   \caption{The polytopes ${\rm Conv}({\mathcal A})$ for sunrise diagrams in the cases that $m_3=0$ (left, where one edge face is a facet) and $m_2=m_3=0$ (right, where no edge faces are facets) and nonzero masses otherwise. }\label{fig:2msunrise-polytope}
\end{center}
\end{figure}

If there are two massless propagators, say $m_2^2=m_3^2=0$, then none of the edge faces are facets. There are five  facets, none of which are resonant. Therefore, none of the edge faces are resonant.
There are no reduction operators. Indeed, from the graphical point of view, any edge contraction results in a massless tadpole, which is identically zero. From the point of view of the Feynman integrals, the one-mass sunrise family spans the solution space of a ${}_2F_1$ differential equation, with no degeneracy resulting in simpler solutions for maximal cuts.

\subsection{Reduction operators for the $L$-loop banana integral} \label{sec:bananas}

The construction for the sunrise integral has a straightforward generalization to $L$ loops. In the following, we will briefly summarize our results. 

\paragraph{GKZ setting and physical locus.} 
Let $n$ be the number of internal edges, so that $n=L+1$.
The GKZ integral for the banana diagram is
    \begin{equation}
        \label{eq:bananaint-gkz}
        I^D_{\mathrm{ban}}(z;\nu)= \kappa_{\rm ban}^{(n)}(\nu)
        \int_{\R_+^n}d^{n}x \;  \bigg( \prod_{i=1}^n x_i^{\nu_i-1} \bigg) \cG(z,x)^{-D/2}\,,
    \end{equation}
    where
      \begin{align}\label{eq:bananapoly}
\cG(z,x) &=   z_0 x_1\cdots x_n+
\sum_{i=1}^n z_{\hat \imath} x_1 \cdots \widehat{x_i} \cdots x_n
+ \sum_{i=1}^n \sum_{j \neq i} z_{i \hat\jmath} x_1 \cdots x_i^2 \cdots \widehat{x_j} \cdots x_n\,,
\end{align}
and $\kappa_{\rm ban}^{(n)}$ is determined from \eqref{def-kappa} with $L+1=n$, $\nu=(D/2,\nu_1,...,\nu_n)$.
The restriction of the GKZ variables to physical variables is implemented by taking
    \begin{equation}
        \begin{array}{rl}
             z_{\widehat\imath}|_{\rm phys} =         & 1\ ,                                              \\
            z_{i \widehat\jmath}|_{\rm phys}=     &   m_i^2 \,, \\
            z_{0}\vert_{\rm phys} = & \sum_{i=1}^n m_i^2- p^2 
             \,.                                       
        \end{array}
    \end{equation}
We assume that there are no further relations among the GKZ variables. 

\paragraph{Reduction operators and contraction relations.}
Again, we find a single reduction operator of order $n-1$ for each edge. For example for the edge 1, we find 
\begin{align}
\begin{split}
    \cQ^1 =& 
    z_0 \partial_{2\hat{1}}\cdots\partial_{n\hat{1}}
    + \sum_{j=2}^n \left( z_{\hat{\jmath}}  \partial_{\hat{1}} + 2 z_{1\hat{\jmath}}  \partial_{0} + \sum_{i \neq 1, j} z_{i\hat{\jmath}}  \partial_{i\hat{1}}
\right)\partial_{2\hat{1}}\cdots\widehat{\partial_{j\hat{1}}}\cdots\partial_{n\hat{1}}\,,
    \end{split}
\end{align}
derived for example as $\cQ^1=\cQ^{F_1}_{\{\hat{1},3\hat{1},4\hat{1},\ldots,n\hat{1}\},\hat{2}}$, i.e.~from observing that 
\beq
\partial_{\hat{1}}\partial_{3\hat{1}}\cdots\partial_{n\hat{1}}\cE_1\simeq\cQ^{F_1}_{\{\hat{1},3\hat{1},4\hat{1},\ldots,n\hat{1}\},\hat{2}}\partial_{\hat{2}}\ ,
\eeq
which fixes $u,v$ labeling the reduction operators. The $\cA$-degree of $\cQ^1$ is $\text{deg}(\cQ^1)=\cA v - \cA u =(-n+2,1,-n+1,\ldots,-n+1).$
 Acting on the banana integral, it gives
\begin{align}\label{eq:bananaQ1action}
    \cQ^1 I^D_{\textrm{ban}}(z;\nu) &=  -
    (3-n-D/2)_{n-2} \kappa_{\rm ban}^{(n)}(\nu)
    \times \\
    &  \begin{cases} \int_{\R^{n-1}_+} d^{n-1}x \, x_2^{\nu_2+n-2} \cdots x_n^{\nu_n+n-2} \left[ \left.\cG(z,x)\right|_{x_1=0}\right]^{2-n-D/2} & \text{if}\ \ \nu_1 =1\,, \\ (\nu_1-1)
  \int_{\R_+^n} d^nx \, x_1^{\nu_1-2}  x_2^{\nu_2+n-2} \cdots x_n^{\nu_n+n-2} \cG(z,x)^{2-n-D/2} & \text{if} \ \ \nu_1>1\ .  \nonumber
    \end{cases}
\end{align}
We can then construct  a dimension-shift operator
\beq \label{widehatQ-banana}
   \widehat \cQ^1 = z_{\hat{1}}^{n-1} \cQ^1\ , \qquad \text{deg}(\widehat \cQ^1) = (1,1,0,...,0)\ ,
\eeq
where we have also displayed its $\cA$-degree.
Recognizing that $z_{\hat 1} x_2 x_3 \cdots x_n = \theta_{\hat 1} \cG(z,x)$, and that  $\theta_{\hat{1}} = n\cE_0-\cE_2-\cdots-\cE_n$ in terms of Euler operators on the contracted space where $x_1=0$, we find that in  the case $\nu_1=1$, we have 
\begin{align}
& \widehat  \cQ^1 I^D_{\textrm{ban}}(z;1,\nu_2,\ldots,\nu_n) 
\\
&=  \frac{2\kappa_{\rm ban}^{(n)}(\nu)}{D-2} \int_{\R^{n-1}_+} d^{n-1}x \, x_2^{\nu_2-1}\cdots  x_n^{\nu_n-1} \left(\theta_{\hat{1}}-n+2 \right) \cdots \left(\theta_{\hat{1}}-1 \right) \theta_{\hat{1}} \cG(z,x)^{-D/2+1} \nonumber \\
&= 
(-1)^{n-1}
I^{D-2}_{(n-1)\textrm{-tad}}(z;\nu_2,\ldots,\nu_n)\,, \nonumber
\end{align}
where $I^{D-2}_{(n-1)\textrm{-tad}}$ is the $(n-1)$-fold tadpole integral in $D-2$ dimensions.

\section{Dimensional resonance and reductions using dimension shifts}
\label{sec:dimension}

In this section, we collect some further implications of our framework. In particular, we show in section \ref{sec:integer-dim} that additional resonance conditions arise 
when the dimension $D$ takes on special values. For $\nu_e \in \bbZ$, this occurs for integer or even integer dimensions depending on the loop level. We recognize some of the arising reduction operators as familiar raising operators. 
In section \ref{reduction-with-dimensionshifts} 
we propose combining reduction operators into dimension-shift operators and discuss how to incorporate these operators into the reduction ladders of sections \ref{sec:bubble}--\ref{sec:generic1loop}. 
It is interesting to see that the GKZ system naturally yields operators connecting Feynman integrals in different dimensions.      

\subsection{Resonance at integer dimensions} \label{sec:integer-dim}

There exists an extensive body of results concerning Feynman integrals in specific spacetime dimensions~$D$. As an illustration, suppose that $D$ is even and $\nu_e \in \bbZ$. The Lee–Pomeransky integral \eqref{eq:LPrep} and 
the GKZ integral \eqref{eq:gengkzint} then reduce to integrals of a rational function.  The latter can be associated with period integrals of an auxiliary complex variety built from $\cG(z,x)$, and its Picard–Fuchs equations governing the period integrals have long been known to follow from toric equations in the GKZ framework \cite{hosono_mirror_1995,hosono_gkzCY_1996,hosono_gkzapp_1996,Bloch:2014qca,bloch_local_2017,muller-stach_picardfuchs_2014,klemm_lloop_2020,bonisch_analytic_2021,bonisch_feynman_2022,lairez_algorithms_2023}. In these situations, resonance also appears, and one can interpret this phenomenon within our more general setting and investigate generic resonances in $D$.

If $D$ is allowed to take integer values, then we can find resonance conditions on facets other than edge facets. In most cases, the polytope of a Feynman integral will have facets $F_{\cU}$ and $F_{\cF}$ corresponding to the polynomials $\cU$ and $\cF$. In this work, we denote them in analogy with the edge facets, as seen in the bubble and triangle examples. That is, the facet $F_{\cU}$ {\em does not} contain the vertices corresponding to the monomials in $\cU$, so that its corresponding Euler operator precisely {\em does} pick up the terms associated to $\cU$. Its reduced subsystem is the one where the $\cU$ terms are absent.

The facet $F_\cF$ is resonant if $LD/2-\nu \in \mathbb{Z}.$ The parameters in the span of the facet satisfy $LD/2-\nu=0$. The
subsystem is generated by the integral where the exponent of $\cF$ is 0 in the Feynman parameter representation. This subsystem has no kinematic dependence. Already for 1-loop integrals, the reduction operators associated to $F_\cF$ are second-order.

The facet $F_\cU$ is resonant if $(L+1)D/2-\nu \in \mathbb{Z}.$ The parameters in the span of the facet satisfy $(L+1)D/2-\nu=0$. The
subsystem is generated by the integral where the exponent of $\cU$ is 0 in the Feynman parameter representation. We can find the reduction operators explicitly as follows. 
Since the $\cU$ polynomial is defined as
\begin{equation}
    \cU = \sum_{T \in \mathcal{T}_1} \prod_{\{f \notin T\}} x_f\,,
\end{equation}
where $\mathcal{T}_1$ denotes the set of spanning trees of the graph, 
we see that the sets $\{f \notin T\}$ can be read as indices of the GKZ variables $z_J$ in the polynomial $\cG(z,x)$.
Then the Euler operator for the face is
\begin{equation}
    E_{\cU} = \sum_{T \in \mathcal{T}_1} z_{\{f \notin T\}} \partial_{\{f \notin T\}}\,.
\end{equation}
The reduction operators associated to $\cU$ are simply the mass derivatives $\partial_{m_e^2}$. To see this, note that in terms of the GKZ variables, we can write
\begin{equation}
    \partial_{m_e^2} = \sum_J \frac{\partial z_J}{\partial m_e^2} \partial_{z_J} 
    = \sum_{T \in \mathcal{T}_1} \partial_{\{f \notin T\}\cup\{e\}}\,,
\end{equation}
where $J$ runs over all the GKZ indices, since the coefficient of $m_e^2$ in $\cG(z,x)$ is precisely $x_e\cU $.
We see that we can solve for the reduction operator $Q^\cU$ in 
\begin{equation}
    \partial_{Je} E_\cU \simeq Q^\cU_{Je,J} \partial_J
\end{equation}
with
\begin{equation}
    Q^\cU_e = \sum_{T \in \mathcal{T}_1} z_{\{f \notin T\}} \partial_{\{f \notin T\} \cup \{e\}}\,,
\end{equation}
which is independent of the choice of $J$.
In the physical restriction $z_{\{f \notin T\}} \to 1$, we recover the mass derivative.
The mass derivatives $\partial_{m_e}^2$ are well known as ``raising operators'': they shift the parameters by $\nu_e \to \nu_e+1$, leaving all others unchanged. This index shift, which is observed directly in the momentum representation, is thus also manifest from the GKZ context.

\subsection{Reduction using dimension shifts} \label{reduction-with-dimensionshifts}

In this section, we observe that a shift of the reduction operators in $\nu_e$ can be turned into a shift in $D$ through a suitable rescaling. This indicates the existence of alternative reduction ladders that incorporate dimension shifts.

\paragraph{Dimension shifts for the bubble.} The simplest case where such a shifting applies is the bubble, where we observed in \eqref{Q-bubble} that the operators $\cQ^e$ shift only the $\nu_e$, keeping $D$ fixed. Introducing $\widehat \cQ_1 = z_2 \cQ_{2,1}$ and $\widehat \cQ_2 = z_1 \cQ_{1,2}$, we realize from \eqref{deg-zdz} and \eqref{order-degr-Q} that $\text{deg}_{f}(\widehat \cQ_e) = (a_{e})_{f}$. Acting on a GKZ solution $I_{\rm bub}^D(z;1,1)$, we conclude that $f_e = \widehat \cQ_eI_{\rm bub}^D(z;1,1)$ must be a tadpole integral in dimension $D-2$, since $\cQ_{e',e}$ reduces to the tadpole topology and the $\cA$-degree of a solution has only shifted $D$.\footnote{Note that there are no toric operators associated to $\cA_F$ after deleting the row of zeros.} Explicitly, one finds that 
\beq
   \widehat \cQ_e I_{\rm bub}^D(z;1,1) = - I_{\rm tad}^{D-2}(z;1)\ .
\eeq
The prefactor can be derived by rewriting the tadpole integral \eqref{eq:tadpolegkzint}. For example, we have  
\begin{equation}
\begin{split}
    z_2x_2 \big( z_2 x_2+z_{2,2} x_2^2\big)^{-D/2}&= \frac{2}{2-D}\theta_2  \big( z_2 x_2+z_{2,2} x_2^2\big)^{1-D/2}\\
    &= \frac{2}{2-D}(2\cE_0-\cE_2)\big( z_2 x_2+z_{2,2} x_2^2\big)^{1-D/2}\,,
\end{split}
\end{equation}
where $\cE_0,\cE_2$ are the Euler operators whose action on the integral is readily evaluated. 

\paragraph{Triangle and general one-loop diagrams.} 
The translation of a $\nu_e$-shift to a shift in $D$ generalizes to all one-loop diagrams with generic kinematics. To see this, let us define an operator
\begin{equation}\label{eq:triangQedef}
    \widehat \cQ_e= \sum_{e'\neq e} z_{e'}\cQ_{e',e}
\end{equation}
for each edge. We now compute the $\cA$-degree of this new operator
\beq \label{d_f=exp}
   \text{deg}_{f}(z_{e'} \cQ_{e',e}) =   \text{deg}_f(z_{e'})+\text{deg}_f(\cQ_{e',e}) =  (a_{e})_{f} = \left(\!\!\begin{array}{c}1\\ \delta_e\end{array}\!\!\right)_f\ ,
\eeq
where we have used \eqref{deg-zdz}, \eqref{order-degr-Q}, and \eqref{genA-oneloop} to give the unit vector $\delta_e$. This implies that all summands in \eqref{eq:triangQedef} have the same $\cA$-degree, and we infer that $\text{deg}_{f}(\widehat \cQ_e) =(a_{e})_{f}$. 
We can now investigate the action of $\widehat \cQ_e$ on the GKZ integrals. A key difference between $\widehat \cQ_e$ and $\cQ_{e',e}$ is that the former do not commute with the 
toric operators. However, the obstruction arises only from derivatives not supported on the face and we find 
\beq \label{commutes-with-LF}
   [\widehat \cQ_e, \cL_{u',v'}^{F_e}] \simeq 0\ , 
\eeq
which holds nontrivially for the sum \eqref{eq:triangQedef} and its summands. 
Let us now assume that we have reduced the integral 
$I^D_{\mathbf{1\cdots n}}(z,\nu)$ to the face $F_e$ using $\widehat\cQ_e$ 
with the general formula \eqref{red-on-1-loop}.
This happens by a single action with $\widehat\cQ_e$ if $\nu_e =1$, e.g.~if we consider 
\beq
  f_e(z):= \widehat \cQ_e I^D_{\mathbf{1\cdots n}}(z;1,1,...,1)\ .
\eeq
We can now inquire about the integral form of $f_e(z)$. 
Using \eqref{commutes-with-LF} we infer that, since $I^D_{\mathbf{1\cdots n}}$ satisfies the full set of toric equations \eqref{toric_GKZ}, $f(z)$ satisfies the toric equations $\cL_{u',v'}^F f_e = 0$. Furthermore, we conclude from \eqref{d_f=exp} that 
\beq
   f_e(z)\ \text{is a solution of the GKZ system}\ \Big(\cA_{F_e}, \nu = \big(\tfrac{D-2}{2},1,\ldots,1 \big) \Big)\ . 
\eeq
The prefactor of $f_e(z)$ compared to the unshifted integrals 
$I^{D-2}_{\mathbf{1\cdots \hat e \cdots n}}(z;1,1,...,1)$ in dimension $D-2$ can be found by direct computation.

\paragraph{General construction of dimension shift operators.} The preceding one-loop examples, together with our construction in section~\ref{sec:higherloops} for the sunrise/banana graphs, indicate that there should exist a natural family of operators that shift the dimension of the integral and a single parameter $\nu_e$. These operators are obtained by compensating the $\cA u$ shift induced by a reduction operator $\cQ_{u,e}$ associated with a face $F_e$, using an appropriate power of $z_i$. In what follows we outline a proposed construction of such operators; however, establishing their existence in full generality and verifying all their desired properties is deferred to future work.

Consider a resonant edge face $F_e$. Associated to this face we introduce vectors $u,u' \in \bbZ^N_{\geq 0}$ via the set
\beq
   \cS^{(e)} = \bigg\{ (u,u'): \cA u'-\cA u + a_e = \left(\!\!\begin{array}{c}1\\ \delta_e\end{array}\!\!\right), \  \cQ_{u,e}\ \text{exists}\bigg\}\ ,
\eeq
where $\delta_e$ is the unit vector with a $1$ in the $e$-th position. 
We then define the operators 
\beq
   \widehat \cQ_e = \sum_{\{(u,u')\} \subset \cS^{(e)}} z^{u'} \cQ_{u,e}\ , 
\eeq
where the sum runs over an appropriate subset of $\cS^{(e)}$ (or possibly the whole set). By construction, all these operators 
have $\cA$-degree
$\text{deg}( \widehat \cQ_e) = (1,\delta_e)$, 
and therefore shift the dimension $D \rightarrow D-2$ and only the $e$-th entry $\nu_e\rightarrow \nu_e-1$. Applied to $I_G^{D}(z;1,...,1)$, they can be used to generate lower-dimensional integrals, which is precisely what we have exploited in \eqref{eq:triangQedef} and in \eqref{widehatQ-banana} for one-loop and banana integrals. In these examples we found a nontrivial set $\cS^{(e)}$, and we were able to show that $[\widehat \cQ_e,\cL^{F_e}_{u',v'}] \simeq  0$ which ensures that the reduced integral is again a GKZ solution  but now on the face $F_e$. It would be desirable to establish this in general. Furthermore, we have seen in our examples that the multiplicative factors $z^{u'}$ always reduce to $1$ on the physical locus, and it would be interesting to check if there always exist such preferred choices.

\section{Summary and Outlook}
\label{sec:conclusions}

We have seen that GKZ systems obtained from Feynman integrals in their Lee-Pomeransky representation typically exhibit resonance associated to the edges of the Feynman graph. 
This is a property of Feynman integrals whose propagators are raised to integer powers.
Based on mathematical results that resonance of GKZ systems implies their reducibility, we showed that if  differential operators can be obtained from GKZ operators according to the relation 
\eqref{eq:PEFsimQd}, then they 
have the graphical effect of contracting edges while shifting other parameters, as shown in equation \eqref{red-on-solutions2}.
By repeated application of reduction operators together with one additional Euler relation, one obtains a system of differential equations called a reduction ladder. 

For one-loop integrals with generic kinematics, there exist linear reduction operators for every pair of edges, and it is straightforward to implement a restriction of GKZ variables to the physical locus. The bubble and triangle examples show how the reduction ladder becomes a closed system. The remaining algebraic relations among the integrals in the ladder are encoded in the toric relations and are thus seen to arise combinatorially from GKZ data. 

For multi-loop sunrise and banana graphs, we found higher-order reduction operators that contract a given edge and reduce the dimension by two. Edge contractions result in multi-tadpole integrals which are particularly simple; however, the physical restriction is more subtle for this class of integrals. We observed that it should be possible to construct dimension-shift versions of reduction operators in general.

A natural next step is to develop algorithms for the present construction. Given a Feynman graph, it is easy to determine its resonant faces. For each resonant face, there is an algorithm to produce one reduction operator. However, we have not outlined a clear method for identifying all possible reduction operators satisfying \eqref{red-on-solutions2}.  From the reduction operators, we would further like an algorithm to construct a reduction ladder and a first-order system. We expect that the remaining algebraic relations can be found from GKZ data, but it may turn out that the most efficient approach combines reduction operators with differential equations obtained by other methods such as IBP. For Feynman integrals, it will be important to be able to achieve a systematic restriction to the physical locus, as well as to handle special kinematics such as vanishing or equal masses.

Conceptually, it would be interesting to explore the full range of reduction operators for a given Feynman integral. Here we have focused on  reduction operators that lower individual propagator powers by 1, but there are many other possibilities for solving the relation \eqref{eq:PEFsimQd}. For example, even for edge facets, one could explore higher-order reduction operators that lower the propagator powers by larger increments and work out their algebra. One could go further and include lower-dimensional resonant faces for consideration, such as intersections of edge facets with stronger resonance conditions. 
As we have seen, there are even more possibilities when the dimension takes integer values. 

Reduction operators give a new perspective on cuts of Feynman integrals, by realizing the idea that when contractions and cuts are applied to the same edge, the result is zero. We have now seen this connection in the parametric representation, with cuts identified by specific integration chains omitting the hyperplane boundaries in the variables of the cut edges, and contractions implemented by reduction operators. The result is a system of differential equations whose solutions include a given cut integral. Just as for uncut integrals, it will be interesting to see if differential approaches to computing cuts can be improved by using reduction operators, both from a practical point of view and in terms of their underlying algebraic structure. The results of this paper already shed light on the observation that the functions involved in cuts are often simpler than in the uncut integrals, and that when they are not (such as in the one-mass sunrise), that fact can be attributed to the absence of resonance. 

Our study of resonance also points to a broader notion of complexity for Feynman integrals. When the GKZ system becomes reducible at special parameter values, the resulting algebraic relations are reflected in the differential equation itself, for instance through invariant subspaces, quotient systems, or a reduction of the effective monodromy and differential Galois group. This connects naturally with existing approaches in which the complexity of tame or Pfaffian functions is anchored in the differential equations that generate them~\cite{gabrielov2004complexity,binyamini_sharply_2022,Grimm:2024elq}. For Feynman integrals, the combinatorial complexity of the defining GKZ data should therefore be part of the story, but resonance must be treated as an additional organizing principle: it identifies loci where the generic solution space decomposes and the effective complexity of the integral family drops. Developing this viewpoint into a systematic theory of amplitude complexity remains an important direction for future work.

\subsection*{Acknowledgments}

We are grateful to Vsevolod Chestnov, Matthew Schwartz, Sid Smith, and Felix Tellander for insightful discussions and valuable comments. In the later stages of this work, some computations  were performed using OpenAI's ChatGPT Pro~5.2--5.5 and verified by the authors.\footnote{Equations \eqref{explicit-dI} and \eqref{algebraic-relation1} were initially produced by ChatGPT. It provided 
checks of several other results and general proofreading.}  The work of RB is funded by the European Union (ERC, MaScAmp, 101167287). Views and opinions expressed are however those of the author(s) only and do not necessarily reflect those of the European Union or the European Research Council Executive Agency. Neither the European Union nor the granting authority can be held responsible for them.  The research of TG and AH is supported, in part, by the Dutch Research
Council (NWO) via a Vici grant.

\appendix

\section{Resonance property for facets}\label{app:Lnutest}
In this appendix we prove that resonance of facets can be tested by applying linear functionals.
For every facet $F$ of an $d\times n$ matrix $\A$, there exists a unique functional $L_F:\Z^d \rightarrow \Z$, called the \emph{primitive integral support function} satisfying
    \begin{itemize}
        \item[(i)] $L_F$ is linear,
        \item[(ii)] $L_F(\Z^{d})=\Z$,
        \item[(iii)] $L_F(\mathbf{a}_i)\ge 0$,
        \item[(iv)] $L_F(\mathbf{a}_i)=0\iff\mathbf{a}_i\in F$.
    \end{itemize}
    
We now want to show that assuming that $\Z \A=\Z^d$ and $F$ is a facet with associated $L_F: \Z^d\rightarrow \Z$, then, for any $\nu \in \C^d$,  $F$ is a resonant face if and only if $L_F(\nu)\in \Z$.

Let us begin by showing that resonance implies $L_F(\nu)\in \Z$. As $\nu$ is resonant for $F$, we have that it can be written as
    \begin{equation}
        \nu = \sum_{i\in F}c_i a_i +\sum_{i=1}^n n_i a_i
    \end{equation}
    with $c_i \in \C$ and $n_i \in \Z$. Applying $L_F$ to this equation we find that
    \begin{equation}
        L_F(\nu)= \sum_{i=1}^n n_i L_F(a_i)
    \end{equation}
    as $L_F$ is linear and $L_F(a_i)=0$ for $i\in F$. $L_F$ maps $\Z^d$ to $\Z$, therefore, the right hand side is an integer.

    For the converse, let $L_F(\nu)=n$ for some $n\in \Z$. As $F$ is a facet, its primitive integral support function has $L_F(\Z^d)=\Z$, and, as $\Z A= \Z^d$, this implies we can find integers $n_i \in \Z$ such that $L_F(\sum_i n_i a_i)=n$. By linearity of $L_F$, we find that
    \begin{equation}
        L_F(\nu-\sum_i n_i a_i)=0
    \end{equation}
    implying that $\nu-\sum_i n_ia_i$ is in the complex span of $F$. As this corresponds to a shift of $\nu$ by integer multiples of the column vectors of $A$, this implies that $F$ is a resonant face for $\nu$.

\section{Tadpoles and bubbles}\label{ap:tadbub}

We list here some results of integrating one-loop tadpole and bubble integrals, together with their cuts. In order to obtain readily recognizable hypergeometric functions, we have chosen to integrate them in the slightly simpler Feynman-parameter representation \begin{align}
    \label{eq:FP-gen}
I^D_G(p,m,\nu) = \frac{\Gamma\left(\nu-\frac{LD}{2}\right)}{\prod_{e=1}^{N_e} \Gamma(\nu_e)}
\int_{x_e \geq 0}\!\! \rd^{N_e} x_e\, 
\delta\Big(1-\sum_{e =1}^{N_e} x_e\Big)
\Big(\prod_{e=1}^{N_e} x_e^{\nu_e-1} \Big)
\frac{\left[\mathcal{U}(x)\right]^{\nu-(L+1)\frac{D}{2}}}
{\left[\mathcal{F}(p,m,x)\right]^{\nu-L \frac{D}{2}}}\,,
\end{align}
shown below, rather than the Lee-Pomeransky representation in the main text, but the results are fully equivalent. The cuts are obtained from parametric contours, as described in \cite{britto_generalized_2023} and shown explicitly below.\footnote{Our convention here differs from \cite{britto_generalized_2023} in that it does not include a prefactor from taking a discontinuity. We take an analytic continuation of the same integrand with a new (unscaled) contour associated to the set of cut propagators.}
\paragraph{Tadpole.} The general Feynman integral for the tadpole in $D$ dimensions evaluates to
\begin{align}
    \label{eq:tadfunction}
      I^D_{\textrm{tad}}(m_1^2;\nu_1) 
      &=\frac{ \Gamma \left( \nu_1-\frac{D}{2} \right) }{\Gamma( \nu_1 ) } \int_0^\infty \rd x\, \delta(1-x) x^{\nu_1}
\left(x^2 m_1^2 \right)^{\frac{D}{2}-\nu_1 } \\
&= \frac{ \Gamma \left( \nu_1-\frac{D}{2} \right) }{ \Gamma( \nu_1 ) }\left(m_1^2 \right)^{\frac{D}{2}-\nu_1 }\ .
\end{align}
Since the curve ${\mathcal F}=0$ is identical to the coordinate hyperplane integration boundary at $x=0$, we declare that in our present convention for cuts,  the cut tadpole is identical to the uncut tadpole,
\begin{equation}
    \mathcal{C}_1 I^D_{\textrm{tad}}(m_1^2;\nu_1) = I^D_{\textrm{tad}}(m_1^2;\nu_1) \,.
\end{equation}

\paragraph{Generic bubble.}
We take $x=x_2$ and define $w, \bar{w}$ such that $\bar{w} \leq w$ and 
\begin{equation}
w\bw = \frac{m_1^2}{p^2}, \quad (1-w)(1-\bw)= \frac{m_2^2}{p^2}\,.
\end{equation}
For brevity of presentation, we now suspend multi-index notation and use $$\nu=\nu_1+\nu_2.$$
Then we find the following results, given in terms of Appell's function $F_1$, equivalent to Lauricella's $F_D^{(2)}$.
These expressions are valid in the Euclidean region ($p^2<0,~m_i^2>0$) for the uncut integral, and for $0<\bw<w<1$ for the cut integrals (positive invariants for the 1-line cuts, negative invariants for the 2-line cut). They may be analytically continued to other kinematic regions of interest. Explicitly, one finds:
\begin{align}
    \label{eq:bubfunctions}
     & I^D_{\textrm{bub}}(p^2,m_1^2,m_2^2;\nu_1,\nu_2) \nonumber \\
      &=\frac{ \Gamma \left( \nu-\frac{D}{2} \right) }{\Gamma(\nu_1)\Gamma(\nu_2) } (-p^2)^{\frac{D}{2}-\nu}\int_0^1 \rd x\,  (1-x)^{\nu_1-1} x^{\nu_2-1}
\left[ (w-x)(x-\bw) \right]^{\frac{D}{2}-\nu } \\
&= (m_1^2)^{\frac{D}{2}-\nu }
\frac{ \Gamma \left( \nu-\frac{D}{2} \right) }{\Gamma(\nu) }
F_1\left(\nu_2,\nu-\tfrac{D}{2},\nu-\tfrac{D}{2},\nu;\tfrac{1}{w},\tfrac{1}{\bw}
\right)\,, \nonumber \\[.5cm]
& \mathcal{C}_1    I^D_{\textrm{bub}}(p^2,m_1^2,m_2^2;\nu_1,\nu_2) \nonumber \\
      &=\frac{ \Gamma \left( \nu-\frac{D}{2} \right) }{\Gamma(\nu_1)\Gamma(\nu_2) } (p^2)^{\frac{D}{2}-\nu}\int_0^{\bw} \rd x\,  (1-x)^{\nu_1-1} x^{\nu_2-1}
\left[ (w-x)(\bw-x) \right]^{\frac{D}{2}-\nu } \\
&= \frac{ \Gamma \left( \nu-\frac{D}{2} \right) \Gamma \left(\frac{D}{2}-\nu+1 \right) }{\Gamma(\nu_1)\Gamma(\frac{D}{2}-\nu_1+1) }
(m_1^2)^{\frac{D}{2}-\nu } \bw^{\nu_2} 
%\\
%& \qquad
F_1\left(\nu_2,1-\nu_1,
\nu-\tfrac{D}{2},\tfrac{D}{2}-\nu_1+1;\bw,\tfrac{\bw}{w}
\right)\,,\nonumber \\[.5cm]
& \mathcal{C}_2    I^D_{\textrm{bub}}(p^2,m_1^2,m_2^2;\nu_1,\nu_2) \nonumber \\
      &=\frac{ \Gamma \left( \nu-\frac{D}{2} \right) }{\Gamma(\nu_1)\Gamma(\nu_2) } (p^2)^{\frac{D}{2}-\nu}\int_w^1 \rd x\,  (1-x)^{\nu_1-1} x^{\nu_2-1}
\left[ (x-w)(\bw-x) \right]^{\frac{D}{2}-\nu } \nonumber \\
&= \frac{ \Gamma \left( \nu-\frac{D}{2} \right) \Gamma \left(\frac{D}{2}-\nu+1 \right) }{\Gamma(\nu_2)\Gamma(\frac{D}{2}-\nu_2+1) }
(m_2^2)^{\frac{D}{2}-\nu } (1-w)^{\nu_1} 
\\
& \qquad
F_1\left(\nu_1,1-\nu_2,
\nu-\tfrac{D}{2},\tfrac{D}{2}-\nu_2+1;1-w,\tfrac{1-w}{1-\bw}
\right)\,, \nonumber \\[.5cm]
& \mathcal{C}_{12}    I^D_{\textrm{bub}}(p^2,m_1^2,m_2^2;\nu_1,\nu_2) \nonumber \\
      &=\frac{ \Gamma \left( \nu-\frac{D}{2} \right) }{\Gamma(\nu_1)\Gamma(\nu_2) } (-p^2)^{\frac{D}{2}-\nu}\int_{\bw}^w \rd x\,  (1-x)^{\nu_1-1} x^{\nu_2-1}
\left[ (w-x)(x-\bw) \right]^{\frac{D}{2}-\nu } \nonumber\\
&= \frac{ \Gamma \left( \nu-\frac{D}{2} \right) \Gamma^2 \left(\frac{D}{2}-\nu+1 \right) }{\Gamma(\nu_1)\Gamma(\nu_2)\Gamma(D-2\nu+2) }
(-p^2)^{\frac{D}{2}-\nu } 
(w-\bw)^{D-2\nu+1} \bw^{\nu_2-1} (1-\bw)^{\nu_1-1}
\nonumber
\\
& \qquad
F_1\left(
\tfrac{D}{2}-\nu+1,
1-\nu_1,1-\nu_2,
D-2\nu+2;\tfrac{w-\bw}{1-\bw},1-\tfrac{w}{\bw}
\right)\,.
\end{align}
Observe that the cut integrals $\mathcal{C}_i I^D_{\textrm{bub}}$ and $\mathcal{C}_{12} I^D_{\textrm{bub}}$ degenerate to expressions in terms of the simpler function ${}_2F_1$ when $\nu_i=1$, and that $\mathcal{C}_{12} I^D_{\textrm{bub}}$ degenerates further to a power function when both $\nu_1=1$ and $\nu_2=1$.

\paragraph{One-mass bubble.}
In the case $m_2=0$, we can follow the same procedure as for the generic bubble. Where the limit is smooth, we can set 
\begin{equation}
    \bw = \frac{m_1^2}{p^2}\,, \quad w=1\,,
\end{equation}
in the expressions above. The results are:  
\begin{align}
    \label{eq:1mbubfunctions}
     & I^D_{\textrm{1m-bub}}(p^2,m_1^2,0;\nu_1,\nu_2) \nonumber \\
      &=\frac{ \Gamma \left( \nu-\frac{D}{2} \right) }{\Gamma(\nu_1)\Gamma(\nu_2) } (m_1^2)^{\frac{D}{2}-\nu}\int_0^1 \rd x\,  (1-x)^{\frac{D}{2}-\nu_2-1} x^{\nu_2-1}
\left[ 1-\frac{p^2}{m_1^2}x \right]^{\frac{D}{2}-\nu }  \\
&= \frac{ \Gamma \left(\nu-\frac{D}{2} \right)\Gamma \left(\frac{D}{2}-\nu_2 \right) }{\Gamma(\nu_1)\Gamma\left(\frac{D}{2}\right) }
(m_1^2)^{\frac{D}{2}-\nu}
{}_2F_1\left(\nu-\tfrac{D}{2},\nu_2,\tfrac{D}{2};\tfrac{p^2}{m_1^2}
\right)\,, \nonumber \\[.5cm]
     & \mathcal{C}_1 I^D_{\textrm{1m-bub}}(p^2,m_1^2,0;\nu_1,\nu_2) \nonumber \\
      &=\frac{ \Gamma \left( \nu-\frac{D}{2} \right) }{\Gamma(\nu_1)\Gamma(\nu_2) } (m_1^2)^{\frac{D}{2}-\nu}\int_0^{\frac{m_1^2}{p^2}} \rd x\,  (1-x)^{\frac{D}{2}-\nu_2-1} x^{\nu_2-1}
\left[ 1-\frac{p^2}{m_1^2}x \right]^{\frac{D}{2}-\nu } \\
&= \frac{ \Gamma \left(\nu-\frac{D}{2} \right)\Gamma \left(\frac{D}{2}-\nu+1 \right) }{\Gamma(\nu_1)\Gamma\left(\frac{D}{2}-\nu_1+1 \right) }
(m_1^2)^{\frac{D}{2}-\nu_1} {(p^2)}^{-\nu_2}
{}_2F_1\!\left(\nu_2,\nu_2-\tfrac{D}{2}+1,\tfrac{D}{2}-\nu_1+1;\tfrac{m_1^2}{p^2}
\right)\!, \nonumber \\[.5cm]
&\mathcal{C}_2 I^D_{\textrm{1m-bub}}(p^2,m_1^2,0;\nu_1,\nu_2) = 0\,,\\[.5cm]
     & \mathcal{C}_{12} I^D_{\textrm{1m-bub}}(p^2,m_1^2,0;\nu_1,\nu_2) \nonumber \\
      &=\frac{ \Gamma \left( \nu-\frac{D}{2} \right) }{\Gamma(\nu_1)\Gamma(\nu_2) } (m_1^2)^{\frac{D}{2}-\nu}\int_{\frac{m_1^2}{p^2}}^1 \rd x\,  (1-x)^{\frac{D}{2}-\nu_2-1} x^{\nu_2-1}
\left[ 1-\frac{p^2}{m_1^2}x \right]^{\frac{D}{2}-\nu } \nonumber \\
&= 
\frac{ \Gamma \left(\nu-\frac{D}{2} \right)\Gamma \left(\frac{D}{2}-\nu +1 \right)\Gamma\left(\frac{D}{2}-\nu_2 \right) }{\Gamma(\nu_1)\Gamma(\nu_2)\Gamma\left(D-\nu_1-2\nu_2+1\right)}
(-p^2)^{1-\frac{D}{2}} (-m_1^2)^{\nu_2-1} (p^2-m_1^2)^{D-\nu_1-2\nu_2}\nonumber \\
& \quad \quad
{}_2F_1\left(1-\nu_2,\tfrac{D}{2}-\nu+1,D-\nu_1-2\nu_2+1;1-\tfrac{p^2}{m_1^2}
\right)\,.
\end{align}
Observe that  $\mathcal{C}_{12} I^D_{\textrm{1m-bub}}$ degenerates to a power function when either $\nu_1=1$ or $\nu_2=1$, but there is no such degeneration for 
 $\mathcal{C}_1 I^D_{\textrm{1m-bub}}$. This property can be explained by the absence of a reduction operator that could contract edge 1.

\section{Resonance and reducibility for the Gauss hypergeometric function}\label{ap:2f1}

We consider Gauss's hypergeometric function as a simple example of a GKZ system that is different from a Feynman integral, to see  how various values of the parameters in  ${}_2F_1(a,b;c;z)$ result in different manifestations of resonance, and to understand the construction and application of the reduction operators.

\begin{figure}[t]
\begin{center}
\begin{tikzpicture}[baseline={([yshift=-.5ex]current bounding box.center)}]
\node[label={below left:$a_2=(1,0,0)$}] (a2) at (0,0)[circle,draw,fill]{};
\node[label={below right:$a_3=(1,1,0)$}] (a3) at (4,0)[circle,draw,fill]{};
\node[label={above right:$a_1=(1,1,1)$}] (a1) at (4,4)[circle,draw,fill]{};
\node[label={above left:$a_4=(1,0,1)$}] (a4) at (0,4)[circle,draw,fill]{};
\draw (a1) --node[above right,text=blue]{$c-b \in \mathbb{Z}$} node[below right,text=purple]{$c-b=1$}(a3);
\draw (a1) -- node[above,text=blue]{$c-a \in \mathbb{Z}$}node[below,text=purple]{$c-a =1$}(a4);
\draw (a2) -- node[above,text=blue]{$b \in \mathbb{Z}$}node[below,text=purple]{$b =0$}(a3);
\draw (a2) -- node[above left,text=blue]{$a \in \mathbb{Z}$} node[below left,text=purple]{$a=0$}(a4);
\end{tikzpicture}
\end{center}
\caption{Resonance conditions for the facets of ${\rm Conv}(\A)$ from the matrix defined in \eqref{eq:2F1-matrix}, and the parameters lying in the span of each facet.}
\label{fig:2F1}
\end{figure}
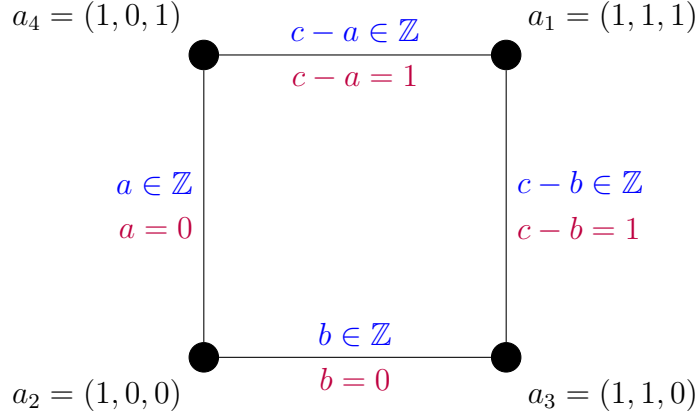

Let us take the GKZ system 
    \begin{equation}
        \label{eq:2F1-matrix}\A =
        \begin{pmatrix}
            1 & 1 & 1 & 1   \\
            1 & 0 & 1 & 0   \\
            1 & 0 & 0 & 1   
        \end{pmatrix}\,,
        \qquad \nu=
        \begin{pmatrix}
           1+a+b-c \\
            a   \\
            b   
        \end{pmatrix}\ .
    \end{equation}
This system is based on the one given in  \cite{stienstra_gkz_2005}, although we have performed row operations in order to make the first row consist of 1's, leading to the following integral representation,
\begin{equation}\label{eq:2F1-gkz-int}
    I(z;a,b,c) = \int_\Gamma \rd x_1 \rd x_2\, x_1^{a-1} x_2^{b-1}p(z,x)^{c-a-b-1}\,,
\end{equation}
where 
\begin{equation}
  p(z,x)=  z_1 x_1 x_2+z_2+z_3 x_1 + z_4 x_2\,.
\end{equation}
We leave the integration contour unspecified, in order to allow for all possible solutions to the system. For any given resonance condition, certain possible boundaries of the contours will become relevant.
Figure \ref{fig:2F1} indicates the  conditions for each of the facets to be resonant for a given $\nu$, along with the specific case in which the parameter $\nu$ is contained in the span of each facet. 
The linear functionals and Euler operators for each facet are given as follows.
\begin{align}
& L_{\{1,3\}}=(-1,0,1), ~~ L_{\{1,4\}}=(-1,1,0), ~~  L_{\{2,3\}}=(0,0,1), ~~  L_{\{2,4\}}=(0,1,0)\,.
\\ & \cE_{\{1,3\}} = \theta_2+\theta_4, \quad 
\cE_{\{1,4\}} = \theta_2+\theta_3, \quad
\cE_{\{2,3\}} = \theta_1+\theta_4, \quad 
\cE_{\{2,4\}} = \theta_1+\theta_3\,.
\end{align}

\paragraph{Construction of reduction operators.}  

Let us begin with  the facet $F=\{2,3\}$. We can solve
the equation $\partial^u \cE_{\{2,3\}} \simeq \cQ^{\{2,3\}}_{u,v} \partial^v$ for 
$u=a_3, v=a_1$ or $u=a_2, v=a_4$ with equivalent results for the reduction operator,
\begin{equation}\label{eq:Q-2F1-23}
    \cQ^{\{2,3\}} = z_1 \partial_3 + z_4 \partial_2\,.
\end{equation}
Similarly, we find
\begin{equation}\label{eq:Q-2F1-13}
    \cQ^{\{1,3\}} = z_2 \partial_3 + z_4 \partial_1\,.
\end{equation}
By the symmetry of the GKZ system under the exchange $3 \leftrightarrow 4,~a \leftrightarrow b$, we see that the reduction operators for the other two facets are
\begin{align}
 \label{eq:Q-2F1-14}   \cQ^{\{1,4\}} &= z_2 \partial_4 + z_3 \partial_1\,, \\
\label{eq:Q-2F1-24}    \cQ^{\{2,4\}} &= z_1 \partial_4 + z_3 \partial_2\,.
\end{align}

\paragraph{Restriction of variables.}
The Gauss hypergeometric function can be recovered from the GKZ system by the restriction 
\begin{equation}\label{eq:2F1-restriction}
    z_1=z, \quad z_2=1,\quad  z_3=z_4=-1 \,.
\end{equation} 
Using the Euler relations to substitute $\partial_2,~\partial_3,~\partial_4$ with their expressions in terms of $\partial_1$,
we find the following restrictions of the reduction operators.
\begin{align}\begin{split}\label{eq:Q-2F1-z}
    \cQ^{\{1,3\}}_{a,b,c} &= a-(1-z)\partial_z\,, \\
    \cQ^{\{1,4\}}_{a,b,c} &= b-(1-z)\partial_z\,, \\
    \cQ^{\{2,3\}}_{a,b,c} &= 1-c+az-(z-z^2)\partial_z\,, \\
    \cQ^{\{2,4\}}_{a,b,c} &= 1-c+bz-(z-z^2)\partial_z \,.
\end{split}\end{align}
One can check that the integrand of (\ref{eq:2F1-gkz-int}), after implementing the restriction, is a solution to the second-order differential operator defining the Gauss hypergeometric function,
\begin{equation}\label{eq:2F1-PA}
    P^{\A}_{a,b,c} = 
    z(z-1) \partial_z^2 + \left[(a+b+1)z-c\right] \partial_z + a b\,.
\end{equation}
Therefore, the restriction of the integral in (\ref{eq:2F1-gkz-int}) is a linear combination of the functions ${}_2F_1(a,b,c;z)$ and $z^{1-c}{}_2F_1(1+a-c,1+b-c,2-c;z)$ or their analytic continuations. For example, we find
\begin{align}\label{eq:2F1-from-gkz}
    {}_2F_1(a,b,c;z) =& \frac{\Gamma(c)}{\Gamma(a)\Gamma(b)\Gamma(c-a-b)} \\
    & \nonumber
  \int_0^1 \rd x_2 \int_0^{\frac{x_2-1}{z x_2-1}} \rd x_1\, x_1^{a-1} x_2^{b-1}\left(z x_1 x_2 +1 - x_1 - x_2\right)^{c-a-b-1}\,.
\end{align}

\paragraph{Example.} The subsystem of the facet $F=\{2,3\}$ is given by
\begin{equation} 
A_{\{2,3\}}=
        \begin{pmatrix}
            1 &  1   \\
            0 &  1   \\
            0 &  0   
        \end{pmatrix}\,,
        \qquad \nu=
        \begin{pmatrix}
            1+a+b-c \\
            a   \\
            b   
        \end{pmatrix}\ .
\end{equation}
 The Euler relations are
$
    \theta_2 + \theta_3 + 1+a+b-c=0, ~ \theta_3+a=0,~ b=0,
$
of which the last is the resonance condition in the span of the face.
There are no toric relations.
Thus we have $\theta_2+1-c=0$, and the  solution to the subsystem for the face is  
\begin{equation}
    g_{\{2,3\}} \sim  z_2^{c-1} z_3^{-a} \,.
\end{equation} 
Another way to arrive at the same result is to set up an integral representation from $A_{\{2,3\}}$ and the parameter, after eliminating the null row, to find \begin{equation} g_{\{2,3\}} \sim \int \rd x\,x^{a-1}(z_3 x+z_2)^{c-a-1}\,.
\end{equation}

This same function also arises as the boundary term in the Euler relation for the face (\ref{inhomog-euler1}). From the observation that $\cE_{\{2,3\}} p(z,x) = x_2 \partial_{x_2} p(z,x)$, it follows  that
\begin{equation}
        \left(\cE_{\{2,3\}} +b \right) I(z;a,b,c)  
=
        \int_\Gamma \rd x_1 \rd x_2 \,\partial_{x_2}\left[x_1^{a-1} x_2^b p(z,x)^{c-a-b-1} \right] 
\end{equation}
We have the integral of a total derivative. If $\Gamma$ contains a component at $x_2=0$, then we find $g_{\{2,3\}}$ as a boundary term.

Now let us examine the effect of the reduction operator $\cQ^{\{2,3\}}=z_1 \partial_3 + z_4 \partial_2$. Since its
degree is $(0,0,1)$, its effect will be to shift the parameters by $(a,b,c) \to (a,b-1,c-1)$.
Since $\cQ^{\{2,3\}} p(z,x) = \partial_{x_2} p(z,x)$, it follows that
\begin{align}
        \cQ^{\{2,3\}} I(z;a,b,c)  
=&
        \int_\Gamma \rd x_1 \rd x_2 \,x_1^{a-1} x_2^{b-1} \partial_{x_2}\left[p(z,x)^{c-a-b-1} \right] \\
        =& 
(1-b) I(z;a,b-1,c-1) +\!  \int_{\partial_{x_2}\Gamma}\! \rd x_1 
\,x_1^{a-1} \left[x_2^{b-1}  p(z,x)^{c-a-b-1} \right]
\nonumber
\end{align}
The first term shows the reduction of the value of $b$ by 1, along with the incidental shift of $c$. If $\partial\Gamma$ contains a component at $x_2=0$, then we find $g_{\{2,3\}}(z;a,b-1,c-1)$ as a boundary term. Other boundary terms may be present if it happens that also  $c-a \in \mathbb{Z}$.
Thus we can find a function $h(z;a,1,c)$ satisfying 
\begin{equation}
\cQ^{\{2,3\}}    h(z;a,1,c) = 0\,,
\end{equation}
by setting $b=1$ in (\ref{eq:2F1-gkz-int}) and selecting an integration contour avoiding $x_2=0$, for example 
\begin{equation}
h(z;a,1,c)    = \int_{-\frac{z_2}{z_4}}^\infty \rd x_2 \int_0^{p(z,x)=0} \rd x_1 \, x_1^{a-1} p(z,x)^{c-a-2}\,.
\end{equation}

In the restriction (\ref{eq:2F1-restriction}), the function $h(z;a,1,c)$ annihilated by $\cQ^{\{2,3\}}_{a,1,c}$ is
\begin{equation}\label{eq:2F1-23-h}
    h(z;a,1,c)  \sim (1-z)^{c-a-1} z^{1-c}.
\end{equation}
The solution to the face subsystem is 
\begin{equation}
    g_{\{2,3\}}  \sim {}_2F_1(a,0,c;z) = {\rm constant}.
\end{equation}
From the fact that the variable $z_1$ is eliminated from the subsystem, we may infer that the restricted differential operator is
\begin{equation}
    P^{\{2,3\}}_{a,0,c}=\partial_z.
\end{equation}
A consequence of the defining relation \eqref{eq:PEFsimQd} for reduction operators is that the functions annihilated by $\cQ^F_{u,v}$ can be obtained by acting with $\partial^v$ on the solutions of the full system where the parameter lies in the span of $F$. In this example, 
the general solution to 
the hypergeometric equation with $b=0$ is 
$k_1+k_2 z^{1-c}{}_2F_1(1+a-c,1-c,2-c;z)$ for constants $k_1, k_2$, and the result of then applying $\partial_z$ is indeed $k_2 h(z;a,1,c)$.

In this simple example where both subsystems have rank 1, we observe the factorization of the GKZ system at resonance, taking the form 
\begin{equation}
    P^{\cA}_{a,1,c} = P^{\{2,3\}}_{a,0,c-1}\cQ^{\{2,3\}}_{a,1,c}\,,
\end{equation}
which, in the restriction, can be checked by noting that
\begin{equation}
 z(z-1) \partial_z^2 + \left[(a+2)z-c\right] \partial_z + a 
=
\partial_z\left[ 1-c+az-(z-z^2)\partial_z\right]\,.
\end{equation}
Solutions to $\cQ^{\{2,3\}}_{a,1,c}=0$ are automatically solutions to the full system $P^{\cA}_{a,1,c}=0$, and the general solution to $P^{\cA}_{a,1,c}=0$ also includes functions of the form   $\cQ^{\{2,3\}}_{a,1,c}f$.

\bibliographystyle{utphys}
\bibliography{references}
\end{document}